\documentclass[aps,prapp,onecolumn,reprint]{revtex4-2}
\usepackage{amsmath}
\usepackage{amssymb}
\usepackage{mathrsfs}
\usepackage{graphicx}
\usepackage{booktabs}  
\usepackage{multirow}  
\usepackage[colorlinks,citecolor=blue, linkcolor=blue,hyperindex,CJKbookmarks]{hyperref}

\begin{document}

\title{Thermal noise cancellation for optomechanically induced
nonreciprocity in a whispering-gallery-mode microresonator}
\author{Zhi-Xiang Tang}
\affiliation{Key Laboratory of Low-Dimensional Quantum Structures and Quantum Control of
Ministry of Education, Key Laboratory for Matter Microstructure and Function
of Hunan Province, Department of Physics and Synergetic Innovation Center
for Quantum Effects and Applications, Hunan Normal University, Changsha
410081, China}
\author{Xun-Wei Xu}
\email{xwxu@hunnu.edu.cn}
\affiliation{Key Laboratory of Low-Dimensional Quantum Structures and Quantum Control of
Ministry of Education, Key Laboratory for Matter Microstructure and Function
of Hunan Province, Department of Physics and Synergetic Innovation Center
for Quantum Effects and Applications, Hunan Normal University, Changsha
410081, China}
\date{\today}

\begin{abstract}
Magnetic-free optomechanically induced nonreciprocity may stimulate a wide range of practical applications in quantum technologies.
However, how to suppress the thermal noise flow from the mechanical reservoir is still a difficulty encountered in achieving optomechanically nonreciprocal effects on a few- and even single-photon level.
Here, we show how to realize thermal noise cancellation by quantum interference for optomechanically induced nonreciprocity
in a whispering-gallery-mode (WGM) microresonator.
We find that both nonreciprocal transmission and amplification can be achieved in the WGM microresonator when it coupled to two coupled mechanical resonators.
More interestingly, the thermal noise can be suppressed when the two coupled mechanical resonators couple to a common thermal reservoir.
The thermal noise cancellation is induced by the destructive quantum interference between the two flow paths of the thermal noises from the common reservoir.
The scheme of quantum interference induced thermal noise cancellation can be applied in both sideband resolved and unresolved regimes, even with strong backscattering taken into account.
Our work provide an effective way to achieve nonreciprocal effects on a few- or single-photon level without precooling the mechanical mode to the ground state.
\end{abstract}

\maketitle

\section{Introduction}

Cavity optomechanics that exploring the effect induced by the interaction
between optical and mechanical resonators has been developed very quickly
in the past two decades (for review, see Ref.~\cite{Aspelmeyer2014RMP}), and many new
topics have been raised in recent years, such as parity-time-symmetry optomechanics~\cite{JingH2014PRL,XuXW2015PRA,LiuZP2016PRL,ZhangJ2018NaPho},
topological optomechanical lattices~\cite{Peano2015PRX,QiL2017OExpr,Lemonde2019NJPh,NiX2021Optic,Ren2022NatCo,Doster2022NatCo,Youssefi2021arXiv,XuXW2022FrP}, and
optomechanical induced nonreciprocity~\cite{Manipatruni2009PRL,Hafezi12OE,Metelmann2015PRX,XuXW2015PRA2,LiBJ19PRJ,XuXW20PRJ,NieWJ2022SCPMA}. Nonreciprocal devices are
indispensable elements in both classical and quantum information processing,
for steering signal transmission and reducing backscattering. Different from
the conventional nonreciprocal devices that based on the magneto-optical
effect, optomechanical systems provide us a magnetic-free platform to
achieve nonreciprocity for potential application in designing nonreciprocal
optical devices on chip.

Optomechanical induced nonreciprocity has made remarkable progress in the
past few years. As a typical scheme, optical nonreciprocity was proposed
based on optomechanical induced transparency~\cite{Agarwal2010PRA,Weis2010Sci,Safavi2011Natur} or amplification~\cite{Massel2011Natur,Safavi2011Natur,Hocke2012NJPh} in a
whispering-gallery-mode (WGM) resonator with single direction optical drive~%
\cite{Hafezi12OE}. Now, isolators~\cite%
{Dong2016NaPho,Ruesink2016NatCo}, directional amplifiers~\cite%
{Dong2016NaPho,Ruesink2016NatCo}, and circulators~\cite%
{Shen2018NatCo,Ruesink2018NatCo} have been realized experimentally. In
addition, there are many other mechanisms that are proposed theoretically or
demonstrated experimentally, such as the intrinsic nonlinearity in
optomechanical system~\cite{Manipatruni2009PRL,Qiu17OE,Xu18PRA,Song19PRA},
Brillouin scattering in WGM optomechanical systems~\cite%
{Junhwan2015nphys,Dong2015ncomms}, synthetic magnetism generated by a closed
loop of optical and mechanical modes~\cite{XuXW2015PRA2,Schmidt2015Optic,Fang2017NatPh,Peterson2017PRX,Bernier2017NatCo,Barzanjeh2017NatCo,Metelmann2017PRA,TianL2017PRA,Malz2018PRL,LiGL2018PRA,JiangC2019PRA,Mercier2019PRAPP,ChenY2021PRL,XuXW2020PRAPP,QianYB2021PRA,Seif2018NatCo,Barzanjeh2018PRL,Habraken2012NJPh,XuHT2019Natur,LaiDG2022PRL}, and edge states in topological
optomechanical lattices~\cite{Sanavio2020PRB,Lemonde2019NJPh}.

Realizing single-photon nonreciprocity is one the most important
orientations of development in optical nonreciprocity. In order to achieve
the nonreciprocal effects on a few- and even single-photon level base on
optomechanical interaction, one requirement is the thermal noise generated
from the mechanical modes should be equal or less than one. It is obvious that the
common way for improving signal-to-noise ratio by enhancing the magnitude of
the signal power~\cite{Ruesink2018NatCo} is unfit for achieving nonreciprocal effects on a
few- and single-photon level. One of the most effective ways to suppress the thermal noise is cooling
the mechanical resonator in a cryogenic environment~\cite{Weis2010Sci} or based on
sideband cooling~\cite{Wilson2007PRL,Marquardt2007PRL}. Nevertheless, how to
suppress the thermal noise without precooling certainly is an interesting
and realistic question.

In this paper, we show how to realize thermal noise cancellation by quantum interference for
optomechanically induced nonreciprocity in a WGM microresonator. We note
that there are quantum interference for interacting qubits in contact with a
common environment, resulting in coherence and entanglement preserving~\cite{DuanLM1997PRL,Contreras2008PRB,WangZH2015PRA}.
Here, we investigate optical nonreciprocity in an optomechanical system
consisting of a WGM microresonator and two mechanical resonators, and the
essential ingredient is the two mechanical resonators coupled to a common
environment. We find that when the two mechanical resonators share the
same thermal reservoir, the thermal noise flow through the two mechanical
resonators may cancel each other by destructive interference. Our work provides
an effective way to achieve nonreciprocal effects on a few- and even single photon level
without precooling the mechanical mode to the ground state. Similar proposal
may also be used to realize thermal noise cancellation for other applications of optomechanics
in quantum technologies~\cite{Barzanjeh2022NatPh}.

The remainder of the paper is organized as follows. In Sec.~II, the
Hamiltonian of a multi-mode optomechanical system is introduced, and the
spectra of the output fields are obtained formally. The optical
nonreciprocal response and thermal noise cancellation in both sideband
resolved and unresolved regimes are discussed in Sec. III. We summarize this
work in Sec.~IV.

\section{Physical model}

\begin{figure}[tbp]
\includegraphics[bb=22 439 635 736, width=8.5cm, clip]{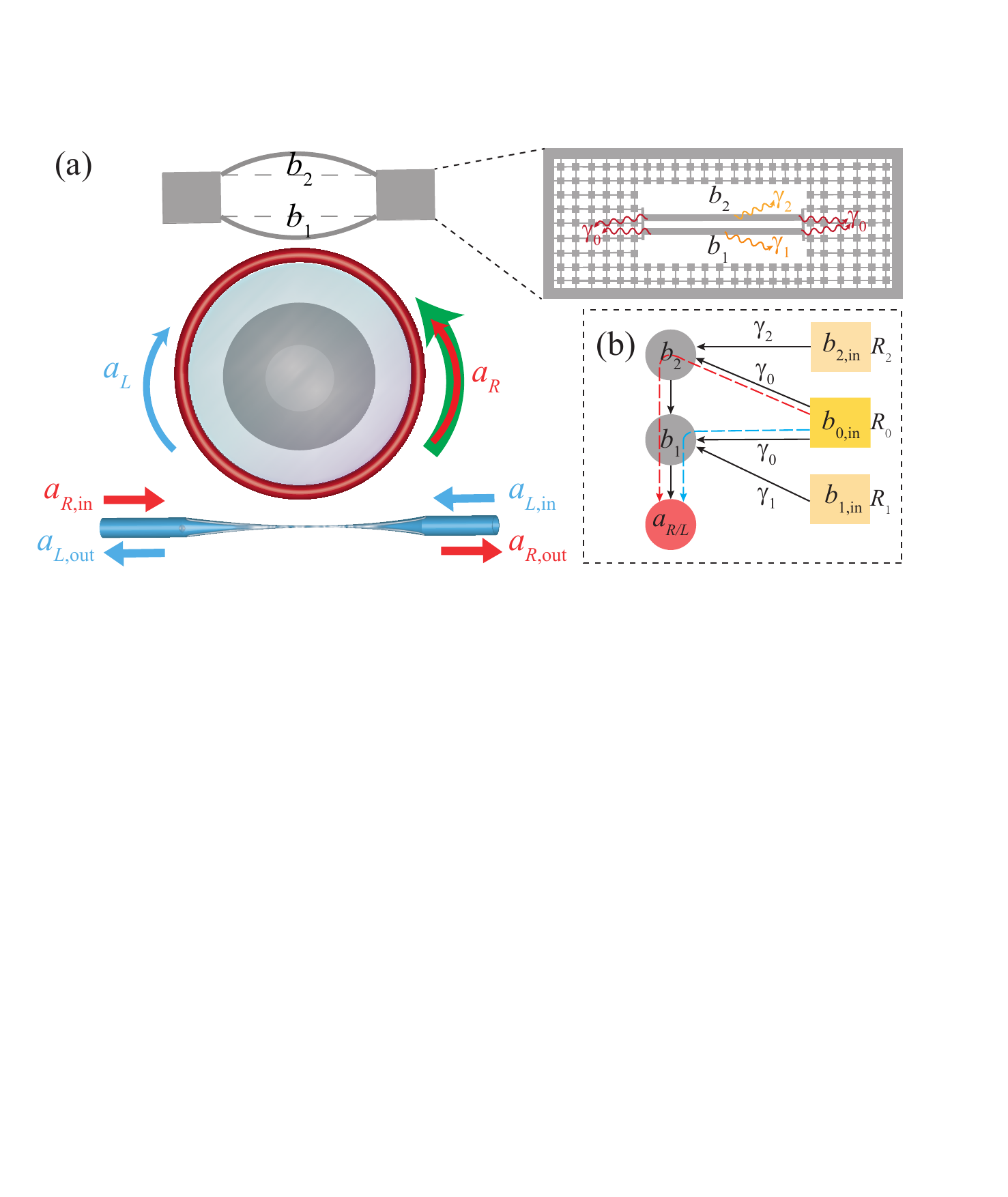}
\caption{(Color online) (a) Schematic diagram of a multimode optomechanical system consisting of a WGM microresonator with two counter-propagating optical modes ($a_{L}$
and $a_{R}$), and two nanomechanical resonators ($b_{1}$ and $b_{2}$) with degenerate frequency $\omega_m$. The optical modes ($a_{L}$ and $a_{R}$) are
coupled to the mechanical resonator $b_{1}$ through their evanescent field, and the two nanomechanical
resonators ($b_{1}$ and $b_{2}$) couple with each other directly via the coupling overhang and indirectly by the phononic crystal (common reservoir $R_0$).
(b) Flow chart depicting the thermal noise ($b_{i,{\rm in}}$, $i=0,1,2$) flow from the three reservoirs ($R_{i}$) to the optical modes $a_{R/L}$. The red and blue dashed curves with an arrowhead depict the two different flow paths from $R_0$ (common reservoir) to $a_{R/L}$.}
\label{fig1}
\end{figure}

We consider a multimode optomechanical system consisting of a WGM microresonator with two degenerate counter-propagating optical modes ($a_{L}$
and $a_{R}$, frequency $\omega_c$), and two nanomechanical
resonators ($b_{1}$ and $b_{2}$, with degenerate frequency $\omega_m$), as schematically
shown in Fig.~\ref{fig1}(a). The optical modes ($a_{L}$ and $a_{R}$) are
coupled to the mechanical resonator $b_{1}$ through their evanescent fields
with single-photon optomechanical coupling rate $g$.
Such near-field optomechanical schemes have already been demonstrated by doubly clamped SiN nanostrings or 2D sheet evanescent coupling to an
optical microresonator in experiments~\cite{Anetsberger2009NatPh,Brawley2016NatCo,LiHK2012PRA}.
The two mechanical resonators are coupled to each other directly via the coupling overhang with strength $J_m$ and indirectly by the phononic crystal~\cite{Safavi2011Natur} (common reservoir $R_0$) supported them on the both ends.
In order to enhancing the optomechanical coupling, a strong pumping field
(strength $\Omega$, frequency $\omega_p$) is input from left through a
tapered fibre evanescently coupled to the microresonator. The system can be
described by a Hamiltonian in the frame rotating at the frequency of the
pumping field $\omega_p$ as ($\hbar =1$)
\begin{eqnarray}\label{Hsys}
H_{\rm sys} &=&\sum_{k=R,L}\left[ \Delta _{0}-g\left( b_{1}^{\dag }+b_{1}\right) %
\right] a_{k}^{\dag }a_{k}  \notag \\
&&+\omega _{m}\left( b_{1}^{\dag }b_{1}+b_{2}^{\dag }b_{2}\right)
+J_{m}\left( b_{1}^{\dag }+b_{1}\right) \left( b_{2}^{\dag }+b_{2}\right)  \notag \\
&&+J_{s}\left( a_{R}^{\dag }a_{L}+a_{L}^{\dag }a_{R}\right) + \left(
\Omega a_{R}^{\dag }+ \Omega^{\ast} a_{R}\right),
\end{eqnarray}%
where $\Delta _{0}=\omega _{c}-\omega _{p}$ is the detuning of cavity mode
from the pumping field, and $J_s$ is the coupling strength between the
counter-propagating optical modes ($a_{L}$ and $a_{R}$) induced by
backscattering.

Before the quantitative calculation, we would like to introduce the physical mechanism that
induces the thermal noise cancellation based on the thermal noise flow chart~\cite{Tsang2010PRL} as shown in Fig.~\ref{fig1}(b).
There are three independent thermal reservoirs ($R_0$, $R_1$, and $R_2$),
where $R_0$ denotes the common reservoir for the two mechanical resonators, and $R_1$ and $R_2$ denote their respective reservoirs.
The common reservoir $R_0$ can either be a phononic crystal~\cite{Safavi2011Natur,Chan2011Natur,MacCabe2020Sci} as shown in Fig~\ref{fig1}(a) or an acoustic waveguide~\cite{Metelmann2015PRX,YuW2019PRL,ZhaoJ2020PRAPP,WangYP2020JAP,Karg2020Sci}.
$\gamma _{0}$ is the decay rate of the two mechanical resonators to the common reservoir with the input thermal noise $b_{0,\mathrm{in}}$.
There are two paths for the thermal noise flow from reservoir $R_0$ to the optical modes denoting by red and blue dashed curves with an arrow-head in Fig.~\ref{fig1}(b).
As the thermal noise coming from the same reservoir, the thermal noise flow through different paths could cancel each other by destructive interference,
which is the physical mechanism for the thermal noise cancellation discussed in this paper.
The phononic decays also can be induced by the scattering effect for the surface roughness and material inhomogeneity or the absorption of the material~\cite{MacCabe2020Sci}, which can be totally described by the coupling of the mechanical resonator $b_i$ to the reservoir $R_i$ with the decay rate $\gamma _{i}$ and input thermal noise $b_{i,\mathrm{in}}$ ($i=1,2$).
Experimentally, the decay rate $\gamma _{i}$ as small as $\omega_{m}/10^{10}$ has been obtained by carefully designing~\cite{MacCabe2020Sci}.
The reservoirs $R_1$ and $R_2$ are independent with other, so the thermal noise flow from them cannot cancel each other by quantum interference.
We will focus on the case that the phononic decay to the phononic crystal is the dominant process, i.e., $\gamma _{0}\gg\gamma _{i}$, so most of the thermal noise (from the common reservoir $R_0$) can be cancelled based on quantum interference.

According to the Heisenberg equation of motion and taking into account the
damping and corresponding input noise
(see Appendix~\ref{AppA} for details),
we get the quantum Langevin equations
(QLEs) as
\begin{eqnarray}\label{nlQLEs1}
\dot{a}_{R} &=&-\frac{\kappa }{2}a_{R}-i\left[ \Delta _{0}-g\left(
b_{1}^{\dag }+b_{1}\right) \right] a_{R}  \notag \\
&&-iJ_{s}a_{L}-i\Omega +\sqrt{\kappa _{\mathrm{ex}}}a_{R,\mathrm{in}}+\sqrt{%
\kappa _{0}}a_{R,\mathrm{vac}},
\end{eqnarray}%
\begin{eqnarray}\label{nlQLEs2}
\dot{a}_{L} &=&-\frac{\kappa }{2}a_{L}-i\left[ \Delta _{0}-g\left(
b_{1}^{\dag }+b_{1}\right) \right] a_{L}  \notag \\
&&-iJ_{s}a_{R}+\sqrt{\kappa _{\mathrm{ex}}}a_{L,\mathrm{in}}+\sqrt{\kappa
_{0}}a_{L,\mathrm{vac}},
\end{eqnarray}%
\begin{eqnarray}\label{nlQLEs3}
\dot{b}_{1} &=&-\left( \frac{\gamma _{m}}{2}+i\omega _{m}\right)
b_{1}+ig\left( a_{R}^{\dag }a_{R}+a_{L}^{\dag }a_{L}\right)- \frac{\gamma_0}{2}b_{2}   \notag \\
&& -iJ_{m}\left( b_{2}+b_{2}^{\dag }\right) +\sqrt{\gamma _{0}}b_{0,\mathrm{in}}+\sqrt{ \gamma _{\mathrm{in}}}b_{1,\mathrm{in}},
\end{eqnarray}%
\begin{eqnarray}\label{nlQLEs4}
\dot{b}_{2} &=&-\left( \frac{\gamma _{m}}{2}+i\omega _{m}\right)
b_{2}-iJ_{m}\left( b_{1}+b_{1}^{\dag }\right)   \notag \\
&& - \frac{\gamma_0}{2}b_{1}+\sqrt{\gamma _{0}}b_{0,\mathrm{in}}+\sqrt{\gamma _{\mathrm{in}}}b_{2,\mathrm{in}}.
\end{eqnarray}%
Here $\kappa =\kappa _{\mathrm{ex}}+\kappa _{0}$ is the total decay rate of
the optical modes; $\kappa _{\mathrm{ex}}$ is external decay rate through the tapered fibre with the input field $a_{k,\mathrm{in}}$,
and $\kappa _{0}$ is the internal decay rate with the input vacuum noise $a_{k,\mathrm{vac}}$; $\gamma _{m}=\gamma _{0} +\gamma _{\mathrm{in}}$ is the total decay rate of the two mechanical resonators with $\gamma _{\mathrm{in}}=\gamma _{1}=\gamma _{2}$.
The input fields (noises) $a_{k,\mathrm{in}}$, $a_{k,%
\mathrm{vac}},\left( k=R,L\right) $, and $b_{i,\mathrm{in}}\left(
i=0,1,2\right) $ satisfy the correlation relations
\begin{equation}
\left\langle a_{k,\mathrm{in}}^{\dag }\left( t^{\prime }\right) a_{k,\mathrm{%
in}}\left( t\right) \right\rangle =S_{k,\mathrm{in}}\left( t\right) \delta
\left( t+t^{\prime }\right) ,
\end{equation}%
\begin{equation}
\left\langle a_{k,\mathrm{vac}}^{\dag }\left( t^{\prime }\right) a_{k,%
\mathrm{vac}}\left( t\right) \right\rangle =0,
\end{equation}%
\begin{equation}
\left\langle b_{i,\mathrm{in}}^{\dag }\left( t^{\prime }\right) b_{i,\mathrm{%
in}}\left( t\right) \right\rangle =N_{\mathrm{th}}\delta \left( t+t^{\prime
}\right) ,
\end{equation}%
where $S_{k,\mathrm{in}}$ is the spectrum of the input quantum field and $N_{%
\mathrm{th}}$ is the mean numbers of the thermal phonons.

To solve the (nonlinear) QLEs, we can write each operator as the sum of its
steady-state value and the quantum noise operator as $a_{k}=\alpha
_{k}+\delta a_{k}$ ($k=R,L$) and $b_{i}=\beta _{i}+\delta b_{i}$ ($i=1,2$).
Here the steady-state values $\alpha _{k}$ and $\beta _{i}$ are obtained
from the QLEs by using a factorization assumption, e.g., $\langle
(b_{i}^{\dag }+b_{i})a_{k}\rangle $=$(\langle b_{i}^{\dag }\rangle +\langle
b_{i}\rangle )\rangle \langle a_{k}\rangle $ and $\langle a_{k}^{\dag
}a_{k}\rangle =\langle a_{k}^{\dag }\rangle \langle a_{k}\rangle $, as
\begin{equation}
\langle a_{R}\rangle \equiv \alpha _{R}=\frac{i\Omega }{-\frac{\kappa }{2}%
-i\omega _{m}+\frac{J_{s}^{2}}{-\frac{\kappa }{2}-i\omega _{m}}},
\end{equation}%
\begin{equation}
\langle a_{L}\rangle \equiv \alpha _{L}=\frac{iJ_{s}}{-\frac{\kappa }{2}%
-i\omega _{m}}\alpha _{R},
\end{equation}%
\begin{equation}
\langle b_{1}^{\dag }\rangle +\langle b_{1}\rangle =\beta _{1}^{\ast }+\beta
_{1}=\frac{\Delta _{0}-\omega _{m}}{g},
\end{equation}%
\begin{equation}
\langle b_{2}\rangle = \beta _{2} \approx -\frac{J_{m}}{\omega _{m}}\frac{\Delta _{0}-\omega _{m}}{g}
\end{equation}%
with setting of $\omega _{m}=\Delta _{0}-g\left( \beta
_{1}^{\ast }+\beta _{1}\right) $. In the strong driving condition $\Omega
\gg \kappa $ with $|\langle a_{k}\rangle |^{2}\gg 1$, the dynamic equations
for quantum noise operators ($\delta a_{k}$ and $\delta b_{l}$) can be
linearized by neglecting the nonlinear terms (e.g., $\delta b_{i}^{\dag
}\delta a_{k}$, $\delta b_{i}\delta a_{k}$, $\delta a_{k}^{\dag }\delta a_{k}
$), and the linearized QLEs for the quantum noise operators are given by
\begin{eqnarray}\label{QLE1}
\frac{d}{dt}\delta a_{R} &=&\left( -i\omega _{m}-\frac{\kappa }{2}\right)
\delta a_{R}+iG_{R}\left( \delta b_{1}^{\dag }+\delta b_{1}\right)   \notag \\
&&-iJ_{s}\delta a_{L}+\sqrt{\kappa _{\mathrm{ex}}}a_{R,\mathrm{in}}+\sqrt{%
\kappa _{0}}a_{R,\mathrm{vac}},
\end{eqnarray}
\begin{eqnarray}\label{QLE2}
\frac{d}{dt}\delta a_{L} &=&\left( -i\omega _{m}-\frac{\kappa }{2}\right)
\delta a_{L}+iG_{L}\left( \delta b_{1}^{\dag }+\delta b_{1}\right)   \notag \\
&&-iJ_{s}\delta a_{R}+\sqrt{\kappa _{\mathrm{ex}}}a_{L,\mathrm{in}}+\sqrt{
\kappa _{0}}a_{L,\mathrm{vac}},
\end{eqnarray}
\begin{eqnarray}\label{QLE3}
\frac{d}{dt}\delta b_{1} &=&\left( -i\omega _{m}-\frac{\gamma _{m}}{2}%
\right) \delta b_{1}+iG_{R}\delta a_{R}^{\dag }+iG_{R}^{\ast }\delta a_{R}
\notag \\
&&+iG_{L}^{\ast }\delta a_{L}+iG_{L}\delta a_{L}^{\dag }-iJ_{m}\left( \delta
b_{2}+\delta b_{2}^{\dag }\right)   \notag \\
&&-\frac{\gamma_0}{2} \delta b_{2} +\sqrt{\gamma _{0}}b_{0,\mathrm{in}}+\sqrt{\gamma _{\mathrm{in}}}b_{1,\mathrm{in}},
\end{eqnarray}
\begin{eqnarray}\label{QLE4}
\frac{d}{dt}\delta b_{2} &=&\left( -i\omega _{m}-\frac{\gamma _{m}}{2}%
\right) \delta b_{2}-iJ_m\left( \delta b_{1}^{\dag }+\delta b_{1}\right)
\notag \\
&&-\frac{\gamma_0}{2} \delta b_{1}+\sqrt{\gamma _{0}}b_{0,\mathrm{in}}+\sqrt{\gamma _{\mathrm{in}}}b_{2,\mathrm{in}},
\end{eqnarray}%
where $G_{R}\equiv g\alpha _{R}$ and $G_{L}\equiv g\alpha _{L}$ are
the effective optomechanical coupling rate.

The linearized QLEs can be concisely expressed as
\begin{equation}
\frac{d}{dt}V\left( t\right) =-MV\left( t\right) +V_{\rm in}\left( t\right) ,
\end{equation}%
where $V\left( t\right) $ is the vector for the quantum noise operators
defined by $V\left( t\right) \equiv \left(
\begin{array}{cccccccc}
\delta a_{R} & \delta a_{L} & \delta b_{1} & \delta b_{2} & \delta
a_{R}^{\dag } & \delta a_{L}^{\dag } & \delta b_{1}^{\dag } & \delta
b_{2}^{\dag }%
\end{array}%
\right) ^{T}$ (superscript $T$ denotes matrix transpose), $V_{\mathrm{in}%
}\left( t\right) $ is the input field operator vectors defined by
\begin{equation}
V_{\rm in}\left( t\right) \equiv \left(
\begin{array}{c}
\sqrt{\kappa _{\mathrm{ex}}}a_{R,\mathrm{in}}+\sqrt{\kappa _{0}}a_{R,\mathrm{%
vac}} \\
\sqrt{\kappa _{\mathrm{ex}}}a_{L,\mathrm{in}}+\sqrt{\kappa _{0}}a_{L,\mathrm{%
vac}} \\
\sqrt{\gamma _{0}}b_{0,\mathrm{in}}+\sqrt{\gamma _{\mathrm{in}}}b_{1,\mathrm{in}} \\
\sqrt{\gamma _{0}}b_{0,\mathrm{in}}+\sqrt{\gamma _{\mathrm{in}}}b_{2,\mathrm{in}} \\
\sqrt{\kappa _{\mathrm{ex}}}a_{R,\mathrm{in}}^{\dag }+\sqrt{\kappa _{0}}a_{R,%
\mathrm{vac}}^{\dag } \\
\sqrt{\kappa _{\mathrm{ex}}}a_{L,\mathrm{in}}^{\dag }+\sqrt{\kappa _{0}}a_{L,%
\mathrm{vac}}^{\dag } \\
\sqrt{\gamma _{0}}b_{0,\mathrm{in}}^{\dag }+\sqrt{\gamma _{\mathrm{in}}}b_{1,\mathrm{in}}^{\dag } \\
\sqrt{\gamma _{0}}b_{0,\mathrm{in}}^{\dag }+\sqrt{\gamma _{\mathrm{in}}}b_{2,\mathrm{in}}^{\dag }%
\end{array}%
\right) ,
\end{equation}%
and the coefficient matrix $M$ is given by
\begin{widetext}
\begin{equation}\label{M-Co}
M=\left(
\begin{array}{cccccccc}
i\omega _{m}+\frac{\kappa }{2} & iJ_{s} & -iG_{R} & 0 & 0 & 0 & -iG_{R} & 0
\\
iJ_{s} & i\omega _{m}+\frac{\kappa }{2} & -iG_{L} & 0 & 0 & 0 & -iG_{L} & 0
\\
-iG_{R}^{\ast } & -iG_{L}^{\ast } & i\omega _{m}+\frac{\gamma _{m}}{2} &
iJ_{m}+\frac{\gamma_0}{2} & -iG_{R} & -iG_{L} & 0 & iJ_{m} \\
0 & 0 & iJ_{m}+\frac{\gamma_0}{2} & i\omega _{m}+\frac{\gamma _{m}}{2} & 0 & 0 & iJ_{m} & 0 \\
0 & 0 & iG_{R}^{\ast } & 0 & \frac{\kappa }{2}-i\omega _{m} & -iJ_{s} &
iG_{R}^{\ast } & 0 \\
0 & 0 & iG_{L}^{\ast } & 0 & -iJ_{s} & \frac{\kappa }{2}-i\omega _{m} &
iG_{L}^{\ast } & 0 \\
iG_{R}^{\ast } & iG_{L}^{\ast } & 0 & -iJ_{m} & iG_{R} & iG_{L} & \frac{%
\gamma _{m}}{2}-i\omega _{m} & -iJ_{m}+\frac{\gamma_0}{2} \\
0 & 0 & -iJ_{m} & 0 & 0 & 0 & -iJ_{m}+\frac{\gamma_0}{2} & \frac{\gamma _{m}}{2}-i\omega _{m}%
\end{array}%
\right).
\end{equation}
\end{widetext}The stability of the system is determined by $M$, and it is
stable only if the real parts of all the eigenvalues of $M$ are positive. In
the following discussions, we will make sure that the stability conditions
are satisfied in all cases.

The linearized QLEs can be solved analytically in the frequency domain with
the method of Fourier transform. By introducing the Fourier transform for operator $O$ as
\begin{equation}\label{FT}
O\left( \omega \right) =\frac{1}{\sqrt{2\pi }}\int^{+\infty}_{-\infty} O\left( t\right) \exp
\left( i\omega t\right) dt,
\end{equation}%
the solution to the linearized QLEs in the frequency domain is obtained as
\begin{equation}
V\left( \omega \right) =U\left( \omega \right) V_{\rm in}\left( \omega \right) ,
\end{equation}%
where $U\left( \omega \right) \equiv \left( M-i\omega I\right) ^{-1}$, and
$I$ denotes the identity matrix. The output fields from the WGM
microresonator are obtained based on the input-output relation~\cite{Gardiner1985PRA}
\begin{equation}
\delta a_{k,\mathrm{out}}+\delta a_{k,\mathrm{in}}=\sqrt{\kappa _{\mathrm{ex}%
}}\delta a_{k},\qquad (k=R,\,L).
\end{equation}%
By the definition of the input and output spectra
\begin{equation}
S_{k,\mathrm{in}}\left( \omega \right) =\int \left\langle \delta a_{k,%
\mathrm{in}}^{\dag }\left( \omega ^{\prime }\right) \delta a_{k,\mathrm{in}%
}\left( \omega \right) \right\rangle d\omega ^{\prime },
\end{equation}%
\begin{equation}
S_{k,\mathrm{out}}\left( \omega \right) =\int \left\langle \delta a_{k,%
\mathrm{out}}^{\dag }\left( \omega ^{\prime }\right) \delta a_{k,\mathrm{out}%
}\left( \omega \right) \right\rangle d\omega ^{\prime },
\end{equation}%
the explicit expressions for the output spectra read
\begin{eqnarray}\label{SRout}
S_{R,\mathrm{out}}\left( \omega \right)  &=&T_{R}\left( \omega \right) S_{R,%
\mathrm{in}}\left( \omega \right) +R_{R}\left( \omega \right) S_{L,\mathrm{in%
}}\left( \omega \right)   \notag \\
&&+S_{R,\mathrm{th}}\left( \omega \right) +S_{R,\mathrm{vac}}\left( \omega
\right) ,
\end{eqnarray}%
\begin{eqnarray}\label{SLout}
S_{L,\mathrm{out}}\left( \omega \right)  &=&T_{L}\left( \omega \right) S_{L,%
\mathrm{in}}\left( \omega \right) +R_{L}\left( \omega \right) S_{R,\mathrm{in%
}}\left( \omega \right)   \notag \\
&&+S_{L,\mathrm{th}}\left( \omega \right) +S_{L,\mathrm{vac}}\left( \omega
\right) ,
\end{eqnarray}%
with the transmission spectra
\begin{equation}
T_{R}\left( \omega \right) =\left\vert \kappa _{\mathrm{ex}}U_{11}\left(
\omega \right) -1\right\vert ^{2}+\kappa _{\mathrm{ex}}^{2}\left\vert
U_{15}\left( \omega \right) \right\vert ^{2},
\end{equation}%
\begin{equation}
T_{L}\left( \omega \right) =\left\vert \kappa _{\mathrm{ex}}U_{22}\left(
\omega \right) -1\right\vert ^{2}+\kappa _{\mathrm{ex}}^{2}\left\vert
U_{26}\left( \omega \right) \right\vert ^{2},
\end{equation}%
the reflection spectra
\begin{equation}
R_{R}\left( \omega \right) =\kappa _{\mathrm{ex}}^{2}\left\vert U_{12}\left(
\omega \right) \right\vert ^{2}+\kappa _{\mathrm{ex}}^{2}\left\vert
U_{16}\left( \omega \right) \right\vert ^{2},
\end{equation}%
\begin{equation}
R_{L}\left( \omega \right) =\kappa _{\mathrm{ex}}^{2}\left\vert U_{21}\left(
\omega \right) \right\vert ^{2}+\kappa _{\mathrm{ex}}^{2}\left\vert
U_{25}\left( \omega \right) \right\vert ^{2},
\end{equation}%
the thermal noise spectra of mechanical modes
\begin{widetext}
\begin{eqnarray}
S_{R,{\rm th}}\left( \omega \right) &=&\kappa _{\mathrm{ex}}\gamma _{0}\left[
\left\vert U_{13}\left( \omega \right) +U_{14}\left( \omega \right)
\right\vert ^{2}+\left\vert U_{17}\left( \omega \right) +U_{18}\left( \omega
\right) \right\vert ^{2}\right]N_{\rm th}  \notag \\
&&+\kappa _{\mathrm{ex}} \gamma _{\mathrm{in}}\left[ \left\vert
U_{13}\left( \omega \right) \right\vert ^{2}+\left\vert U_{14}\left( \omega
\right) \right\vert ^{2}+\left\vert U_{17}\left( \omega \right) \right\vert
^{2}+\left\vert U_{18}\left( \omega \right) \right\vert ^{2}\right]N_{\rm th},
\end{eqnarray}%
\begin{eqnarray}
S_{L,{\rm th}}\left( \omega \right) &=&\kappa _{\mathrm{ex}}\gamma _{0}\left[
\left\vert U_{23}\left( \omega \right) +U_{24}\left( \omega \right)
\right\vert ^{2}+\left\vert U_{27}\left( \omega \right) +U_{28}\left( \omega
\right) \right\vert ^{2}\right] N_{\rm th} \notag \\
&&+\kappa _{\mathrm{ex}} \gamma _{\mathrm{in}}\left[ \left\vert
U_{23}\left( \omega \right) \right\vert ^{2}+\left\vert U_{24}\left( \omega
\right) \right\vert ^{2}+\left\vert U_{27}\left( \omega \right) \right\vert
^{2}+\left\vert U_{28}\left( \omega \right) \right\vert ^{2}\right]N_{\rm th},
\end{eqnarray}%
and the vacuum quantum noise spectra,
\begin{eqnarray}
S_{R,{\rm vac}}\left( \omega \right) &=&\kappa _{\mathrm{ex}}^{2}\left\vert U_{15}\left(
\omega \right) \right\vert ^{2}+\kappa _{\mathrm{ex}}\kappa _{0}\left\vert
U_{15}\left( \omega \right) \right\vert ^{2}+\kappa _{\mathrm{ex}}^{2}\left\vert
U_{16}\left( \omega \right) \right\vert ^{2}+\kappa _{\mathrm{ex}}\kappa
_{0}\left\vert U_{16}\left( \omega \right) \right\vert ^{2}  \notag \\
&&+\kappa _{\mathrm{ex}}\gamma _{0}\left\vert U_{17}\left( \omega \right)
+U_{18}\left( \omega \right) \right\vert ^{2}+\kappa _{\mathrm{ex}}\gamma _{\mathrm{in}}\left[ \left\vert U_{17}\left( \omega
\right) \right\vert ^{2}+\left\vert U_{18}\left( \omega \right) \right\vert
^{2}\right],
\end{eqnarray}%
\begin{eqnarray}
S_{L,{\rm vac}}\left( \omega \right) &=&\kappa _{\mathrm{ex}}^{2}\left\vert U_{25}\left(
\omega \right) \right\vert ^{2}+\kappa _{\mathrm{ex}}\kappa _{0}\left\vert
U_{25}\left( \omega \right) \right\vert ^{2}+\kappa _{\mathrm{ex}}^{2}\left\vert
U_{26}\left( \omega \right) \right\vert ^{2}+\kappa _{\mathrm{ex}}\kappa
_{0}\left\vert U_{26}\left( \omega \right) \right\vert ^{2}  \notag \\
&&+\kappa _{\mathrm{ex}}\gamma _{0}\left\vert U_{27}\left( \omega \right)
+U_{28}\left( \omega \right) \right\vert ^{2}+\kappa _{\mathrm{ex}} \gamma _{\mathrm{in}}\left[ \left\vert U_{27}\left( \omega
\right) \right\vert ^{2}+\left\vert U_{28}\left( \omega \right) \right\vert
^{2}\right].
\end{eqnarray}%
Here, $U_{nm}\left( \omega \right) $ (for $n,m=1,...,8$) represents the
element at the $n$th row and $m$th column of the matrix $U\left( \omega
\right) $.
\end{widetext}

\section{Results and Discussions}

In this section, we will study the optical nonreciprocal response in both sideband resolved and unresolved
regimes.
In order to show the effect of thermal noise cancellation, we compare the thermal noise spectra for a WGM microresonator coupling to two coupled mechanical resonators with the case of the WGM microresonator coupling to only one mechanical resonator.
Moreover, we will discuss the effects of backscattering in the WGM microresonator on the
optical nonreciprocal response and thermal noise cancellation before the end of this section.

\subsection{Nonreciprocal transmission in sideband resolved regime}

\begin{figure}[htp]
\centering
\includegraphics[bb=101 383 466 725, width=8.5cm, clip]{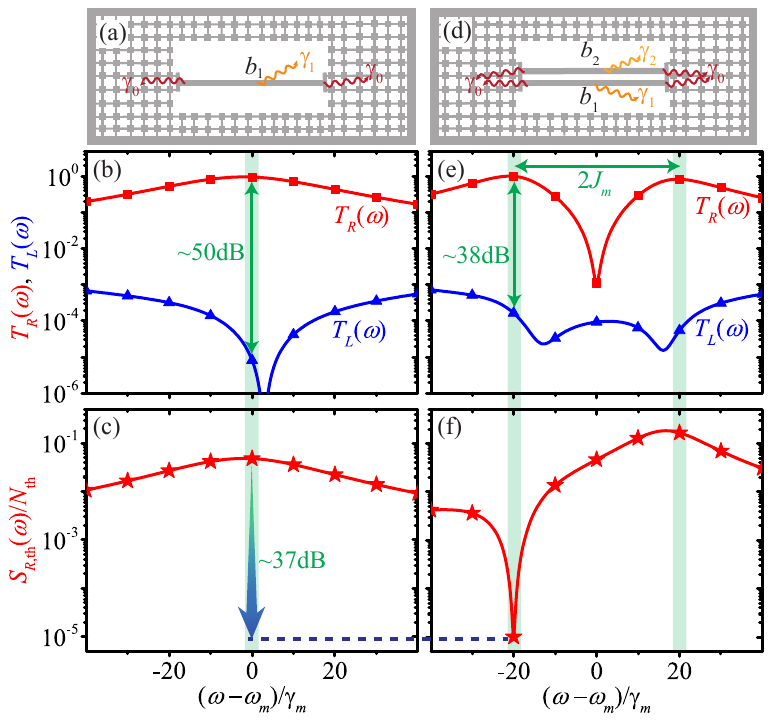}
\caption{(Color online) (Color online) Schematic diagrams of (a) one nanomechanical resonator and (d) two coupled nanomechanical resonators. (b) The transmission spectra $T_R(\omega)$ and $T_L(\omega)$, and (c) the thermal noise spectrum $S_{R,{\rm out}}(\omega)/N_{\rm th}$ for the WGM microresonator coupling to one mechanical resonator, (e) $T_R(\omega)$ and $T_L(\omega)$, and (f) $S_{R,{\rm out}}(\omega)/N_{\rm th}$ for the WGM microresonator coupling to two coupled mechanical resonators, in the sideband resolved regime ($\omega _{m}= 5 \kappa _{0}$). The other parameters are $\Delta =\omega _{m}$, $G_{R}=0.1\kappa _{0}$, and $J_{m}=0.01\protect\kappa _{0}$, $J_{s}=0.1\kappa _{0}$, $\kappa _{\mathrm{ex}}=\kappa _{0}$, $G_{L}=iJ_{s}G_{R}/(-\protect\kappa /2-i\protect\omega _{m})$, $\gamma_{0}=\omega _{m}/10^{4}$, and $\gamma_{1}=\gamma_{2}=\omega _{m}/10^{8}$.}
\label{fig2}
\end{figure}

In this subsection, we assume that the system is working in the sideband resolved
regime with mechanical frequency $\omega _{m} > \kappa _{0}$, which is within the reach of recent experiments.
As reported in Ref.~\cite{Anetsberger2009NatPh}, the doubly clamped SiN nanostrings with fundamental resonance frequencies $\Omega_m/ 2\pi= 6.5-16$ MHz and mechanical quality factors of $Q_m=10^4-10^5$ are coupled to the optical modes of toroid silica microcavity (unloaded optical linewidth of $4.9$ MHz) through evanescent field.
Furthermore, the mechanical resonators with high frequency ranging from 100 MHz to a few GHz have been realized in the mechanical devices with smaller size or greater stiffness~\cite{Adrian2023RMP,SongX2014PRL,Weber2016NatCo,Singh2014NatNa}.

\begin{figure}[htp]
\centering
\includegraphics[bb=148 469 432 737, width=8.5cm, clip]{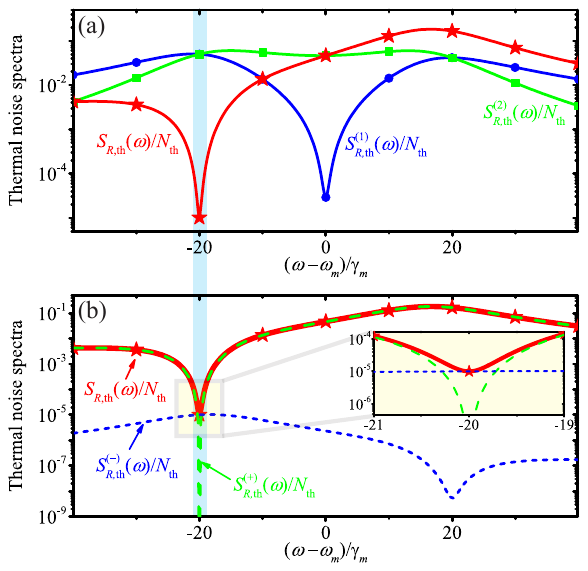}
\caption{(Color online) (a) The thermal noise spectra $S^{(1)}_{R,{\rm out}}(\omega)/N_{\rm th}$, $S^{(2)}_{R,{\rm out}}(\omega)/N_{\rm th}$, and $S_{R,{\rm out}}(\omega)/N_{\rm th}$, (b) $S^{(+)}_{R,{\rm out}}(\omega)/N_{\rm th}$, $S^{(-)}_{R,{\rm out}}(\omega)/N_{\rm th}$, and $S_{R,{\rm out}}(\omega)/N_{\rm th}$, are plotted as functions of the normalized detuning $(\omega - \omega_m)/\gamma_m$. The other parameters are the same as those given in Fig.~\ref{fig2}(d).}
\label{fig3}
\end{figure}

For comparison, we consider a WGM microresonator coupling to one nanomechanical resonator (denoted by $b_1$) first, with the schematic diagram of $b_1$ shown in Fig.~\ref{fig2}(a).
$b_1$ couples to reservoir $R_0$ (i.e., the phononic crystal) for phonon loss from the both ends ($\gamma_0$), and to the reservoir $R_1$ for the surface roughness and material inhomogeneity or the absorption of the material ($\gamma_1$), simultaneously.
This model can be described by the total Hamiltonian~(\ref{A1}) in Appendix~\ref{AppA} with all the parameters related to $b_2$ setting to zero.
The corresponding transmission spectra $T_R(\omega)$ and $T_L(\omega)$, and the thermal noise spectrum $S_{R,{\rm out}}(\omega)/N_{\rm th}$ are shown in Figs.~\ref{fig2}(b) and \ref{fig2}(c).
As predicted theoretically in Ref.~\cite{Hafezi12OE} and experimentally demonstrated in Refs.~\cite{Dong2016NaPho,Ruesink2016NatCo,Shen2018NatCo,Ruesink2018NatCo},
an optical isolator with the isolation $I_{R/L}\equiv T_R(\omega)/T_L(\omega)$ as high as $50$~dB can be achieved at the resonant frequency $\omega=\omega_m$. However, to achieve the nonreciprocal effects on a few- and even single-photon level, the thermal noise should be considered simultaneously. From Fig.~\ref{fig2}(b), we get the thermal noise spectrum $S_{R,{\rm out}}(\omega)/N_{\rm th}=0.048$ for the optical isolator working at the frequency $\omega=\omega_m$. According to the Bose-Einstein statistics, the mean numbers of the thermal phonons is given by $N_{\rm th}=[\exp(\hbar \omega_m /k_B T)-1]^{-1}$, where $k_B$ is the Boltzmann constant, and $T$ is the temperature of the reservoir at the thermal equilibrium. For example, a mechanical resonator with frequency $\omega_m=60$ MHz at the room temperature $T=300$K, contains about $N_{\rm th}\approx 10^5$ thermal phonons, then we get the thermal noise spectrum $S_{R,{\rm out}}(\omega) \approx 4800$.
In order to improve the signal-to-noise ratio, we need to enhance the magnitude of
the signal power~\cite{Ruesink2018NatCo}, or suppress the thermal noise by cooling the mechanical resonator in a cryogenic environment~\cite{Weis2010Sci} or based on
sideband cooling~\cite{Wilson2007PRL,Marquardt2007PRL}.

Here, we show that the thermal noise can be suppressed based on quantum interference for the WGM microresonator coupling to two coupled mechanical resonators [Fig.~\ref{fig2}(d)], with the transmission spectra $T_R(\omega)$ and $T_L(\omega)$, and thermal noise spectrum $S_{R,{\rm out}}(\omega)/N_{\rm th}$ shown in Figs.~\ref{fig2}(e) and \ref{fig2}(f).
Different from the nonreciprocity in the WGM microresonator coupling to one mechanical mode with one transmission window at $\omega=\omega_m$~\cite{Hafezi12OE,Dong2016NaPho,Ruesink2016NatCo,Shen2018NatCo,Ruesink2018NatCo}, there are two peaks around frequencies $\omega _{m}\pm J_m$ in the transmission spectrum of $T_R(\omega)$ for fields from left to right [see Fig.~\ref{fig2}(e)].
At the same time, there are two dips in the transmission spectrum of $T_L(\omega)$ for fields from right to left with the optomechanical interaction $G_L$ enhanced by the backscattering $J_s$.
The optical nonreciprocity with an isolation about $38$ dB is achieved at $\omega=\omega_m\pm J_m$ with backscattering effect $J_s=0.1 \kappa _{0}$ taken into account.
More importantly, there is a dip in the thermal noise spectrum $S_{R,{\rm th}}/N_{\rm th}$ at the frequency $\omega=\omega_m-J_m$, which is about $37$ dB lower than the case for the WGM microresonator coupling to one mechanical resonator [see Figs.~\ref{fig2}(c) and \ref{fig2}(f)].
In this case, the thermal noise spectrum for a mechanical resonator with frequency $\omega_m=60$ MHz at the room temperature $T=300$K is suppressed to $S_{R,{\rm out}}(\omega) \approx 1$, which means that we can achieve nonreciprocal transmission with signal-to-noise ratio $s/n \approx1$ around $\omega=\omega_m- J_m$ on a single-photon level without precooling the mechanical mode to the ground state.

Physically, the thermal noise suppression (enhancement) is induced by the quantum interference between the two flow paths of the thermal noises from the common reservoir $R_0$ to the optical modes $a_{R/L}$, as shown by the flow chart in Fig.~\ref{fig1}(b).
Here, the thermal noise flow from the common reservoir $R_0$ to the optical modes $a_{R/L}$ through the mechanical resonator $b_1$ [blue dashed curves with an arrowhead in Fig.~\ref{fig1}(b)] is given by
\begin{equation}
S^{(1)}_{R,{\rm th}}\left( \omega \right)=\kappa _{\mathrm{ex}}\gamma _{m}\left[
\left\vert U_{13}\left( \omega \right) \right\vert ^{2}+\left\vert U_{17}\left( \omega \right)\right\vert ^{2}\right]N_{\rm th},
\end{equation}
and the thermal noise flow through the mechanical resonator $b_2$ [red dashed curves with an arrowhead Fig.~\ref{fig1}(b)] is given by
\begin{equation}
S^{(2)}_{R,{\rm th}}\left( \omega \right)=\kappa _{\mathrm{ex}}\gamma _{m}\left[
\left\vert U_{14}\left( \omega \right) \right\vert ^{2}+\left\vert U_{18}\left( \omega \right)\right\vert ^{2}\right]N_{\rm th}.
\end{equation}
To reveal the quantum interference more quantitatively, we show the thermal noises flow for different paths in Fig.~\ref{fig3}(a).
The strength of the thermal noises flow by different paths $S^{(1)}_{R,{\rm th}}\left( \omega \right)$ and $S^{(2)}_{R,{\rm th}}\left( \omega \right)$ are almost the same value
around the frequency $\omega=\omega_m-J_m$, which one of the critical point for achieving high visibility in destructive interference.
The dip in the thermal noise spectrum at $\omega=\omega_m-J_m$ is induced by the destructive interference with $S^{(1)}_{R,{\rm th}}\left( \omega \right) \simeq  S^{(2)}_{R,{\rm th}}\left( \omega \right)$.

The thermal noise suppression can also be understood by introducing the bright and dark modes~\cite{DongCH2012Sci,WangYD2012PRL,TianL2012PRL,LaiDG2020PRA,ZhaoHM2022PRA} (normal modes) $b_{\pm}\equiv (b_1 \pm b_2)/\sqrt{2}$.
Based on some analytical derivations and Table~\ref{Tab1} in Appendix~\ref{DarkMode}, we find that the dark mode $b_{-}$ with eigenvalue $\omega_m-J_m$ is decoupled from the common reservoir $R_0$. Meanwhile, the bright mode $b_{+}$ with eigenvalue $\omega_m+J_m$ is coupled to the common reservoir $R_0$ with the strength enhanced by a factor of $\sqrt{2}$.
The thermal noise spectra $S^{(+)}_{R,{\rm th}}\left( \omega \right)$ and $S^{(-)}_{R,{\rm th}}\left( \omega \right)$ coming from the bright and dark modes are shown in Fig.~\ref{fig3}(b).
Clearly, the thermal noise $S^{(-)}_{R,{\rm th}}\left( \omega \right)$ from the dark mode is much lower than the thermal noise $S^{(+)}_{R,{\rm th}}\left( \omega \right)$ from the bright mode, except that the $S^{(+)}_{R,{\rm th}}\left( \omega \right)$ is dramatically suppressed by quantum interference around the frequency $\omega_m-J_m$ for the dark and bright modes $b_{\pm}$ coupling to the reservoirs $R_1$ and $R_2$ simultaneously [see the inset of Fig.~\ref{fig3}(b) and Table~\ref{Tab1}].

\subsection{Nonreciprocal amplification in sideband unresolved regime}

\begin{figure}[htp]
\centering
\includegraphics[bb=101 383 466 657, width=8.5cm, clip]{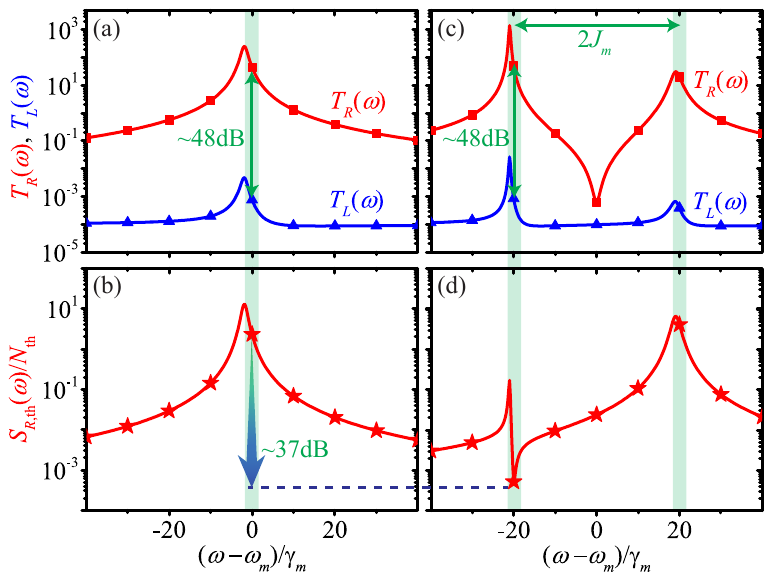}
\caption{(Color online) (a) The transmission spectra $T_R(\omega)$ and $T_L(\omega)$, and (b) the thermal noise spectrum $S_{R,{\rm out}}(\omega)/N_{\rm th}$ for the WGM microresonator coupling to one mechanical resonator, (c) $T_R(\omega)$ and $T_L(\omega)$, and (d) $S_{R,{\rm out}}(\omega)/N_{\rm th}$ for the WGM microresonator coupling to two coupled mechanical resonators, in the sideband unresolved regime ($\omega _{m}= 0.1 \kappa _{0}$). The other parameters are $\Delta =\omega _{m}$, $G_{R}=0.01\kappa _{0}$, and $J_{m}=0.0002\protect\kappa _{0}$, $J_{s}=0.1\kappa _{0}$, $\kappa _{\mathrm{ex}}=\kappa _{0}$, $G_{L}=iJ_{s}G_{R}/(-\protect\kappa /2-i\protect\omega _{m})$, $\gamma_{0}=\omega _{m}/10^{4}$, and $\gamma_{1}=\gamma_{2}=\omega _{m}/10^{8}$.}
\label{fig4}
\end{figure}

In this subsection, we will discuss the optical nonreciprocal response of the system in the sideband unresolved
regime with mechanical frequency $\omega _{m} < \kappa _{0}$, which is also a common situation in recent experiments~\cite{Brawley2016NatCo,Doolin2014PRA,Asano2020CmPhy}.
Difference from the case in the sideband resolved regime, the blue sideband transitions play an important role in the sideband unresolved regime, which may induced  significantly different results in the transmission spectrum.

The transmission spectra in the sideband unresolved regime are shown in Figs.~\ref{fig4}(a) and \ref{fig4}(c).
Similarly, there is one peak in the transmission spectrum $T_R(\omega)$ around the frequency $\omega _{m}$ for the WGM microresonator coupling to one mechanical resonator, and two peaks in the transmission spectrum $T_R(\omega)$ around the frequencies $\omega _{m}\pm J_m$ for the WGM microresonator coupling to two coupled mechanical resonators.
The optical nonreciprocity with a isolation about $48$ dB is achieved for both of these systems.
Moreover, the values of the peaks in $T_R(\omega)$ are greater than one, i.e. nonreciprocal amplification is achieved in sideband unresolved regime.
Different from the ampliation achieving with blue-detuned external driving~\cite{Massel2011Natur,Safavi2011Natur,Hocke2012NJPh}, here the nonreciprocal amplification is achieved with a red-detuned external driving field in sideband unresolved regime.

Physically, nonreciprocal amplification appearing in the sideband unresolved regime is induced by the blue-sideband transitions.
There are both red-sideband transitions for the terms $iG_R \delta b_1$ in Eq.~(\ref{QLE1}) and $iG_R^{\ast } \delta a_R$ in Eq.~(\ref{QLE3}), and blue-sideband transitions for the terms $iG_R \delta b^{\dagger}_1$ in Eq.~(\ref{QLE1}) and $iG_R \delta a^{\dagger}_R$ in Eq.~(\ref{QLE3}).
According to the quantum theory for sideband cooling in optomechanical systems~\cite{Wilson2007PRL,Marquardt2007PRL}, the red-sideband transitions correspond to the excitation changing between the optical and mechanical modes, and blue-sideband transitions correspond to the excitation of both optical and mechanical modes simultaneously.
The theoretical analysis and experimental results show that the blue-sideband transitions can induced signal absorption and ampliation~\cite{Massel2011Natur,Safavi2011Natur,Hocke2012NJPh}.
In the sideband resolved regime, the blue-sideband transitions are suppressed for large detuning.
However, in the sideband unresolved regime, the blue-sideband transitions become significant even with a red-detuned external driving field as discussed here.

The blue-sideband transitions can amplify not only the signal field but also the input thermal noises, which is the origin of optical induced heating in optomechanical systems~\cite{Grudinin2010PRL,WangH2014PRA}.
As shown in Fig.~\ref{fig4}(b), there is a peak in the thermal noise spectrum around $\omega=\omega_m$ for the WGM microresonator coupling to one mechanical resonator.
In contrast, the thermal noise around the frequency $\omega=\omega_m-J_m$ is suppressed dramatically by quantum interference for the WGM microresonator coupling to two coupled mechanical resonators, which is about $37$ dB lower than the case for the WGM microresonator coupling to one mechanical resonator [see Fig.~\ref{fig4}(d)].
This suggests that the scheme of thermal noise cancellation for two mechanical resonators coupled to a common environment also works in the sideband unresolved regimes for nonreciprocal amplification.

\subsection{The effects of backscattering}

\begin{widetext}
\begin{figure*}[htbp]
\centering
\includegraphics[bb=60 369 552 600, width=17cm, clip]{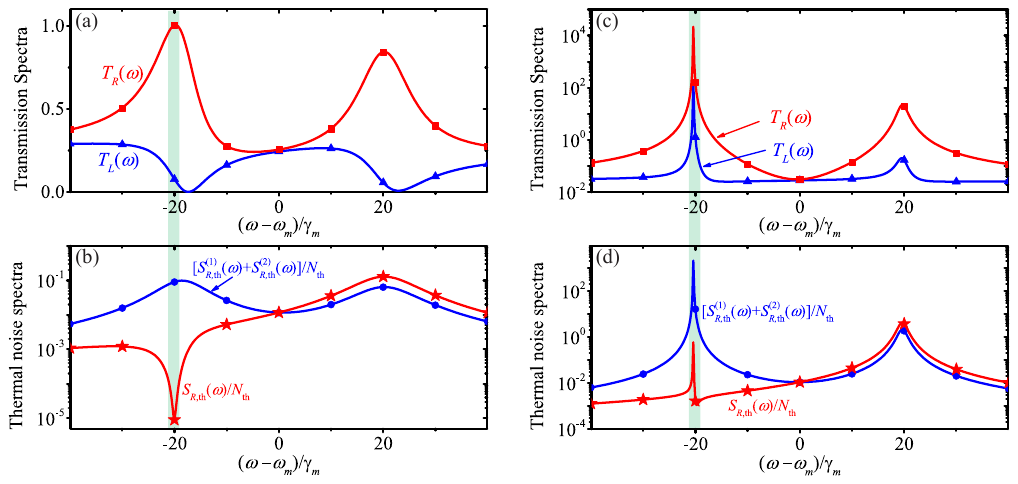}
\caption{(Color online) (a) The transmission spectra $T_R(\omega)$ and $T_L(\omega)$, and (b) the thermal noise spectra $S_{R,{\rm out}}(\omega)/N_{\rm th}$ and $[S^{(1)}_{R,{\rm out}}(\omega)+S^{(2)}_{R,{\rm out}}(\omega)]/N_{\rm th}$ are plotted as functions of the normalized detuning $(\omega - \omega_m)/\gamma_m$ for $\Delta =\omega _{m}=5 \kappa _{0}$, $J_{s}=\kappa _{0}$, $G_{R}=0.1 \kappa _{0}$, and $J_{m}=0.01 \kappa _{0}$.
(c) $T_R(\omega)$ and $T_L(\omega)$, and (d) $S_{R,{\rm out}}(\omega)/N_{\rm th}$ and  $[S^{(1)}_{R,{\rm out}}(\omega)+S^{(2)}_{R,{\rm out}}(\omega)]/N_{\rm th}$ are plotted as functions of the normalized detuning $(\omega - \omega_m)/\gamma_m$ for $\Delta =\omega _{m}= 0.1 \kappa _{0}$, $J_{s}= 0.45\kappa _{0}$, $G_{R}=0.01\kappa _{0}$, and $J_{m}=0.0002\protect\kappa _{0}$.
The other parameters are $G_{L}=iJ_{s}G_{R}/(-\protect\kappa /2-i\protect\omega _{m})$, $\kappa _{\mathrm{ex}}=\kappa _{0}$, $\gamma_{0}=\omega _{m}/10^{4}$, and $\gamma_{1}=\gamma_{2}=\omega _{m}/10^{8}$.}
\label{fig5}
\end{figure*}
\end{widetext}

Before the end of this section, let us discuss the effect of backscattering $J_s$ between the two degenerate counter-propagating optical modes ($a_{L}$
and $a_{R}$) on the optical nonreciprocal response and thermal noise cancellation in both sideband resolved and unresolved regimes.
In the preceding discussions, the weak backscattering with $J_s=0.1\kappa_0$ is considered, and non-reciprocal effects are obvious for $|G_R|\gg |G_L|$.

When the backscattering strength $J_s$ becomes larger, the backscattering may strongly redistributes
the signal field.
In the sideband resolved regime, the non-reciprocal effect with isolation $I_{R/L}\approx 11$ dB is obtained as shown in Fig.~\ref{fig5}(a), where the condition of $|G_R|\gg |G_L|$ is still satisfied for $J_s=k_0$.
In contrast, we have $|G_R| \sim |G_L|$ for $J_s=k_0$ in the sideband unresolved regime, and the system becomes instable for the negative real parts of the eigenvalues of $M$ in Eq.~(\ref{M-Co}).
To guarantee the stability of the system in the sideband unresolved regime, we take the backscattering strength $J_s=0.45 \kappa _{0}$ as an example, and the transmission spectra with isolation $I_{R/L}\approx 21$ dB is shown in Fig.~\ref{fig5}(c).

The scheme of thermal noise suppression for the two mechanical resonators couple to a common reservoir still works in the strong backscattering condition.
In Figs.~\ref{fig5}(b) and \ref{fig5}(d), the thermal noise spectra are shown in both the sideband resolved and unresolved regimes.
By comparison with results of $[S^{(1)}_{R,{\rm out}}(\omega)+S^{(2)}_{R,{\rm out}}(\omega)]/N_{\rm th}$, the thermal noise $S_{R,{\rm out}}(\omega)/N_{\rm th}$ around the frequency $\omega=\omega_m-J_m$ is dramatically suppressed in both the sideband resolved and unresolved regimes.
These results indicate that the scheme of quantum interference induced thermal noise cancellation is a robust method against backscattering.

\section{Conclusions}

In summary, we proposed a scheme to realize thermal noise cancellation for optomechanically induced nonreciprocity in a WGM microresonator by coupling two mechanical resonators to a common reservoir.
We found that nonreciprocal transmission and amplification with high isolation can be achieved in the WGM microresonator for optomechanical system working in the sideband resolved and unresolved regimes.
More interestingly, the thermal noise can be cancelled by the destructive quantum interference between the two flow paths of the thermal noises from the common reservoir to the optical modes.
Numerical results indicate that the scheme of quantum interference induced thermal noise cancellation can be applied in both sideband resolved and unresolved regimes, even with strong backscattering taken into account.
Optomechanically induced nonreciprocity with thermal noise cancellation may stimulate a wide range of practical applications of optomechanical devices in chiral quantum networks~\cite{Mahmoodian2016PRL} and chiral quantum optics~\cite{Lodahl2017Natur}.

The scheme of quantum interference induced thermal noise cancellation is a general method for thermal noise suppression in optomechanics with wide applications in quantum technologies~\cite{Barzanjeh2022NatPh}.
It can be applied to suppress the thermal noise in the microwave-to-optical conversion for optomechanical transducers~\cite{Hill2012NatCo,Bochmann2013NatPh,Andrews2014NatPh,Bagci2014Natur,Dong2015AnP,Tian2015AnP,Forsch2020NatPh}.
Similar schemes can also be applied to suppress the thermal noise in quantum sensing via optomechanical interactions~\cite{HuYW2013FrPhy,Metcalfe2014ApPRv,LiBB2021Nanop}.

\vskip 2pc \leftline{\bf Acknowledgement}

This work is by the National Natural Science Foundation of China (Grants No.~12064010 and  No.~12247105),
Natural Science Foundation of Hunan Province of China (Grant
No.~2021JJ20036), and the Science and Technology Innovation Program of Hunan Province (Grant No.~2022RC1203).

\appendix

\begin{widetext}

\section{Derivation of the quantum Langevin equations}\label{AppA}

The total Hamiltonian of the WGM optomechanical system coupling with the reservoirs is described by
\begin{equation}\label{A1}
H_{\mathrm{tot}}=H_{\mathrm{sys}}+H_{\mathrm{re}}+H_{\mathrm{int}},
\end{equation}%
where $H_{\mathrm{sys}}$ is the optomechanical Hamiltonian Eq.~(\ref{Hsys}) in the main text, $H_{%
\mathrm{re}}$ is the free Hamiltonian of the optical and mechanical
reservoirs,%
\begin{equation}
H_{\mathrm{re}}=\sum_{k=R,L}\int_{0}^{+\infty }d\omega \omega \left(
a_{k,0,\omega }^{\dag }a_{k,0,\omega }+a_{k,\mathrm{ex},\omega }^{\dag }a_{k,%
\mathrm{ex},\omega }\right) +\sum_{i=0,1,2}\int_{0}^{+\infty }d\omega \omega
b_{i,\omega }^{\dag }b_{i,\omega },
\end{equation}%
and $H_{\mathrm{int}}$\ describes the interaction between the reservoirs and
the optomechanical system,%
\begin{eqnarray}
H_{\mathrm{int}} &=&i\sum_{k=R,L}\int_{0}^{+\infty }d\omega h_{k\mathrm{,ex}%
}\left( \omega \right) \left[ a_{k,\mathrm{ex},\omega }^{\dag
}a_{k}-a_{k}^{\dag }a_{k,\mathrm{ex},\omega }\right] +i\sum_{k=R,L}%
\int_{0}^{+\infty }d\omega h_{k,0}\left( \omega \right) \left[ a_{k,0,\omega
}^{\dag }a_{k}-a_{k}^{\dag }a_{k,0,\omega }\right]   \nonumber \\
&&+i\sum_{i=1,2}\int_{0 }^{+\infty }d\omega g_{i,\mathrm{in}}\left(
\omega \right) \left[ b_{i,\omega }^{\dag }b_{i}-b_{i}^{\dag }b_{i,\omega }%
\right] +i\sum_{i=1,2}\int_{0 }^{+\infty }d\omega g_{i,0}\left( \omega
\right) \left[ b_{0,\omega }^{\dag }b_{i}-b_{i}^{\dag }b_{0,\omega }\right]
\end{eqnarray}%
with the coupling strengths $h_{k\mathrm{,ex}}\left( \omega \right) $, $%
h_{k,0}\left( \omega \right) $, $g_{i,\mathrm{in}}\left( \omega \right) $,
and $g_{i,0}\left( \omega \right) $. Here, $a_{k,\mathrm{ex},\omega }$, $%
a_{k,\mathrm{0},\omega }$, and $b_{i,\omega }$ are the boson annihilation
operators for the reservoirs with $\left[ a_{k,\mathrm{ex},\omega },a_{k,%
\mathrm{ex},\omega ^{\prime }}^{\dag }\right] =\delta \left( \omega -\omega
^{\prime }\right) $, $\left[ a_{k,\mathrm{0},\omega },a_{k,\mathrm{0},\omega
^{\prime }}^{\dag }\right] =\delta \left( \omega -\omega ^{\prime }\right) $%
, and $\left[ b_{i,\omega },b_{i,\omega ^{\prime }}^{\dag }\right] =\delta
\left( \omega -\omega ^{\prime }\right) $.

The Heisenberg equations for the operators $a_{k}$, $b_{i}$, $a_{k,0,\omega }
$, $a_{k,\mathrm{ex},\omega }$, $b_{0,\omega }$, and $b_{i,\omega }$ read%
\begin{equation}\label{A4}
\frac{da_{R}}{dt}=-i\left[ \Delta _{0}-g\left( b_{1}^{\dag
}+b_{1}\right) \right] a_{R}-iJ_{s}a_{L}-i\Omega -\int_{0}^{+\infty }d\omega
h_{R\mathrm{,ex}}\left( \omega \right) a_{R,\mathrm{ex},\omega
}-\int_{0}^{+\infty }d\omega h_{R,0}\left( \omega \right) a_{R,0,\omega },
\end{equation}
\begin{equation}\label{A5}
\frac{da_{L}}{dt}=-i\left[ \Delta _{0}-g\left( b_{1}^{\dag
}+b_{1}\right) \right] a_{L}-iJ_{s}a_{R}-\int_{0}^{+\infty }d\omega h_{L%
\mathrm{,ex}}\left( \omega \right) a_{L,\mathrm{ex},\omega
}-\int_{0}^{+\infty }d\omega h_{L,0}\left( \omega \right) a_{L,0,\omega },
\end{equation}
\begin{equation}\label{A6}
\frac{db_{1}}{dt}=-i\omega _{m}b_{1}+ig\sum_{k=R,L}a_{k}^{\dag
}a_{k}-iJ_{m}\left( b_{2}^{\dag }+b_{2}\right) -\int_{0}^{+\infty }d\omega
g_{1,\mathrm{in}}\left( \omega \right) b_{1,\omega }-\int_{0}^{+\infty
}d\omega g_{1,0}\left( \omega \right) b_{0,\omega },
\end{equation}
\begin{equation}\label{A7}
\frac{db_{2}}{dt}=-i\omega _{m}b_{2}-iJ_{m}\left( b_{1}^{\dag }+b_{1}\right)
-\int_{0}^{+\infty }d\omega g_{2,\mathrm{in}}\left( \omega \right)
b_{2,\omega }-\int_{0}^{+\infty }d\omega g_{2,0}\left( \omega \right)
b_{0,\omega },
\end{equation}
\begin{equation}\label{A8}
\frac{da_{k,0,\omega }}{dt}=-i\omega a_{k,0,\omega }+h_{k,0}\left( \omega
\right) a_{k},
\end{equation}
\begin{equation}\label{A9}
\frac{da_{k,\mathrm{ex},\omega }}{dt}=-i\omega a_{k,\mathrm{ex},\omega
}+h_{k,\mathrm{ex}}\left( \omega \right) a_{k},
\end{equation}%
\begin{equation}\label{A10}
\frac{db_{0,\omega }}{dt}=-i\omega b_{0,\omega }+\sum_{i=1,2}g_{i,0}\left(
\omega \right) b_{i},
\end{equation}%
\begin{equation}\label{A11}
\frac{db_{i,\omega }}{dt}=-i\omega b_{i,\omega }+g_{i,\mathrm{in}}\left(
\omega \right) b_{i}.
\end{equation}
The Eqs.~(\ref{A8})-(\ref{A11}) can be solved analytically as%
\begin{equation}
a_{k,0,\omega }=a_{k,0,\omega }\left( t_{0}\right) e^{-i\omega \left(
t-t_{0}\right) }+h_{k,0}\left( \omega \right) \int_{t_{0}}^{t}a_{k}\left(
t^{\prime }\right) e^{-i\omega \left( t-t^{\prime }\right) }dt^{\prime },
\end{equation}%
\begin{equation}
a_{k,\mathrm{ex},\omega }=a_{k,\mathrm{ex},\omega }\left( t_{0}\right)
e^{-i\omega \left( t-t_{0}\right) }+h_{k,\mathrm{ex}}\left( \omega \right)
\int_{t_{0}}^{t}a_{k}\left( t^{\prime }\right) e^{-i\omega \left(
t-t^{\prime }\right) }dt^{\prime },
\end{equation}%
\begin{equation}
b_{0,\omega }=b_{0,\omega }\left( t_{0}\right) e^{i\omega \left(
t_{0}-t\right) }+\sum_{i=1,2}\int_{t_{0}}^{t}g_{i,0}\left( \omega \right)
b_{i}\left( t^{\prime }\right) e^{i\omega \left( t^{\prime }-t\right)
}dt^{\prime },
\end{equation}%
\begin{equation}
b_{i,\omega }=b_{i,\omega }\left( t_{0}\right) e^{i\omega \left(
t_{0}-t\right) }+\int_{t_{0}}^{t}g_{i,\mathrm{in}}\left( \omega \right)
b_{i}\left( t^{\prime }\right) e^{i\omega \left( t^{\prime }-t\right)
}dt^{\prime }.
\end{equation}
Substitute these solutions into Eqs.~(\ref{A4})-(\ref{A7}), we get
\begin{equation}
\frac{da_{R}}{dt}=-i\left[ \Delta _{0}-g\left( b_{1}^{\dag
}+b_{1}\right) \right] a_{R}-iJ_{s}a_{L}-i\Omega -\frac{\kappa _{R\mathrm{,ex%
}}+\kappa _{R,0}}{2}a_{R}+\sqrt{\kappa _{R\mathrm{,ex}}}a_{R\mathrm{,in}}+%
\sqrt{\kappa _{R,0}}a_{R\mathrm{,vac}},
\end{equation}%
\begin{equation}
\frac{da_{L}}{dt}=-i\left[ \Delta _{0}-g\left( b_{1}^{\dag
}+b_{1}\right) \right] a_{L}-iJ_{s}a_{R}-\frac{\kappa _{L\mathrm{,ex}%
}+\kappa _{L,0}}{2}a_{L}+\sqrt{\kappa _{L\mathrm{,ex}}}a_{L\mathrm{,in}}+%
\sqrt{\kappa _{L,0}}a_{L\mathrm{,vac}},
\end{equation}%
\begin{equation}
\frac{db_{1}}{dt}=-i\omega _{m}b_{1}+ig\sum_{k=R,L}a_{k}^{\dag
}a_{k}-iJ_{m}\left( b_{2}^{\dag }+b_{2}\right) -\frac{\sqrt{\gamma
_{1,0}\gamma _{2,0}}}{2}b_{2}-\frac{\gamma _{1,0}+\gamma _{1,\mathrm{in}}}{2}%
b_{1}+\sqrt{\gamma _{1,\mathrm{in}}}b_{1,\mathrm{in}}+\sqrt{\gamma _{1,0}}b_{0,\mathrm{in%
}},
\end{equation}%
\begin{equation}
\frac{db_{2}}{dt}=-i\omega _{m}b_{2}-iJ_{m}\left( b_{1}^{\dag }+b_{1}\right)
-\frac{\sqrt{\gamma _{1,0}\gamma _{2,0}}}{2}b_{1}-\frac{\gamma _{2,0}+\gamma
_{2,\mathrm{in}}}{2}b_{2}+\sqrt{\gamma _{2,\mathrm{in}}}b_{2,\mathrm{in}}+\sqrt{\gamma
_{2,0}}b_{0,\mathrm{in}},
\end{equation}%
based on the Markov approximation $h_{k\mathrm{,ex}}\left( \omega \right) =%
\sqrt{\kappa _{k\mathrm{,ex}}/2\pi }$, $h_{k,0}\left( \omega \right) =\sqrt{%
\kappa _{k,0}/2\pi }$, $g_{i,\mathrm{in}}\left( \omega \right) =\sqrt{\gamma
_{i,\mathrm{in}}/2\pi }$, $g_{i,0}\left( \omega \right) =\sqrt{\gamma
_{i,0}/2\pi }$, and the input noise operators are defined by%
\begin{equation}
a_{k\mathrm{,in}}\equiv -\frac{1}{\sqrt{2\pi }}\int_{0 }^{+\infty
}d\omega a_{k,\mathrm{ex},\omega }\left( t_{0}\right) e^{-i\omega \left(
t-t_{0}\right) },
\end{equation}%
\begin{equation}
a_{k\mathrm{,vac}}\equiv -\frac{1}{\sqrt{2\pi }}\int_{0 }^{+\infty
}d\omega a_{k,0,\omega }\left( t_{0}\right) e^{-i\omega \left(
t-t_{0}\right) },
\end{equation}%
\begin{equation}
b_{0,\mathrm{in}}\equiv -\frac{1}{\sqrt{2\pi }}\int_{0 }^{+\infty }d\omega
b_{0,\omega }\left( t_{0}\right) e^{i\omega \left( t_{0}-t\right) },
\end{equation}%
\begin{equation}
b_{i,\mathrm{in}}\equiv -\frac{1}{\sqrt{2\pi }}\int_{0}^{+\infty }d\omega
b_{i,\omega }\left( t_{0}\right) e^{i\omega \left( t_{0}-t\right) }.
\end{equation}%
Under the conditions $\kappa _{\mathrm{ex}}=\kappa _{R\mathrm{,ex}}=\kappa _{L\mathrm{,ex}}$, $\kappa _{0}=\kappa _{R,0}=\kappa _{L,0}$, $\gamma _{\mathrm{in}}=\gamma_{1,\mathrm{in}}=\gamma _{2,\mathrm{in}}$, and $\gamma _{0}=\gamma _{1,0}=\gamma_{2,0}$, we get the QLEs.~(\ref{nlQLEs1})-(\ref{nlQLEs4}) in the main text with $\kappa =\kappa _{\mathrm{ex}}+\kappa _{0}$ and $\gamma _{m}=\gamma _{0}+\gamma _{\mathrm{in}}$.

\section{Thermal noise spectra based on bright and dark modes}\label{DarkMode}

In order to understand the thermal noise cancellation around the frequency $%
\omega =\omega _{m}-J_{m}$, let us introduce the bright and dark modes $%
b_{\pm }$, i.e., the supermodes of the two coupled mechanical modes, as%
\begin{equation}
b_{\pm }=\frac{1}{\sqrt{2}}\left( b_{1}\pm b_{2}\right),
\end{equation}%
then the Hamiltonian $H_{\mathrm{sys}}$ for optomechanical system under the rotating-wave approximation can be
rewritten as
\begin{eqnarray}
H_{\mathrm{sys}} &=&\sum_{k=R,L}\left[ \Delta _{0}-g^{\prime }\left(
b_{+}^{\dag }+b_{+}\right) \right] a_{k}^{\dag }a_{k}+\sum_{k=R,L}\left[
\Delta _{0}-g^{\prime }\left( b_{-}^{\dag }+b_{-}\right) \right] a_{k}^{\dag
}a_{k}  \nonumber \\
&&+\left( \omega _{m}+J_{m}\right) b_{+}^{\dag }b_{+}+\left( \omega
_{m}-J_{m}\right) b_{-}^{\dag }b_{-}+J_{s}\left( a_{R}^{\dag
}a_{L}+a_{L}^{\dag }a_{R}\right) +\left( \Omega a_{R}^{\dag }+\Omega ^{\ast
}a_{R}\right)
\end{eqnarray}%
with $g^{\prime }=g/\sqrt{2}$. Similarly,
the free Hamiltonian of the optical and mechanical reservoirs $H_{\mathrm{re}%
}$ and the interaction Hamilton between the reservoirs and the optomechanical system $%
H_{\mathrm{int}}$ are rewritten as%
\begin{equation}
H_{\mathrm{re}}=\sum_{k=R,L}\int_{0}^{+\infty }d\omega \omega \left(
a_{k,0,\omega }^{\dag }a_{k,0,\omega }+a_{k,\mathrm{ex},\omega }^{\dag }a_{k,%
\mathrm{ex},\omega }\right) +\int_{0}^{+\infty }d\omega \omega \left(
B_{+,\omega }^{\dag }B_{+,\omega }+B_{-,\omega }^{\dag }B_{-,\omega }+ b_{0,\omega }^{\dag }b_{0,\omega }\right) ,
\end{equation}%
and%
\begin{eqnarray}
H_{\mathrm{int}} &=&i\sum_{k=R,L}\int_{0}^{+\infty }d\omega h_{k\mathrm{,ex}%
}\left( \omega \right) \left[ a_{k,\mathrm{ex},\omega }^{\dag
}a_{k}-a_{k}^{\dag }a_{k,\mathrm{ex},\omega }\right] +i\sum_{k=R,L}%
\int_{0}^{+\infty }d\omega h_{k,0}\left( \omega \right) \left[ a_{k,0,\omega
}^{\dag }a_{k}-a_{k}^{\dag }a_{k,0,\omega }\right]   \nonumber \\
&&+i\int_{-\infty }^{+\infty }d\omega g_{\mathrm{in}}\left( \omega \right) %
\left[ B_{+,\omega }^{\dag }b_{+}-b_{+}^{\dag }B_{+,\omega }\right]
+i\int_{-\infty }^{+\infty }d\omega \sqrt{2}g_0\left( \omega \right) \left[
b_{0,\omega }^{\dag }b_{+}-b_{+}^{\dag }b_{0,\omega }\right]   \nonumber \\
&&+i\int_{-\infty }^{+\infty }d\omega g_{\mathrm{in}}\left( \omega \right) %
\left[ B_{-,\omega }^{\dag }b_{-}-b_{-}^{\dag }B_{-,\omega }\right] ,
\end{eqnarray}%
where $B_{\pm ,\omega }=\left( b_{1,\omega }\pm b_{2,\omega }\right) /\sqrt{2%
}$, $g_0\left( \omega \right) =g_{2,0}\left( \omega \right) =g_{1,0}\left(
\omega \right) $, and $g_{\mathrm{in}}\left( \omega \right) =g_{1\mathrm{,in}%
}\left( \omega \right) =g_{2\mathrm{,in}}\left( \omega \right) $.
The coupling strengths between the nanomechanical resonators and the reservoirs ($R_0$, $R_1$, and $R_2$) are summarized in Table~\ref{Tab1}. We can see that the dark mode $b_{-}$ with eigenvalue $\omega_m-J_m$ is decoupled from the common reservoir $R_0$, but the bright mode $b_{+}$ with eigenvalue $\omega_m+J_m$ is coupled to the common reservoir $R_0$ with the strength enhanced by a factor of $\sqrt{2}$. For this reason, the thermal noise for the dark mode $b_{-}$ ($\omega =\omega_m-J_m$) is suppressed dramatically, while thermal noise for the bright mode $b_{+}$ ($\omega=\omega_m+J_m$) is enhanced.
Although the dark mode $b_{-}$ is coupled to both the reservoirs $R_1$ and $R_2$, the coupling strengths are suppressed by a factor of $\sqrt{2}$. On the whole, the reservoir $R_2$ does not induce more thermal noise effect.

\begin{table*}[tbp]

\scriptsize

\centering

\caption{
Coupling strengths (CSs) between the nanomechanical resonators (NMRs) and the reservoirs ($R_0$, $R_1$, and $R_2$), for $g_0\left( \omega \right) =g_{2,0}\left( \omega \right) =g_{1,0}\left(
\omega \right) $ and $g_{\mathrm{in}}\left( \omega \right) =g_{1\mathrm{,in}}\left( \omega \right) =g_{2\mathrm{,in}}\left( \omega \right) $.
}

\label{Tab1}

\begin{tabular}{c|c|cc|cc}

\toprule

\multirow{2}{*}{Reservoirs} & {One NMR} & \multicolumn{4}{c}{Two NMRs}
\\
\cmidrule(r){2-6}

&  $b_1(\omega_m)$
&  $b_1\;(\omega_m)$        &  $b_2\;(\omega_m)$
&  $b_+\;(\omega_m+J_m)$    &  $b_-\;(\omega_m-J_m)$
\\

\midrule

CSs to $R_0$             &$g_0$                          & $g_0$                    & $g_0$                   & $\sqrt{2}g_0$                   & $0$     \\

CSs to $R_1$             &$g_{\rm in}$                   & $g_{\rm in}$             & $0$                     & $g_{\rm in}/\sqrt{2}$           & $g_{\rm in}/\sqrt{2}$     \\

CSs to $R_2$             &$0$                            & $0$                      & $g_{\rm in}$            & $g_{\rm in}/\sqrt{2}$           & $g_{\rm in}/\sqrt{2}$     \\

\bottomrule

\end{tabular}

\end{table*}

Under the Markov approximation for $h_{k\mathrm{,ex}}\left( \omega \right) =%
\sqrt{\kappa _{k\mathrm{,ex}}/2\pi }$, $h_{k,0}\left( \omega \right) =\sqrt{%
\kappa _{k,0}/2\pi }$, $g_{\mathrm{in}}\left( \omega \right) =\sqrt{\gamma _{%
\mathrm{in}}/2\pi }$, and $g_0\left( \omega \right) =\sqrt{\gamma
_{0}/2\pi }$, the dynamic equations for the operators $a_{k}$ and $b_{\pm }$
are obtained as%
\begin{equation}\label{LEQ1app}
\frac{da_{R}}{dt}=-i\left[ \Delta _{0}-g^{\prime }\left( b_{+}^{\dag
}+b_{+}\right) -g^{\prime }\left( b_{-}^{\dag }+b_{-}\right) \right]
a_{R}-iJ_{s}a_{L}-i\Omega -\frac{\kappa }{2}a_{R}+\sqrt{\kappa _{\mathrm{ex}}%
}a_{R\mathrm{,in}}+\sqrt{\kappa _{0}}a_{R\mathrm{,vac}},
\end{equation}%
\begin{equation}\label{LEQ2app}
\frac{da_{L}}{dt}=-i\left[ \Delta _{0}-g^{\prime }\left( b_{+}^{\dag
}+b_{+}\right) -g^{\prime }\left( b_{-}^{\dag }+b_{-}\right) \right]
a_{L}-iJ_{s}a_{R}-\frac{\kappa }{2}a_{L}+\sqrt{\kappa _{\mathrm{ex}}}a_{L%
\mathrm{,in}}+\sqrt{\kappa _{0}}a_{L\mathrm{,vac}},
\end{equation}%
\begin{equation}\label{LEQ3app}
\frac{db_{+}}{dt}=-\frac{\gamma _{\mathrm{in}}+2\gamma _{0}}{2}b_{+}\left(
t\right) -i\left( \omega _{m}+J_{m}\right) b_{+}+ig^{\prime }\left(
a_{R}^{\dag }a_{R}+a_{L}^{\dag }a_{L}\right) +\sqrt{\gamma _{\mathrm{in}}}%
B_{+,\mathrm{in}}+\sqrt{2\gamma _{0}}b_{0,\mathrm{in}},
\end{equation}%
\begin{equation}\label{LEQ4app}
\frac{db_{-}}{dt}=-\frac{\gamma _{\mathrm{in}}}{2}b_{-}\left( t\right)
-i\left( \omega _{m}-J_{m}\right) b_{-}+ig^{\prime }\left( a_{R}^{\dag
}a_{R}+a_{L}^{\dag }a_{L}\right) +\sqrt{\gamma _{\mathrm{in}}}B_{-,\mathrm{in%
}},
\end{equation}%
where $B_{\pm ,\mathrm{in}}\left( t\right) $ is the input quantum noise
defined by
\begin{equation}
B_{\pm ,\mathrm{in}}\left( t\right) \equiv -\sqrt{\frac{1}{2\pi }}\int_{0
}^{+\infty }d\omega B_{\pm ,\omega }\left( t_{0}\right) e^{i\omega \left(
t_{0}-t\right) }
\end{equation}
with the correlation function
\begin{equation}
\left\langle B_{\pm ,\mathrm{in}}^{\dag }\left( \omega ^{\prime }\right)
B_{\pm ,\mathrm{in}}\left( \omega \right) \right\rangle =N_{\mathrm{th}%
}\delta \left( \omega +\omega ^{\prime }\right) .
\end{equation}

To calculate the spectra of the output fields, we linearize the nonlinear
QLEs~(\ref{LEQ1app})-(\ref{LEQ4app}) by rewriting the operators as the sum of the mean values and
the small quantum fluctuation terms as $a_{k}=\alpha _{k}+\delta a_{k}$ and $%
b_{\pm }=\beta _{\pm }+\delta b_{\pm }$. Keep only the first-order terms in
the small quantum fluctuation terms $\delta a_{k}$ and $\delta b_{\pm }$, the linearized QLEs read
\begin{equation}\label{AppLEQ}
\frac{d}{dt}\widetilde{V}=-\widetilde{M}\widetilde{V}+\widetilde{V}_{\mathrm{%
in}},
\end{equation}%
where we have the vector of the quantum fluctuation operators $\widetilde{V}=\left(
\begin{array}{cccccccc}
\delta a_{R} & \delta a_{L} & \delta b_{+} & \delta
b_{-} & \delta a_{R}^{\dag } & \delta a_{L}^{\dag } & \delta b_{+}^{\dag } & \delta b_{-}^{\dag }\end{array}%
\right)^{T} $, the vector of the input noise operators%
\begin{equation}
\widetilde{V}_{\mathrm{in}}=\left(
\begin{array}{c}
\sqrt{\kappa _{\mathrm{ex}}}a_{R,\mathrm{in}}+\sqrt{\kappa _{0}}a_{R,\mathrm{%
vac}} \\
\sqrt{\kappa _{\mathrm{ex}}}a_{L,\mathrm{in}}+\sqrt{\kappa _{0}}a_{L,\mathrm{%
vac}} \\
\sqrt{\gamma _{\mathrm{in}}}B_{+,\mathrm{in}}+\sqrt{2\gamma _{0}}b_{0,%
\mathrm{in}} \\
\sqrt{\gamma _{\mathrm{in}}}B_{-,\mathrm{in}} \\
\sqrt{\kappa _{\mathrm{ex}}}a_{R,\mathrm{in}}^{\dag }+\sqrt{\kappa _{0}}a_{R,%
\mathrm{vac}}^{\dag } \\
\sqrt{\kappa _{\mathrm{ex}}}a_{L,\mathrm{in}}^{\dag }+\sqrt{\kappa _{0}}a_{L,%
\mathrm{vac}}^{\dag } \\
\sqrt{\gamma _{\mathrm{in}}}B_{+,\mathrm{in}}^{\dag }+\sqrt{2\gamma _{0}}%
b_{0,\mathrm{in}}^{\dag } \\
\sqrt{\gamma _{\mathrm{in}}}B_{-,\mathrm{in}}^{\dag }%
\end{array}%
\right) ,
\end{equation}%
and the matrix of the coefficients%
\[
\widetilde{M}=\left(
\begin{array}{cccccccc}
\frac{\kappa }{2}+i\omega _{m} & iJ_{s} & -iG_{R}^{\prime } &
-iG_{R}^{\prime } & 0 & 0 & -iG_{R}^{\prime } & -iG_{R}^{\prime } \\
iJ_{s} & \frac{\kappa }{2}+i\omega _{m} & -iG_{L}^{\prime } &
-iG_{L}^{\prime } & 0 & 0 & -iG_{L}^{\prime } & -iG_{L}^{\prime } \\
-iG_{R}^{\prime \ast } & -iG_{L}^{\prime \ast } & \frac{\gamma _{+}}{2}%
+i\omega _{+} & 0 & -iG_{R}^{\prime } & -iG_{L}^{\prime } & 0 & 0 \\
-iG_{R}^{\prime \ast } & -iG_{L}^{\prime \ast } & 0 & \frac{\gamma _{\mathrm{%
in}}}{2}+i\omega _{-} & -iG_{R}^{\prime } & -iG_{L}^{\prime } & 0 & 0 \\
0 & 0 & iG_{R}^{\prime \ast } & iG_{R}^{\prime \ast } & \frac{\kappa }{2}%
-i\omega _{m} & -iJ_{s} & iG_{R}^{\prime \ast } & iG_{R}^{\prime \ast } \\
0 & 0 & iG_{L}^{\prime \ast } & iG_{L}^{\prime \ast } & -iJ_{s} & \frac{%
\kappa }{2}-i\omega _{m} & iG_{L}^{\prime \ast } & iG_{L}^{\prime \ast } \\
iG_{R}^{\prime \ast } & iG_{L}^{\prime \ast } & 0 & 0 & iG_{R}^{\prime } &
iG_{L}^{\prime } & \frac{\gamma _{+}}{2}-i\omega _{+} & 0 \\
iG_{R}^{\prime \ast } & iG_{L}^{\prime \ast } & 0 & 0 & iG_{R}^{\prime } &
iG_{L}^{\prime } & 0 & \frac{\gamma _{\mathrm{in}}}{2}-i\omega _{-}%
\end{array}%
\right)
\]%
with $G_{j}^{\prime }=g^{\prime }\alpha _{j}$, $\gamma _{+}=\gamma _{\mathrm{%
in}}+2\gamma _{0}$, and $\omega _{\pm }=\omega _{m}\pm J_{m}$.

Based on the Fourier transform~(\ref{FT}), the linearized QLEs (\ref{AppLEQ}) can be solved in the frequency domain as
\begin{equation}
\widetilde{V}\left( \omega \right) =\widetilde{U}\left( \omega \right)
\widetilde{V}_{\mathrm{in}}\left( \omega \right) ,
\end{equation}%
where $\widetilde{U}\left( \omega \right) =\left( \widetilde{M}-i\omega I\right)
^{-1}$. The output spectra have the same form as Eqs.~(\ref{SRout}) and (\ref{SLout}), where the transmission spectra are replaced by%
\begin{equation}
T_{R}\left( \omega \right) =\left\vert \widetilde{U}_{11}\left( \omega
\right) \kappa _{\mathrm{ex}}-1\right\vert ^{2}+\kappa _{\mathrm{ex}%
}^{2}\left\vert \widetilde{U}_{15}\left( \omega \right) \right\vert ^{2},
\end{equation}%
\begin{equation}
T_{L}\left( \omega \right) =\left\vert \kappa _{\mathrm{ex}}\widetilde{U}%
_{22}\left( \omega \right) -1\right\vert ^{2}+\kappa _{\mathrm{ex}%
}^{2}\left\vert \widetilde{U}_{26}\left( \omega \right) \right\vert ^{2},
\end{equation}%
the reflection spectra are replaced by%
\begin{equation}
R_{R}\left( \omega \right) =\kappa _{\mathrm{ex}}^{2}\left\vert \widetilde{U}%
_{12}\left( \omega \right) \right\vert ^{2}+\kappa _{\mathrm{ex}%
}^{2}\left\vert \widetilde{U}_{16}\left( \omega \right) \right\vert ^{2},
\end{equation}%
\begin{equation}
R_{L}\left( \omega \right) =\kappa _{\mathrm{ex}}^{2}\left\vert \widetilde{U}%
_{21}\left( \omega \right) \right\vert ^{2}+\kappa _{\mathrm{ex}%
}^{2}\left\vert \widetilde{U}_{25}\left( \omega \right) \right\vert ^{2},
\end{equation}%
the thermal noise spectra are replaced by%
\begin{equation}
S_{R,\mathrm{th}}\left( \omega \right) =S_{R,\mathrm{th}}^{(+)}\left( \omega
\right) +S_{R,\mathrm{th}}^{(-)}\left( \omega \right) ,
\end{equation}%
\begin{equation}
S_{L,\mathrm{th}}\left( \omega \right) =S_{L,\mathrm{th}}^{(+)}\left( \omega
\right) +S_{L,\mathrm{th}}^{(-)}\left( \omega \right) ,
\end{equation}%
with%
\begin{equation}
S_{R,\mathrm{th}}^{(+)}\left( \omega \right) =\kappa _{\mathrm{ex}}\gamma
_{+}\left[ \left\vert \widetilde{U}_{13}\left( \omega \right) \right\vert
^{2}+\left\vert \widetilde{U}_{17}\left( \omega \right) \right\vert ^{2}%
\right] N_{\mathrm{th}},
\end{equation}%
\begin{equation}
S_{R,\mathrm{th}}^{(-)}\left( \omega \right) =\kappa _{\mathrm{ex}}\gamma _{%
\mathrm{in}}\left[ \left\vert \widetilde{U}_{14}\left( \omega \right)
\right\vert ^{2}+\left\vert \widetilde{U}_{18}\left( \omega \right)
\right\vert ^{2}\right] N_{\mathrm{th}},
\end{equation}%
\begin{equation}
S_{L,\mathrm{th}}^{(+)}\left( \omega \right) =\kappa _{\mathrm{ex}}\gamma
_{+}\left[ \left\vert \widetilde{U}_{23}\left( \omega \right) \right\vert
^{2}+\left\vert \widetilde{U}_{27}\left( \omega \right) \right\vert ^{2}%
\right] N_{\mathrm{th}},
\end{equation}%
\begin{equation}
S_{L,\mathrm{th}}^{(-)}\left( \omega \right) =\kappa _{\mathrm{ex}}\gamma _{%
\mathrm{in}}\left[ \left\vert \widetilde{U}_{24}\left( \omega \right)
\right\vert ^{2}+\left\vert \widetilde{U}_{28}\left( \omega \right)
\right\vert ^{2}\right] N_{\mathrm{th}},
\end{equation}%
and the contribution from the incoming vacuum fields read%
\begin{eqnarray}
S_{R,\mathrm{vac}}\left( \omega \right)  &=&\kappa _{\mathrm{ex}%
}^{2}\left\vert \widetilde{U}_{15}\left( \omega \right) \right\vert
^{2}+\kappa _{\mathrm{ex}}^{2}\left\vert \widetilde{U}_{16}\left( \omega
\right) \right\vert ^{2}+\kappa _{\mathrm{ex}}\kappa _{0}\left\vert
\widetilde{U}_{15}\left( \omega \right) \right\vert ^{2}+\kappa _{\mathrm{ex}%
}\kappa _{0}\left\vert \widetilde{U}_{16}\left( \omega \right) \right\vert
^{2}  \nonumber \\
&&+\kappa _{\mathrm{ex}}\gamma _{+}\left\vert \widetilde{U}_{17}\left(
\omega \right) \right\vert ^{2}+\kappa _{\mathrm{ex}}\gamma _{\mathrm{in}%
}\left\vert \widetilde{U}_{18}\left( \omega \right) \right\vert ^{2},
\end{eqnarray}%
\begin{eqnarray}
S_{L,\mathrm{vac}}\left( \omega \right)  &=&\kappa _{\mathrm{ex}%
}^{2}\left\vert \widetilde{U}_{25}\left( \omega \right) \right\vert
^{2}+\kappa _{\mathrm{ex}}\kappa _{0}\left\vert \widetilde{U}_{25}\left(
\omega \right) \right\vert ^{2}+\kappa _{\mathrm{ex}}^{2}\left\vert
\widetilde{U}_{26}\left( \omega \right) \right\vert ^{2}+\kappa _{\mathrm{ex}%
}\kappa _{0}\left\vert \widetilde{U}_{26}\left( \omega \right) \right\vert
^{2}  \nonumber \\
&&+\kappa _{\mathrm{ex}}\gamma _{+}\left\vert \widetilde{U}_{27}\left(
\omega \right) \right\vert ^{2}+\kappa _{\mathrm{ex}}\gamma _{\mathrm{in}%
}\left\vert \widetilde{U}_{28}\left( \omega \right) \right\vert ^{2}.
\end{eqnarray}
Here, $\widetilde{U}_{nm}\left( \omega \right) $ (for $n,m=1,...,8$) denotes the
element at the $n$th row and $m$th column of the matrix $\widetilde{U}\left( \omega
\right) $.

\end{widetext}

\bibliography{ref}

\begin{thebibliography}{96}%
\makeatletter
\providecommand \@ifxundefined [1]{%
 \@ifx{#1\undefined}
}%
\providecommand \@ifnum [1]{%
 \ifnum #1\expandafter \@firstoftwo
 \else \expandafter \@secondoftwo
 \fi
}%
\providecommand \@ifx [1]{%
 \ifx #1\expandafter \@firstoftwo
 \else \expandafter \@secondoftwo
 \fi
}%
\providecommand \natexlab [1]{#1}%
\providecommand \enquote  [1]{``#1''}%
\providecommand \bibnamefont  [1]{#1}%
\providecommand \bibfnamefont [1]{#1}%
\providecommand \citenamefont [1]{#1}%
\providecommand \href@noop [0]{\@secondoftwo}%
\providecommand \href [0]{\begingroup \@sanitize@url \@href}%
\providecommand \@href[1]{\@@startlink{#1}\@@href}%
\providecommand \@@href[1]{\endgroup#1\@@endlink}%
\providecommand \@sanitize@url [0]{\catcode `\\12\catcode `\$12\catcode
  `\&12\catcode `\#12\catcode `\^12\catcode `\_12\catcode `\%12\relax}%
\providecommand \@@startlink[1]{}%
\providecommand \@@endlink[0]{}%
\providecommand \url  [0]{\begingroup\@sanitize@url \@url }%
\providecommand \@url [1]{\endgroup\@href {#1}{\urlprefix }}%
\providecommand \urlprefix  [0]{URL }%
\providecommand \Eprint [0]{\href }%
\providecommand \doibase [0]{https://doi.org/}%
\providecommand \selectlanguage [0]{\@gobble}%
\providecommand \bibinfo  [0]{\@secondoftwo}%
\providecommand \bibfield  [0]{\@secondoftwo}%
\providecommand \translation [1]{[#1]}%
\providecommand \BibitemOpen [0]{}%
\providecommand \bibitemStop [0]{}%
\providecommand \bibitemNoStop [0]{.\EOS\space}%
\providecommand \EOS [0]{\spacefactor3000\relax}%
\providecommand \BibitemShut  [1]{\csname bibitem#1\endcsname}%
\let\auto@bib@innerbib\@empty
\bibitem [{\citenamefont {Aspelmeyer}\ \emph {et~al.}(2014)\citenamefont
  {Aspelmeyer}, \citenamefont {Kippenberg},\ and\ \citenamefont
  {Marquardt}}]{Aspelmeyer2014RMP}%
  \BibitemOpen
  \bibfield  {author} {\bibinfo {author} {\bibfnamefont {M.}~\bibnamefont
  {Aspelmeyer}}, \bibinfo {author} {\bibfnamefont {T.~J.}\ \bibnamefont
  {Kippenberg}},\ and\ \bibinfo {author} {\bibfnamefont {F.}~\bibnamefont
  {Marquardt}},\ }\bibfield  {title} {\bibinfo {title} {Cavity optomechanics},\
  }\href {https://doi.org/10.1103/RevModPhys.86.1391} {\bibfield  {journal}
  {\bibinfo  {journal} {Rev. Mod. Phys.}\ }\textbf {\bibinfo {volume} {86}},\
  \bibinfo {pages} {1391} (\bibinfo {year} {2014})}\BibitemShut {NoStop}%
\bibitem [{\citenamefont {Jing}\ \emph {et~al.}(2014)\citenamefont {Jing},
  \citenamefont {\"Ozdemir}, \citenamefont {L\"u}, \citenamefont {Zhang},
  \citenamefont {Yang},\ and\ \citenamefont {Nori}}]{JingH2014PRL}%
  \BibitemOpen
  \bibfield  {author} {\bibinfo {author} {\bibfnamefont {H.}~\bibnamefont
  {Jing}}, \bibinfo {author} {\bibfnamefont {S.~K.}\ \bibnamefont {\"Ozdemir}},
  \bibinfo {author} {\bibfnamefont {X.-Y.}\ \bibnamefont {L\"u}}, \bibinfo
  {author} {\bibfnamefont {J.}~\bibnamefont {Zhang}}, \bibinfo {author}
  {\bibfnamefont {L.}~\bibnamefont {Yang}},\ and\ \bibinfo {author}
  {\bibfnamefont {F.}~\bibnamefont {Nori}},\ }\bibfield  {title} {\bibinfo
  {title} {$\mathcal{PT}$-symmetric phonon laser},\ }\href
  {https://doi.org/10.1103/PhysRevLett.113.053604} {\bibfield  {journal}
  {\bibinfo  {journal} {Phys. Rev. Lett.}\ }\textbf {\bibinfo {volume} {113}},\
  \bibinfo {pages} {053604} (\bibinfo {year} {2014})}\BibitemShut {NoStop}%
\bibitem [{\citenamefont {Xu}\ \emph {et~al.}(2015)\citenamefont {Xu},
  \citenamefont {Liu}, \citenamefont {Sun},\ and\ \citenamefont
  {Li}}]{XuXW2015PRA}%
  \BibitemOpen
  \bibfield  {author} {\bibinfo {author} {\bibfnamefont {X.-W.}\ \bibnamefont
  {Xu}}, \bibinfo {author} {\bibfnamefont {Y.-x.}\ \bibnamefont {Liu}},
  \bibinfo {author} {\bibfnamefont {C.-P.}\ \bibnamefont {Sun}},\ and\ \bibinfo
  {author} {\bibfnamefont {Y.}~\bibnamefont {Li}},\ }\bibfield  {title}
  {\bibinfo {title} {Mechanical $\mathcal{PT}$ symmetry in coupled
  optomechanical systems},\ }\href {https://doi.org/10.1103/PhysRevA.92.013852}
  {\bibfield  {journal} {\bibinfo  {journal} {Phys. Rev. A}\ }\textbf {\bibinfo
  {volume} {92}},\ \bibinfo {pages} {013852} (\bibinfo {year}
  {2015})}\BibitemShut {NoStop}%
\bibitem [{\citenamefont {Liu}\ \emph {et~al.}(2016)\citenamefont {Liu},
  \citenamefont {Zhang}, \citenamefont {\"Ozdemir}, \citenamefont {Peng},
  \citenamefont {Jing}, \citenamefont {L\"u}, \citenamefont {Li}, \citenamefont
  {Yang}, \citenamefont {Nori},\ and\ \citenamefont {Liu}}]{LiuZP2016PRL}%
  \BibitemOpen
  \bibfield  {author} {\bibinfo {author} {\bibfnamefont {Z.-P.}\ \bibnamefont
  {Liu}}, \bibinfo {author} {\bibfnamefont {J.}~\bibnamefont {Zhang}}, \bibinfo
  {author} {\bibfnamefont {S.~K.}\ \bibnamefont {\"Ozdemir}}, \bibinfo {author}
  {\bibfnamefont {B.}~\bibnamefont {Peng}}, \bibinfo {author} {\bibfnamefont
  {H.}~\bibnamefont {Jing}}, \bibinfo {author} {\bibfnamefont {X.-Y.}\
  \bibnamefont {L\"u}}, \bibinfo {author} {\bibfnamefont {C.-W.}\ \bibnamefont
  {Li}}, \bibinfo {author} {\bibfnamefont {L.}~\bibnamefont {Yang}}, \bibinfo
  {author} {\bibfnamefont {F.}~\bibnamefont {Nori}},\ and\ \bibinfo {author}
  {\bibfnamefont {Y.-x.}\ \bibnamefont {Liu}},\ }\bibfield  {title} {\bibinfo
  {title} {Metrology with $\mathcal{PT}$-symmetric cavities: Enhanced
  sensitivity near the $\mathcal{PT}$-phase transition},\ }\href
  {https://doi.org/10.1103/PhysRevLett.117.110802} {\bibfield  {journal}
  {\bibinfo  {journal} {Phys. Rev. Lett.}\ }\textbf {\bibinfo {volume} {117}},\
  \bibinfo {pages} {110802} (\bibinfo {year} {2016})}\BibitemShut {NoStop}%
\bibitem [{\citenamefont {{Zhang}}\ \emph {et~al.}(2018)\citenamefont
  {{Zhang}}, \citenamefont {{Peng}}, \citenamefont {{{\"O}zdemir}},
  \citenamefont {{Pichler}}, \citenamefont {{Krimer}}, \citenamefont {{Zhao}},
  \citenamefont {{Nori}}, \citenamefont {{Liu}}, \citenamefont {{Rotter}},\
  and\ \citenamefont {{Yang}}}]{ZhangJ2018NaPho}%
  \BibitemOpen
  \bibfield  {author} {\bibinfo {author} {\bibfnamefont {J.}~\bibnamefont
  {{Zhang}}}, \bibinfo {author} {\bibfnamefont {B.}~\bibnamefont {{Peng}}},
  \bibinfo {author} {\bibfnamefont {{\c{S}}.~K.}\ \bibnamefont
  {{{\"O}zdemir}}}, \bibinfo {author} {\bibfnamefont {K.}~\bibnamefont
  {{Pichler}}}, \bibinfo {author} {\bibfnamefont {D.~O.}\ \bibnamefont
  {{Krimer}}}, \bibinfo {author} {\bibfnamefont {G.}~\bibnamefont {{Zhao}}},
  \bibinfo {author} {\bibfnamefont {F.}~\bibnamefont {{Nori}}}, \bibinfo
  {author} {\bibfnamefont {Y.-x.}\ \bibnamefont {{Liu}}}, \bibinfo {author}
  {\bibfnamefont {S.}~\bibnamefont {{Rotter}}},\ and\ \bibinfo {author}
  {\bibfnamefont {L.}~\bibnamefont {{Yang}}},\ }\bibfield  {title} {\bibinfo
  {title} {{A phonon laser operating at an exceptional point}},\ }\href
  {https://doi.org/10.1038/s41566-018-0213-5} {\bibfield  {journal} {\bibinfo
  {journal} {Nature Photonics}\ }\textbf {\bibinfo {volume} {12}},\ \bibinfo
  {pages} {479} (\bibinfo {year} {2018})}\BibitemShut {NoStop}%
\bibitem [{\citenamefont {Peano}\ \emph {et~al.}(2015)\citenamefont {Peano},
  \citenamefont {Brendel}, \citenamefont {Schmidt},\ and\ \citenamefont
  {Marquardt}}]{Peano2015PRX}%
  \BibitemOpen
  \bibfield  {author} {\bibinfo {author} {\bibfnamefont {V.}~\bibnamefont
  {Peano}}, \bibinfo {author} {\bibfnamefont {C.}~\bibnamefont {Brendel}},
  \bibinfo {author} {\bibfnamefont {M.}~\bibnamefont {Schmidt}},\ and\ \bibinfo
  {author} {\bibfnamefont {F.}~\bibnamefont {Marquardt}},\ }\bibfield  {title}
  {\bibinfo {title} {Topological phases of sound and light},\ }\href
  {https://doi.org/10.1103/PhysRevX.5.031011} {\bibfield  {journal} {\bibinfo
  {journal} {Phys. Rev. X}\ }\textbf {\bibinfo {volume} {5}},\ \bibinfo {pages}
  {031011} (\bibinfo {year} {2015})}\BibitemShut {NoStop}%
\bibitem [{\citenamefont {{Qi}}\ \emph {et~al.}(2017)\citenamefont {{Qi}},
  \citenamefont {{Xing}}, \citenamefont {{Wang}}, \citenamefont {{Zhu}},\ and\
  \citenamefont {{Zhang}}}]{QiL2017OExpr}%
  \BibitemOpen
  \bibfield  {author} {\bibinfo {author} {\bibfnamefont {L.}~\bibnamefont
  {{Qi}}}, \bibinfo {author} {\bibfnamefont {Y.}~\bibnamefont {{Xing}}},
  \bibinfo {author} {\bibfnamefont {H.-F.}\ \bibnamefont {{Wang}}}, \bibinfo
  {author} {\bibfnamefont {A.-D.}\ \bibnamefont {{Zhu}}},\ and\ \bibinfo
  {author} {\bibfnamefont {S.}~\bibnamefont {{Zhang}}},\ }\bibfield  {title}
  {\bibinfo {title} {{Simulating Z\_2 topological insulators via a
  one-dimensional cavity optomechanical cells array}},\ }\href
  {https://doi.org/10.1364/OE.25.017948} {\bibfield  {journal} {\bibinfo
  {journal} {Optics Express}\ }\textbf {\bibinfo {volume} {25}},\ \bibinfo
  {pages} {17948} (\bibinfo {year} {2017})}\BibitemShut {NoStop}%
\bibitem [{\citenamefont {{Lemonde}}\ \emph {et~al.}(2019)\citenamefont
  {{Lemonde}}, \citenamefont {{Peano}}, \citenamefont {{Rabl}},\ and\
  \citenamefont {{Angelakis}}}]{Lemonde2019NJPh}%
  \BibitemOpen
  \bibfield  {author} {\bibinfo {author} {\bibfnamefont {M.-A.}\ \bibnamefont
  {{Lemonde}}}, \bibinfo {author} {\bibfnamefont {V.}~\bibnamefont {{Peano}}},
  \bibinfo {author} {\bibfnamefont {P.}~\bibnamefont {{Rabl}}},\ and\ \bibinfo
  {author} {\bibfnamefont {D.~G.}\ \bibnamefont {{Angelakis}}},\ }\bibfield
  {title} {\bibinfo {title} {{Quantum state transfer via acoustic edge states
  in a 2D optomechanical array}},\ }\href
  {https://doi.org/10.1088/1367-2630/ab51f5} {\bibfield  {journal} {\bibinfo
  {journal} {New Journal of Physics}\ }\textbf {\bibinfo {volume} {21}},\
  \bibinfo {eid} {113030} (\bibinfo {year} {2019})}\BibitemShut {NoStop}%
\bibitem [{\citenamefont {{Ni}}\ \emph {et~al.}(2021)\citenamefont {{Ni}},
  \citenamefont {{Kim}},\ and\ \citenamefont {{Al{\`u}}}}]{NiX2021Optic}%
  \BibitemOpen
  \bibfield  {author} {\bibinfo {author} {\bibfnamefont {X.}~\bibnamefont
  {{Ni}}}, \bibinfo {author} {\bibfnamefont {S.}~\bibnamefont {{Kim}}},\ and\
  \bibinfo {author} {\bibfnamefont {A.}~\bibnamefont {{Al{\`u}}}},\ }\bibfield
  {title} {\bibinfo {title} {{Topological insulator in two synthetic dimensions
  based on an optomechanical resonator}},\ }\href
  {https://doi.org/10.1364/OPTICA.430821} {\bibfield  {journal} {\bibinfo
  {journal} {Optica}\ }\textbf {\bibinfo {volume} {8}},\ \bibinfo {pages}
  {1024} (\bibinfo {year} {2021})}\BibitemShut {NoStop}%
\bibitem [{\citenamefont {{Ren}}\ \emph {et~al.}(2022)\citenamefont {{Ren}},
  \citenamefont {{Shah}}, \citenamefont {{Pfeifer}}, \citenamefont {{Brendel}},
  \citenamefont {{Peano}}, \citenamefont {{Marquardt}},\ and\ \citenamefont
  {{Painter}}}]{Ren2022NatCo}%
  \BibitemOpen
  \bibfield  {author} {\bibinfo {author} {\bibfnamefont {H.}~\bibnamefont
  {{Ren}}}, \bibinfo {author} {\bibfnamefont {T.}~\bibnamefont {{Shah}}},
  \bibinfo {author} {\bibfnamefont {H.}~\bibnamefont {{Pfeifer}}}, \bibinfo
  {author} {\bibfnamefont {C.}~\bibnamefont {{Brendel}}}, \bibinfo {author}
  {\bibfnamefont {V.}~\bibnamefont {{Peano}}}, \bibinfo {author} {\bibfnamefont
  {F.}~\bibnamefont {{Marquardt}}},\ and\ \bibinfo {author} {\bibfnamefont
  {O.}~\bibnamefont {{Painter}}},\ }\bibfield  {title} {\bibinfo {title}
  {{Topological phonon transport in an optomechanical system}},\ }\href
  {https://doi.org/10.1038/s41467-022-30941-0} {\bibfield  {journal} {\bibinfo
  {journal} {Nature Communications}\ }\textbf {\bibinfo {volume} {13}},\
  \bibinfo {eid} {3476} (\bibinfo {year} {2022})}\BibitemShut {NoStop}%
\bibitem [{\citenamefont {{Doster}}\ \emph {et~al.}(2022)\citenamefont
  {{Doster}}, \citenamefont {{Shah}}, \citenamefont {{F{\"o}sel}},
  \citenamefont {{Paulitschke}}, \citenamefont {{Marquardt}},\ and\
  \citenamefont {{Weig}}}]{Doster2022NatCo}%
  \BibitemOpen
  \bibfield  {author} {\bibinfo {author} {\bibfnamefont {J.}~\bibnamefont
  {{Doster}}}, \bibinfo {author} {\bibfnamefont {T.}~\bibnamefont {{Shah}}},
  \bibinfo {author} {\bibfnamefont {T.}~\bibnamefont {{F{\"o}sel}}}, \bibinfo
  {author} {\bibfnamefont {P.}~\bibnamefont {{Paulitschke}}}, \bibinfo {author}
  {\bibfnamefont {F.}~\bibnamefont {{Marquardt}}},\ and\ \bibinfo {author}
  {\bibfnamefont {E.~M.}\ \bibnamefont {{Weig}}},\ }\bibfield  {title}
  {\bibinfo {title} {{Observing polarization patterns in the collective motion
  of nanomechanical arrays}},\ }\href
  {https://doi.org/10.1038/s41467-022-30024-0} {\bibfield  {journal} {\bibinfo
  {journal} {Nature Communications}\ }\textbf {\bibinfo {volume} {13}},\
  \bibinfo {eid} {2478} (\bibinfo {year} {2022})}\BibitemShut {NoStop}%
\bibitem [{\citenamefont {{Youssefi}}\ \emph {et~al.}(2022)\citenamefont
  {{Youssefi}}, \citenamefont {{Kono}}, \citenamefont {{Bancora}},
  \citenamefont {{Chegnizadeh}}, \citenamefont {{Pan}}, \citenamefont
  {{Vovk}},\ and\ \citenamefont {{Kippenberg}}}]{Youssefi2021arXiv}%
  \BibitemOpen
  \bibfield  {author} {\bibinfo {author} {\bibfnamefont {A.}~\bibnamefont
  {{Youssefi}}}, \bibinfo {author} {\bibfnamefont {S.}~\bibnamefont {{Kono}}},
  \bibinfo {author} {\bibfnamefont {A.}~\bibnamefont {{Bancora}}}, \bibinfo
  {author} {\bibfnamefont {M.}~\bibnamefont {{Chegnizadeh}}}, \bibinfo {author}
  {\bibfnamefont {J.}~\bibnamefont {{Pan}}}, \bibinfo {author} {\bibfnamefont
  {T.}~\bibnamefont {{Vovk}}},\ and\ \bibinfo {author} {\bibfnamefont {T.~J.}\
  \bibnamefont {{Kippenberg}}},\ }\bibfield  {title} {\bibinfo {title}
  {Topological lattices realized in superconducting circuit optomechanics},\
  }\href {https://doi.org/10.1038/s41586-022-05367-9} {\bibfield  {journal}
  {\bibinfo  {journal} {\nat}\ }\textbf {\bibinfo {volume} {612}},\ \bibinfo
  {pages} {666} (\bibinfo {year} {2022})}\BibitemShut {NoStop}%
\bibitem [{\citenamefont {{Xu}}\ \emph {et~al.}(2022)\citenamefont {{Xu}},
  \citenamefont {{Zhao}}, \citenamefont {{Wang}}, \citenamefont {{Chen}},\ and\
  \citenamefont {{Liu}}}]{XuXW2022FrP}%
  \BibitemOpen
  \bibfield  {author} {\bibinfo {author} {\bibfnamefont {X.-W.}\ \bibnamefont
  {{Xu}}}, \bibinfo {author} {\bibfnamefont {Y.-J.}\ \bibnamefont {{Zhao}}},
  \bibinfo {author} {\bibfnamefont {H.}~\bibnamefont {{Wang}}}, \bibinfo
  {author} {\bibfnamefont {A.-X.}\ \bibnamefont {{Chen}}},\ and\ \bibinfo
  {author} {\bibfnamefont {Y.-X.}\ \bibnamefont {{Liu}}},\ }\bibfield  {title}
  {\bibinfo {title} {{Generalized Su-Schrieffer-Heeger model in one dimensional
  optomechanical arrays}},\ }\href {https://doi.org/10.3389/fphy.2021.813801}
  {\bibfield  {journal} {\bibinfo  {journal} {Frontiers in Physics}\ }\textbf
  {\bibinfo {volume} {9}},\ \bibinfo {eid} {813801} (\bibinfo {year}
  {2022})}\BibitemShut {NoStop}%
\bibitem [{\citenamefont {Manipatruni}\ \emph {et~al.}(2009)\citenamefont
  {Manipatruni}, \citenamefont {Robinson},\ and\ \citenamefont
  {Lipson}}]{Manipatruni2009PRL}%
  \BibitemOpen
  \bibfield  {author} {\bibinfo {author} {\bibfnamefont {S.}~\bibnamefont
  {Manipatruni}}, \bibinfo {author} {\bibfnamefont {J.~T.}\ \bibnamefont
  {Robinson}},\ and\ \bibinfo {author} {\bibfnamefont {M.}~\bibnamefont
  {Lipson}},\ }\bibfield  {title} {\bibinfo {title} {Optical nonreciprocity in
  optomechanical structures},\ }\href
  {https://doi.org/10.1103/PhysRevLett.102.213903} {\bibfield  {journal}
  {\bibinfo  {journal} {Phys. Rev. Lett.}\ }\textbf {\bibinfo {volume} {102}},\
  \bibinfo {pages} {213903} (\bibinfo {year} {2009})}\BibitemShut {NoStop}%
\bibitem [{\citenamefont {Hafezi}\ and\ \citenamefont
  {Rabl}(2012)}]{Hafezi12OE}%
  \BibitemOpen
  \bibfield  {author} {\bibinfo {author} {\bibfnamefont {M.}~\bibnamefont
  {Hafezi}}\ and\ \bibinfo {author} {\bibfnamefont {P.}~\bibnamefont {Rabl}},\
  }\bibfield  {title} {\bibinfo {title} {Optomechanically induced
  non-reciprocity in microring resonators},\ }\href
  {https://doi.org/10.1364/OE.20.007672} {\bibfield  {journal} {\bibinfo
  {journal} {Opt. Express}\ }\textbf {\bibinfo {volume} {20}},\ \bibinfo
  {pages} {7672} (\bibinfo {year} {2012})}\BibitemShut {NoStop}%
\bibitem [{\citenamefont {Metelmann}\ and\ \citenamefont
  {Clerk}(2015)}]{Metelmann2015PRX}%
  \BibitemOpen
  \bibfield  {author} {\bibinfo {author} {\bibfnamefont {A.}~\bibnamefont
  {Metelmann}}\ and\ \bibinfo {author} {\bibfnamefont {A.~A.}\ \bibnamefont
  {Clerk}},\ }\bibfield  {title} {\bibinfo {title} {Nonreciprocal photon
  transmission and amplification via reservoir engineering},\ }\href
  {https://doi.org/10.1103/PhysRevX.5.021025} {\bibfield  {journal} {\bibinfo
  {journal} {Phys. Rev. X}\ }\textbf {\bibinfo {volume} {5}},\ \bibinfo {pages}
  {021025} (\bibinfo {year} {2015})}\BibitemShut {NoStop}%
\bibitem [{\citenamefont {Xu}\ and\ \citenamefont {Li}(2015)}]{XuXW2015PRA2}%
  \BibitemOpen
  \bibfield  {author} {\bibinfo {author} {\bibfnamefont {X.-W.}\ \bibnamefont
  {Xu}}\ and\ \bibinfo {author} {\bibfnamefont {Y.}~\bibnamefont {Li}},\
  }\bibfield  {title} {\bibinfo {title} {Optical nonreciprocity and
  optomechanical circulator in three-mode optomechanical systems},\ }\href
  {https://doi.org/10.1103/PhysRevA.91.053854} {\bibfield  {journal} {\bibinfo
  {journal} {Phys. Rev. A}\ }\textbf {\bibinfo {volume} {91}},\ \bibinfo
  {pages} {053854} (\bibinfo {year} {2015})}\BibitemShut {NoStop}%
\bibitem [{\citenamefont {Li}\ \emph {et~al.}(2019)\citenamefont {Li},
  \citenamefont {Huang}, \citenamefont {Xu}, \citenamefont {Miranowicz},\ and\
  \citenamefont {Jing}}]{LiBJ19PRJ}%
  \BibitemOpen
  \bibfield  {author} {\bibinfo {author} {\bibfnamefont {B.}~\bibnamefont
  {Li}}, \bibinfo {author} {\bibfnamefont {R.}~\bibnamefont {Huang}}, \bibinfo
  {author} {\bibfnamefont {X.}~\bibnamefont {Xu}}, \bibinfo {author}
  {\bibfnamefont {A.}~\bibnamefont {Miranowicz}},\ and\ \bibinfo {author}
  {\bibfnamefont {H.}~\bibnamefont {Jing}},\ }\bibfield  {title} {\bibinfo
  {title} {Nonreciprocal unconventional photon blockade in a spinning
  optomechanical system},\ }\href {https://doi.org/10.1364/PRJ.7.000630}
  {\bibfield  {journal} {\bibinfo  {journal} {Photon. Res.}\ }\textbf {\bibinfo
  {volume} {7}},\ \bibinfo {pages} {630} (\bibinfo {year} {2019})}\BibitemShut
  {NoStop}%
\bibitem [{\citenamefont {Xu}\ \emph {et~al.}(2020{\natexlab{a}})\citenamefont
  {Xu}, \citenamefont {Zhao}, \citenamefont {Wang}, \citenamefont {Jing},\ and\
  \citenamefont {Chen}}]{XuXW20PRJ}%
  \BibitemOpen
  \bibfield  {author} {\bibinfo {author} {\bibfnamefont {X.}~\bibnamefont
  {Xu}}, \bibinfo {author} {\bibfnamefont {Y.}~\bibnamefont {Zhao}}, \bibinfo
  {author} {\bibfnamefont {H.}~\bibnamefont {Wang}}, \bibinfo {author}
  {\bibfnamefont {H.}~\bibnamefont {Jing}},\ and\ \bibinfo {author}
  {\bibfnamefont {A.}~\bibnamefont {Chen}},\ }\bibfield  {title} {\bibinfo
  {title} {Quantum nonreciprocality in quadratic optomechanics},\ }\href
  {https://doi.org/10.1364/PRJ.8.000143} {\bibfield  {journal} {\bibinfo
  {journal} {Photon. Res.}\ }\textbf {\bibinfo {volume} {8}},\ \bibinfo {pages}
  {143} (\bibinfo {year} {2020}{\natexlab{a}})}\BibitemShut {NoStop}%
\bibitem [{\citenamefont {{Nie}}\ \emph {et~al.}(2022)\citenamefont {{Nie}},
  \citenamefont {{Wang}}, \citenamefont {{Wu}}, \citenamefont {{Chen}},\ and\
  \citenamefont {{Lan}}}]{NieWJ2022SCPMA}%
  \BibitemOpen
  \bibfield  {author} {\bibinfo {author} {\bibfnamefont {W.}~\bibnamefont
  {{Nie}}}, \bibinfo {author} {\bibfnamefont {L.}~\bibnamefont {{Wang}}},
  \bibinfo {author} {\bibfnamefont {Y.}~\bibnamefont {{Wu}}}, \bibinfo {author}
  {\bibfnamefont {A.}~\bibnamefont {{Chen}}},\ and\ \bibinfo {author}
  {\bibfnamefont {Y.}~\bibnamefont {{Lan}}},\ }\bibfield  {title} {\bibinfo
  {title} {{Optomechanical ratchet resonators}},\ }\href
  {https://doi.org/10.1007/s11433-021-1831-y} {\bibfield  {journal} {\bibinfo
  {journal} {Sci. China-Phys. Mech. Astron.}\ }\textbf {\bibinfo {volume}
  {65}},\ \bibinfo {eid} {230311} (\bibinfo {year} {2022})}\BibitemShut
  {NoStop}%
\bibitem [{\citenamefont {Agarwal}\ and\ \citenamefont
  {Huang}(2010)}]{Agarwal2010PRA}%
  \BibitemOpen
  \bibfield  {author} {\bibinfo {author} {\bibfnamefont {G.~S.}\ \bibnamefont
  {Agarwal}}\ and\ \bibinfo {author} {\bibfnamefont {S.}~\bibnamefont
  {Huang}},\ }\bibfield  {title} {\bibinfo {title} {Electromagnetically induced
  transparency in mechanical effects of light},\ }\href
  {https://doi.org/10.1103/PhysRevA.81.041803} {\bibfield  {journal} {\bibinfo
  {journal} {Phys. Rev. A}\ }\textbf {\bibinfo {volume} {81}},\ \bibinfo
  {pages} {041803} (\bibinfo {year} {2010})}\BibitemShut {NoStop}%
\bibitem [{\citenamefont {{Weis}}\ \emph {et~al.}(2010)\citenamefont {{Weis}},
  \citenamefont {{Rivi{\`e}re}}, \citenamefont {{Del{\'e}glise}}, \citenamefont
  {{Gavartin}}, \citenamefont {{Arcizet}}, \citenamefont {{Schliesser}},\ and\
  \citenamefont {{Kippenberg}}}]{Weis2010Sci}%
  \BibitemOpen
  \bibfield  {author} {\bibinfo {author} {\bibfnamefont {S.}~\bibnamefont
  {{Weis}}}, \bibinfo {author} {\bibfnamefont {R.}~\bibnamefont
  {{Rivi{\`e}re}}}, \bibinfo {author} {\bibfnamefont {S.}~\bibnamefont
  {{Del{\'e}glise}}}, \bibinfo {author} {\bibfnamefont {E.}~\bibnamefont
  {{Gavartin}}}, \bibinfo {author} {\bibfnamefont {O.}~\bibnamefont
  {{Arcizet}}}, \bibinfo {author} {\bibfnamefont {A.}~\bibnamefont
  {{Schliesser}}},\ and\ \bibinfo {author} {\bibfnamefont {T.~J.}\ \bibnamefont
  {{Kippenberg}}},\ }\bibfield  {title} {\bibinfo {title} {{Optomechanically
  Induced Transparency}},\ }\href {https://doi.org/10.1126/science.1195596}
  {\bibfield  {journal} {\bibinfo  {journal} {Science}\ }\textbf {\bibinfo
  {volume} {330}},\ \bibinfo {pages} {1520} (\bibinfo {year}
  {2010})}\BibitemShut {NoStop}%
\bibitem [{\citenamefont {{Safavi-Naeini}}\ \emph {et~al.}(2011)\citenamefont
  {{Safavi-Naeini}}, \citenamefont {{Alegre}}, \citenamefont {{Chan}},
  \citenamefont {{Eichenfield}}, \citenamefont {{Winger}}, \citenamefont
  {{Lin}}, \citenamefont {{Hill}}, \citenamefont {{Chang}},\ and\ \citenamefont
  {{Painter}}}]{Safavi2011Natur}%
  \BibitemOpen
  \bibfield  {author} {\bibinfo {author} {\bibfnamefont {A.~H.}\ \bibnamefont
  {{Safavi-Naeini}}}, \bibinfo {author} {\bibfnamefont {T.~P.~M.}\ \bibnamefont
  {{Alegre}}}, \bibinfo {author} {\bibfnamefont {J.}~\bibnamefont {{Chan}}},
  \bibinfo {author} {\bibfnamefont {M.}~\bibnamefont {{Eichenfield}}}, \bibinfo
  {author} {\bibfnamefont {M.}~\bibnamefont {{Winger}}}, \bibinfo {author}
  {\bibfnamefont {Q.}~\bibnamefont {{Lin}}}, \bibinfo {author} {\bibfnamefont
  {J.~T.}\ \bibnamefont {{Hill}}}, \bibinfo {author} {\bibfnamefont {D.~E.}\
  \bibnamefont {{Chang}}},\ and\ \bibinfo {author} {\bibfnamefont
  {O.}~\bibnamefont {{Painter}}},\ }\bibfield  {title} {\bibinfo {title}
  {{Electromagnetically induced transparency and slow light with
  optomechanics}},\ }\href {https://doi.org/10.1038/nature09933} {\bibfield
  {journal} {\bibinfo  {journal} {\nat}\ }\textbf {\bibinfo {volume} {472}},\
  \bibinfo {pages} {69} (\bibinfo {year} {2011})}\BibitemShut {NoStop}%
\bibitem [{\citenamefont {{Massel}}\ \emph {et~al.}(2011)\citenamefont
  {{Massel}}, \citenamefont {{Heikkil{\"a}}}, \citenamefont {{Pirkkalainen}},
  \citenamefont {{Cho}}, \citenamefont {{Saloniemi}}, \citenamefont
  {{Hakonen}},\ and\ \citenamefont {{Sillanp{\"a}{\"a}}}}]{Massel2011Natur}%
  \BibitemOpen
  \bibfield  {author} {\bibinfo {author} {\bibfnamefont {F.}~\bibnamefont
  {{Massel}}}, \bibinfo {author} {\bibfnamefont {T.~T.}\ \bibnamefont
  {{Heikkil{\"a}}}}, \bibinfo {author} {\bibfnamefont {J.~M.}\ \bibnamefont
  {{Pirkkalainen}}}, \bibinfo {author} {\bibfnamefont {S.~U.}\ \bibnamefont
  {{Cho}}}, \bibinfo {author} {\bibfnamefont {H.}~\bibnamefont {{Saloniemi}}},
  \bibinfo {author} {\bibfnamefont {P.~J.}\ \bibnamefont {{Hakonen}}},\ and\
  \bibinfo {author} {\bibfnamefont {M.~A.}\ \bibnamefont
  {{Sillanp{\"a}{\"a}}}},\ }\bibfield  {title} {\bibinfo {title} {{Microwave
  amplification with nanomechanical resonators}},\ }\href
  {https://doi.org/10.1038/nature10628} {\bibfield  {journal} {\bibinfo
  {journal} {\nat}\ }\textbf {\bibinfo {volume} {480}},\ \bibinfo {pages} {351}
  (\bibinfo {year} {2011})}\BibitemShut {NoStop}%
\bibitem [{\citenamefont {{Hocke}}\ \emph {et~al.}(2012)\citenamefont
  {{Hocke}}, \citenamefont {{Zhou}}, \citenamefont {{Schliesser}},
  \citenamefont {{Kippenberg}}, \citenamefont {{Huebl}},\ and\ \citenamefont
  {{Gross}}}]{Hocke2012NJPh}%
  \BibitemOpen
  \bibfield  {author} {\bibinfo {author} {\bibfnamefont {F.}~\bibnamefont
  {{Hocke}}}, \bibinfo {author} {\bibfnamefont {X.}~\bibnamefont {{Zhou}}},
  \bibinfo {author} {\bibfnamefont {A.}~\bibnamefont {{Schliesser}}}, \bibinfo
  {author} {\bibfnamefont {T.~J.}\ \bibnamefont {{Kippenberg}}}, \bibinfo
  {author} {\bibfnamefont {H.}~\bibnamefont {{Huebl}}},\ and\ \bibinfo {author}
  {\bibfnamefont {R.}~\bibnamefont {{Gross}}},\ }\bibfield  {title} {\bibinfo
  {title} {{Electromechanically induced absorption in a circuit
  nano-electromechanical system}},\ }\href
  {https://doi.org/10.1088/1367-2630/14/12/123037} {\bibfield  {journal}
  {\bibinfo  {journal} {New Journal of Physics}\ }\textbf {\bibinfo {volume}
  {14}},\ \bibinfo {eid} {123037} (\bibinfo {year} {2012})}\BibitemShut
  {NoStop}%
\bibitem [{\citenamefont {{Shen}}\ \emph {et~al.}(2016)\citenamefont {{Shen}},
  \citenamefont {{Zhang}}, \citenamefont {{Chen}}, \citenamefont {{Zou}},
  \citenamefont {{Xiao}}, \citenamefont {{Zou}}, \citenamefont {{Sun}},
  \citenamefont {{Guo}},\ and\ \citenamefont {{Dong}}}]{Dong2016NaPho}%
  \BibitemOpen
  \bibfield  {author} {\bibinfo {author} {\bibfnamefont {Z.}~\bibnamefont
  {{Shen}}}, \bibinfo {author} {\bibfnamefont {Y.-L.}\ \bibnamefont {{Zhang}}},
  \bibinfo {author} {\bibfnamefont {Y.}~\bibnamefont {{Chen}}}, \bibinfo
  {author} {\bibfnamefont {C.-L.}\ \bibnamefont {{Zou}}}, \bibinfo {author}
  {\bibfnamefont {Y.-F.}\ \bibnamefont {{Xiao}}}, \bibinfo {author}
  {\bibfnamefont {X.-B.}\ \bibnamefont {{Zou}}}, \bibinfo {author}
  {\bibfnamefont {F.-W.}\ \bibnamefont {{Sun}}}, \bibinfo {author}
  {\bibfnamefont {G.-C.}\ \bibnamefont {{Guo}}},\ and\ \bibinfo {author}
  {\bibfnamefont {C.-H.}\ \bibnamefont {{Dong}}},\ }\bibfield  {title}
  {\bibinfo {title} {{Experimental realization of optomechanically induced
  non-reciprocity}},\ }\href {https://doi.org/10.1038/nphoton.2016.161}
  {\bibfield  {journal} {\bibinfo  {journal} {Nature Photonics}\ }\textbf
  {\bibinfo {volume} {10}},\ \bibinfo {pages} {657} (\bibinfo {year}
  {2016})}\BibitemShut {NoStop}%
\bibitem [{\citenamefont {{Ruesink}}\ \emph {et~al.}(2016)\citenamefont
  {{Ruesink}}, \citenamefont {{Miri}}, \citenamefont {{Al{\`{u}}}},\ and\
  \citenamefont {{Verhagen}}}]{Ruesink2016NatCo}%
  \BibitemOpen
  \bibfield  {author} {\bibinfo {author} {\bibfnamefont {F.}~\bibnamefont
  {{Ruesink}}}, \bibinfo {author} {\bibfnamefont {M.-A.}\ \bibnamefont
  {{Miri}}}, \bibinfo {author} {\bibfnamefont {A.}~\bibnamefont
  {{Al{\`{u}}}}},\ and\ \bibinfo {author} {\bibfnamefont {E.}~\bibnamefont
  {{Verhagen}}},\ }\bibfield  {title} {\bibinfo {title} {{Nonreciprocity and
  magnetic-free isolation based on optomechanical interactions}},\ }\href
  {https://doi.org/10.1038/ncomms13662} {\bibfield  {journal} {\bibinfo
  {journal} {Nature Communications}\ }\textbf {\bibinfo {volume} {7}},\
  \bibinfo {eid} {13662} (\bibinfo {year} {2016})}\BibitemShut {NoStop}%
\bibitem [{\citenamefont {{Shen}}\ \emph {et~al.}(2018)\citenamefont {{Shen}},
  \citenamefont {{Zhang}}, \citenamefont {{Chen}}, \citenamefont {{Sun}},
  \citenamefont {{Zou}}, \citenamefont {{Guo}}, \citenamefont {{Zou}},\ and\
  \citenamefont {{Dong}}}]{Shen2018NatCo}%
  \BibitemOpen
  \bibfield  {author} {\bibinfo {author} {\bibfnamefont {Z.}~\bibnamefont
  {{Shen}}}, \bibinfo {author} {\bibfnamefont {Y.-L.}\ \bibnamefont {{Zhang}}},
  \bibinfo {author} {\bibfnamefont {Y.}~\bibnamefont {{Chen}}}, \bibinfo
  {author} {\bibfnamefont {F.-W.}\ \bibnamefont {{Sun}}}, \bibinfo {author}
  {\bibfnamefont {X.-B.}\ \bibnamefont {{Zou}}}, \bibinfo {author}
  {\bibfnamefont {G.-C.}\ \bibnamefont {{Guo}}}, \bibinfo {author}
  {\bibfnamefont {C.-L.}\ \bibnamefont {{Zou}}},\ and\ \bibinfo {author}
  {\bibfnamefont {C.-H.}\ \bibnamefont {{Dong}}},\ }\bibfield  {title}
  {\bibinfo {title} {{Reconfigurable optomechanical circulator and directional
  amplifier}},\ }\href {https://doi.org/10.1038/s41467-018-04187-8} {\bibfield
  {journal} {\bibinfo  {journal} {Nature Communications}\ }\textbf {\bibinfo
  {volume} {9}},\ \bibinfo {eid} {1797} (\bibinfo {year} {2018})}\BibitemShut
  {NoStop}%
\bibitem [{\citenamefont {{Ruesink}}\ \emph {et~al.}(2018)\citenamefont
  {{Ruesink}}, \citenamefont {{Mathew}}, \citenamefont {{Miri}}, \citenamefont
  {{Al{\`{u}}}},\ and\ \citenamefont {{Verhagen}}}]{Ruesink2018NatCo}%
  \BibitemOpen
  \bibfield  {author} {\bibinfo {author} {\bibfnamefont {F.}~\bibnamefont
  {{Ruesink}}}, \bibinfo {author} {\bibfnamefont {J.~P.}\ \bibnamefont
  {{Mathew}}}, \bibinfo {author} {\bibfnamefont {M.-A.}\ \bibnamefont
  {{Miri}}}, \bibinfo {author} {\bibfnamefont {A.}~\bibnamefont
  {{Al{\`{u}}}}},\ and\ \bibinfo {author} {\bibfnamefont {E.}~\bibnamefont
  {{Verhagen}}},\ }\bibfield  {title} {\bibinfo {title} {{Optical circulation
  in a multimode optomechanical resonator}},\ }\href
  {https://doi.org/10.1038/s41467-018-04202-y} {\bibfield  {journal} {\bibinfo
  {journal} {Nature Communications}\ }\textbf {\bibinfo {volume} {9}},\
  \bibinfo {eid} {1798} (\bibinfo {year} {2018})}\BibitemShut {NoStop}%
\bibitem [{\citenamefont {Qiu}\ \emph {et~al.}(2017)\citenamefont {Qiu},
  \citenamefont {Dong}, \citenamefont {Liu},\ and\ \citenamefont
  {Zhang}}]{Qiu17OE}%
  \BibitemOpen
  \bibfield  {author} {\bibinfo {author} {\bibfnamefont {H.}~\bibnamefont
  {Qiu}}, \bibinfo {author} {\bibfnamefont {J.}~\bibnamefont {Dong}}, \bibinfo
  {author} {\bibfnamefont {L.}~\bibnamefont {Liu}},\ and\ \bibinfo {author}
  {\bibfnamefont {X.}~\bibnamefont {Zhang}},\ }\bibfield  {title} {\bibinfo
  {title} {Energy-efficient on-chip optical diode based on the optomechanical
  effect},\ }\href {https://doi.org/10.1364/OE.25.008975} {\bibfield  {journal}
  {\bibinfo  {journal} {Opt. Express}\ }\textbf {\bibinfo {volume} {25}},\
  \bibinfo {pages} {8975} (\bibinfo {year} {2017})}\BibitemShut {NoStop}%
\bibitem [{\citenamefont {Xu}\ \emph {et~al.}(2018)\citenamefont {Xu},
  \citenamefont {Song}, \citenamefont {Zheng}, \citenamefont {Wang},\ and\
  \citenamefont {Li}}]{Xu18PRA}%
  \BibitemOpen
  \bibfield  {author} {\bibinfo {author} {\bibfnamefont {X.-W.}\ \bibnamefont
  {Xu}}, \bibinfo {author} {\bibfnamefont {L.~N.}\ \bibnamefont {Song}},
  \bibinfo {author} {\bibfnamefont {Q.}~\bibnamefont {Zheng}}, \bibinfo
  {author} {\bibfnamefont {Z.~H.}\ \bibnamefont {Wang}},\ and\ \bibinfo
  {author} {\bibfnamefont {Y.}~\bibnamefont {Li}},\ }\bibfield  {title}
  {\bibinfo {title} {Optomechanically induced nonreciprocity in a three-mode
  optomechanical system},\ }\href {https://doi.org/10.1103/PhysRevA.98.063845}
  {\bibfield  {journal} {\bibinfo  {journal} {Phys. Rev. A}\ }\textbf {\bibinfo
  {volume} {98}},\ \bibinfo {pages} {063845} (\bibinfo {year}
  {2018})}\BibitemShut {NoStop}%
\bibitem [{\citenamefont {Song}\ \emph {et~al.}(2019)\citenamefont {Song},
  \citenamefont {Zheng}, \citenamefont {Xu}, \citenamefont {Jiang},\ and\
  \citenamefont {Li}}]{Song19PRA}%
  \BibitemOpen
  \bibfield  {author} {\bibinfo {author} {\bibfnamefont {L.~N.}\ \bibnamefont
  {Song}}, \bibinfo {author} {\bibfnamefont {Q.}~\bibnamefont {Zheng}},
  \bibinfo {author} {\bibfnamefont {X.-W.}\ \bibnamefont {Xu}}, \bibinfo
  {author} {\bibfnamefont {C.}~\bibnamefont {Jiang}},\ and\ \bibinfo {author}
  {\bibfnamefont {Y.}~\bibnamefont {Li}},\ }\bibfield  {title} {\bibinfo
  {title} {Optimal unidirectional amplification induced by optical gain in
  optomechanical systems},\ }\href
  {https://doi.org/10.1103/PhysRevA.100.043835} {\bibfield  {journal} {\bibinfo
   {journal} {Phys. Rev. A}\ }\textbf {\bibinfo {volume} {100}},\ \bibinfo
  {pages} {043835} (\bibinfo {year} {2019})}\BibitemShut {NoStop}%
\bibitem [{\citenamefont {{Kim}}\ \emph {et~al.}(2015)\citenamefont {{Kim}},
  \citenamefont {{Kuzyk}}, \citenamefont {{Han}}, \citenamefont {{Wang}},\ and\
  \citenamefont {{Bahl}}}]{Junhwan2015nphys}%
  \BibitemOpen
  \bibfield  {author} {\bibinfo {author} {\bibfnamefont {J.}~\bibnamefont
  {{Kim}}}, \bibinfo {author} {\bibfnamefont {M.~C.}\ \bibnamefont {{Kuzyk}}},
  \bibinfo {author} {\bibfnamefont {K.}~\bibnamefont {{Han}}}, \bibinfo
  {author} {\bibfnamefont {H.}~\bibnamefont {{Wang}}},\ and\ \bibinfo {author}
  {\bibfnamefont {G.}~\bibnamefont {{Bahl}}},\ }\bibfield  {title} {\bibinfo
  {title} {{Non-reciprocal Brillouin scattering induced transparency}},\ }\href
  {https://doi.org/10.1038/nphys3236} {\bibfield  {journal} {\bibinfo
  {journal} {Nature Physics}\ }\textbf {\bibinfo {volume} {11}},\ \bibinfo
  {pages} {275} (\bibinfo {year} {2015})}\BibitemShut {NoStop}%
\bibitem [{\citenamefont {{Dong}}\ \emph
  {et~al.}(2015{\natexlab{a}})\citenamefont {{Dong}}, \citenamefont {{Shen}},
  \citenamefont {{Zou}}, \citenamefont {{Zhang}}, \citenamefont {{Fu}},\ and\
  \citenamefont {{Guo}}}]{Dong2015ncomms}%
  \BibitemOpen
  \bibfield  {author} {\bibinfo {author} {\bibfnamefont {C.-H.}\ \bibnamefont
  {{Dong}}}, \bibinfo {author} {\bibfnamefont {Z.}~\bibnamefont {{Shen}}},
  \bibinfo {author} {\bibfnamefont {C.-L.}\ \bibnamefont {{Zou}}}, \bibinfo
  {author} {\bibfnamefont {Y.-L.}\ \bibnamefont {{Zhang}}}, \bibinfo {author}
  {\bibfnamefont {W.}~\bibnamefont {{Fu}}},\ and\ \bibinfo {author}
  {\bibfnamefont {G.-C.}\ \bibnamefont {{Guo}}},\ }\bibfield  {title} {\bibinfo
  {title} {{Brillouin-scattering-induced transparency and non-reciprocal light
  storage}},\ }\href {https://doi.org/10.1038/ncomms7193} {\bibfield  {journal}
  {\bibinfo  {journal} {Nature Communications}\ }\textbf {\bibinfo {volume}
  {6}},\ \bibinfo {eid} {6193} (\bibinfo {year}
  {2015}{\natexlab{a}})}\BibitemShut {NoStop}%
\bibitem [{\citenamefont {{Schmidt}}\ \emph {et~al.}(2015)\citenamefont
  {{Schmidt}}, \citenamefont {{Kessler}}, \citenamefont {{Peano}},
  \citenamefont {{Painter}},\ and\ \citenamefont
  {{Marquardt}}}]{Schmidt2015Optic}%
  \BibitemOpen
  \bibfield  {author} {\bibinfo {author} {\bibfnamefont {M.}~\bibnamefont
  {{Schmidt}}}, \bibinfo {author} {\bibfnamefont {S.}~\bibnamefont
  {{Kessler}}}, \bibinfo {author} {\bibfnamefont {V.}~\bibnamefont {{Peano}}},
  \bibinfo {author} {\bibfnamefont {O.}~\bibnamefont {{Painter}}},\ and\
  \bibinfo {author} {\bibfnamefont {F.}~\bibnamefont {{Marquardt}}},\
  }\bibfield  {title} {\bibinfo {title} {{Optomechanical creation of magnetic
  fields for photons on a lattice}},\ }\href
  {https://doi.org/10.1364/OPTICA.2.000635} {\bibfield  {journal} {\bibinfo
  {journal} {Optica}\ }\textbf {\bibinfo {volume} {2}},\ \bibinfo {pages} {635}
  (\bibinfo {year} {2015})}\BibitemShut {NoStop}%
\bibitem [{\citenamefont {{Fang}}\ \emph {et~al.}(2017)\citenamefont {{Fang}},
  \citenamefont {{Luo}}, \citenamefont {{Metelmann}}, \citenamefont
  {{Matheny}}, \citenamefont {{Marquardt}}, \citenamefont {{Clerk}},\ and\
  \citenamefont {{Painter}}}]{Fang2017NatPh}%
  \BibitemOpen
  \bibfield  {author} {\bibinfo {author} {\bibfnamefont {K.}~\bibnamefont
  {{Fang}}}, \bibinfo {author} {\bibfnamefont {J.}~\bibnamefont {{Luo}}},
  \bibinfo {author} {\bibfnamefont {A.}~\bibnamefont {{Metelmann}}}, \bibinfo
  {author} {\bibfnamefont {M.~H.}\ \bibnamefont {{Matheny}}}, \bibinfo {author}
  {\bibfnamefont {F.}~\bibnamefont {{Marquardt}}}, \bibinfo {author}
  {\bibfnamefont {A.~A.}\ \bibnamefont {{Clerk}}},\ and\ \bibinfo {author}
  {\bibfnamefont {O.}~\bibnamefont {{Painter}}},\ }\bibfield  {title} {\bibinfo
  {title} {{Generalized non-reciprocity in an optomechanical circuit via
  synthetic magnetism and reservoir engineering}},\ }\href
  {https://doi.org/10.1038/nphys4009} {\bibfield  {journal} {\bibinfo
  {journal} {Nature Physics}\ }\textbf {\bibinfo {volume} {13}},\ \bibinfo
  {pages} {465} (\bibinfo {year} {2017})}\BibitemShut {NoStop}%
\bibitem [{\citenamefont {Peterson}\ \emph {et~al.}(2017)\citenamefont
  {Peterson}, \citenamefont {Lecocq}, \citenamefont {Cicak}, \citenamefont
  {Simmonds}, \citenamefont {Aumentado},\ and\ \citenamefont
  {Teufel}}]{Peterson2017PRX}%
  \BibitemOpen
  \bibfield  {author} {\bibinfo {author} {\bibfnamefont {G.~A.}\ \bibnamefont
  {Peterson}}, \bibinfo {author} {\bibfnamefont {F.}~\bibnamefont {Lecocq}},
  \bibinfo {author} {\bibfnamefont {K.}~\bibnamefont {Cicak}}, \bibinfo
  {author} {\bibfnamefont {R.~W.}\ \bibnamefont {Simmonds}}, \bibinfo {author}
  {\bibfnamefont {J.}~\bibnamefont {Aumentado}},\ and\ \bibinfo {author}
  {\bibfnamefont {J.~D.}\ \bibnamefont {Teufel}},\ }\bibfield  {title}
  {\bibinfo {title} {Demonstration of efficient nonreciprocity in a microwave
  optomechanical circuit},\ }\href {https://doi.org/10.1103/PhysRevX.7.031001}
  {\bibfield  {journal} {\bibinfo  {journal} {Phys. Rev. X}\ }\textbf {\bibinfo
  {volume} {7}},\ \bibinfo {pages} {031001} (\bibinfo {year}
  {2017})}\BibitemShut {NoStop}%
\bibitem [{\citenamefont {{Bernier}}\ \emph {et~al.}(2017)\citenamefont
  {{Bernier}}, \citenamefont {{T{\'o}th}}, \citenamefont {{Koottandavida}},
  \citenamefont {{Ioannou}}, \citenamefont {{Malz}}, \citenamefont
  {{Nunnenkamp}}, \citenamefont {{Feofanov}},\ and\ \citenamefont
  {{Kippenberg}}}]{Bernier2017NatCo}%
  \BibitemOpen
  \bibfield  {author} {\bibinfo {author} {\bibfnamefont {N.~R.}\ \bibnamefont
  {{Bernier}}}, \bibinfo {author} {\bibfnamefont {L.~D.}\ \bibnamefont
  {{T{\'o}th}}}, \bibinfo {author} {\bibfnamefont {A.}~\bibnamefont
  {{Koottandavida}}}, \bibinfo {author} {\bibfnamefont {M.~A.}\ \bibnamefont
  {{Ioannou}}}, \bibinfo {author} {\bibfnamefont {D.}~\bibnamefont {{Malz}}},
  \bibinfo {author} {\bibfnamefont {A.}~\bibnamefont {{Nunnenkamp}}}, \bibinfo
  {author} {\bibfnamefont {A.~K.}\ \bibnamefont {{Feofanov}}},\ and\ \bibinfo
  {author} {\bibfnamefont {T.~J.}\ \bibnamefont {{Kippenberg}}},\ }\bibfield
  {title} {\bibinfo {title} {{Nonreciprocal reconfigurable microwave
  optomechanical circuit}},\ }\href
  {https://doi.org/10.1038/s41467-017-00447-1} {\bibfield  {journal} {\bibinfo
  {journal} {Nature Communications}\ }\textbf {\bibinfo {volume} {8}},\
  \bibinfo {eid} {604} (\bibinfo {year} {2017})}\BibitemShut {NoStop}%
\bibitem [{\citenamefont {{Barzanjeh}}\ \emph {et~al.}(2017)\citenamefont
  {{Barzanjeh}}, \citenamefont {{Wulf}}, \citenamefont {{Peruzzo}},
  \citenamefont {{Kalaee}}, \citenamefont {{Dieterle}}, \citenamefont
  {{Painter}},\ and\ \citenamefont {{Fink}}}]{Barzanjeh2017NatCo}%
  \BibitemOpen
  \bibfield  {author} {\bibinfo {author} {\bibfnamefont {S.}~\bibnamefont
  {{Barzanjeh}}}, \bibinfo {author} {\bibfnamefont {M.}~\bibnamefont {{Wulf}}},
  \bibinfo {author} {\bibfnamefont {M.}~\bibnamefont {{Peruzzo}}}, \bibinfo
  {author} {\bibfnamefont {M.}~\bibnamefont {{Kalaee}}}, \bibinfo {author}
  {\bibfnamefont {P.~B.}\ \bibnamefont {{Dieterle}}}, \bibinfo {author}
  {\bibfnamefont {O.}~\bibnamefont {{Painter}}},\ and\ \bibinfo {author}
  {\bibfnamefont {J.~M.}\ \bibnamefont {{Fink}}},\ }\bibfield  {title}
  {\bibinfo {title} {{Mechanical on-chip microwave circulator}},\ }\href
  {https://doi.org/10.1038/s41467-017-01304-x} {\bibfield  {journal} {\bibinfo
  {journal} {Nature Communications}\ }\textbf {\bibinfo {volume} {8}},\
  \bibinfo {eid} {953} (\bibinfo {year} {2017})}\BibitemShut {NoStop}%
\bibitem [{\citenamefont {Metelmann}\ and\ \citenamefont
  {Clerk}(2017)}]{Metelmann2017PRA}%
  \BibitemOpen
  \bibfield  {author} {\bibinfo {author} {\bibfnamefont {A.}~\bibnamefont
  {Metelmann}}\ and\ \bibinfo {author} {\bibfnamefont {A.~A.}\ \bibnamefont
  {Clerk}},\ }\bibfield  {title} {\bibinfo {title} {Nonreciprocal quantum
  interactions and devices via autonomous feedforward},\ }\href
  {https://doi.org/10.1103/PhysRevA.95.013837} {\bibfield  {journal} {\bibinfo
  {journal} {Phys. Rev. A}\ }\textbf {\bibinfo {volume} {95}},\ \bibinfo
  {pages} {013837} (\bibinfo {year} {2017})}\BibitemShut {NoStop}%
\bibitem [{\citenamefont {Tian}\ and\ \citenamefont {Li}(2017)}]{TianL2017PRA}%
  \BibitemOpen
  \bibfield  {author} {\bibinfo {author} {\bibfnamefont {L.}~\bibnamefont
  {Tian}}\ and\ \bibinfo {author} {\bibfnamefont {Z.}~\bibnamefont {Li}},\
  }\bibfield  {title} {\bibinfo {title} {Nonreciprocal quantum-state conversion
  between microwave and optical photons},\ }\href
  {https://doi.org/10.1103/PhysRevA.96.013808} {\bibfield  {journal} {\bibinfo
  {journal} {Phys. Rev. A}\ }\textbf {\bibinfo {volume} {96}},\ \bibinfo
  {pages} {013808} (\bibinfo {year} {2017})}\BibitemShut {NoStop}%
\bibitem [{\citenamefont {Malz}\ \emph {et~al.}(2018)\citenamefont {Malz},
  \citenamefont {T\'oth}, \citenamefont {Bernier}, \citenamefont {Feofanov},
  \citenamefont {Kippenberg},\ and\ \citenamefont {Nunnenkamp}}]{Malz2018PRL}%
  \BibitemOpen
  \bibfield  {author} {\bibinfo {author} {\bibfnamefont {D.}~\bibnamefont
  {Malz}}, \bibinfo {author} {\bibfnamefont {L.~D.}\ \bibnamefont {T\'oth}},
  \bibinfo {author} {\bibfnamefont {N.~R.}\ \bibnamefont {Bernier}}, \bibinfo
  {author} {\bibfnamefont {A.~K.}\ \bibnamefont {Feofanov}}, \bibinfo {author}
  {\bibfnamefont {T.~J.}\ \bibnamefont {Kippenberg}},\ and\ \bibinfo {author}
  {\bibfnamefont {A.}~\bibnamefont {Nunnenkamp}},\ }\bibfield  {title}
  {\bibinfo {title} {Quantum-limited directional amplifiers with
  optomechanics},\ }\href {https://doi.org/10.1103/PhysRevLett.120.023601}
  {\bibfield  {journal} {\bibinfo  {journal} {Phys. Rev. Lett.}\ }\textbf
  {\bibinfo {volume} {120}},\ \bibinfo {pages} {023601} (\bibinfo {year}
  {2018})}\BibitemShut {NoStop}%
\bibitem [{\citenamefont {Li}\ \emph {et~al.}(2018)\citenamefont {Li},
  \citenamefont {Xiao}, \citenamefont {Li},\ and\ \citenamefont
  {Wang}}]{LiGL2018PRA}%
  \BibitemOpen
  \bibfield  {author} {\bibinfo {author} {\bibfnamefont {G.}~\bibnamefont
  {Li}}, \bibinfo {author} {\bibfnamefont {X.}~\bibnamefont {Xiao}}, \bibinfo
  {author} {\bibfnamefont {Y.}~\bibnamefont {Li}},\ and\ \bibinfo {author}
  {\bibfnamefont {X.}~\bibnamefont {Wang}},\ }\bibfield  {title} {\bibinfo
  {title} {Tunable optical nonreciprocity and a phonon-photon router in an
  optomechanical system with coupled mechanical and optical modes},\ }\href
  {https://doi.org/10.1103/PhysRevA.97.023801} {\bibfield  {journal} {\bibinfo
  {journal} {Phys. Rev. A}\ }\textbf {\bibinfo {volume} {97}},\ \bibinfo
  {pages} {023801} (\bibinfo {year} {2018})}\BibitemShut {NoStop}%
\bibitem [{\citenamefont {Jiang}\ \emph {et~al.}(2019)\citenamefont {Jiang},
  \citenamefont {Song},\ and\ \citenamefont {Li}}]{JiangC2019PRA}%
  \BibitemOpen
  \bibfield  {author} {\bibinfo {author} {\bibfnamefont {C.}~\bibnamefont
  {Jiang}}, \bibinfo {author} {\bibfnamefont {L.~N.}\ \bibnamefont {Song}},\
  and\ \bibinfo {author} {\bibfnamefont {Y.}~\bibnamefont {Li}},\ }\bibfield
  {title} {\bibinfo {title} {Directional phase-sensitive amplifier between
  microwave and optical photons},\ }\href
  {https://doi.org/10.1103/PhysRevA.99.023823} {\bibfield  {journal} {\bibinfo
  {journal} {Phys. Rev. A}\ }\textbf {\bibinfo {volume} {99}},\ \bibinfo
  {pages} {023823} (\bibinfo {year} {2019})}\BibitemShut {NoStop}%
\bibitem [{\citenamefont {Mercier~de L\'epinay}\ \emph
  {et~al.}(2019)\citenamefont {Mercier~de L\'epinay}, \citenamefont
  {Damsk\"agg}, \citenamefont {Ockeloen-Korppi},\ and\ \citenamefont
  {Sillanp\"a\"a}}]{Mercier2019PRAPP}%
  \BibitemOpen
  \bibfield  {author} {\bibinfo {author} {\bibfnamefont {L.}~\bibnamefont
  {Mercier~de L\'epinay}}, \bibinfo {author} {\bibfnamefont {E.}~\bibnamefont
  {Damsk\"agg}}, \bibinfo {author} {\bibfnamefont {C.~F.}\ \bibnamefont
  {Ockeloen-Korppi}},\ and\ \bibinfo {author} {\bibfnamefont {M.~A.}\
  \bibnamefont {Sillanp\"a\"a}},\ }\bibfield  {title} {\bibinfo {title}
  {Realization of directional amplification in a microwave optomechanical
  device},\ }\href {https://doi.org/10.1103/PhysRevApplied.11.034027}
  {\bibfield  {journal} {\bibinfo  {journal} {Phys. Rev. Applied}\ }\textbf
  {\bibinfo {volume} {11}},\ \bibinfo {pages} {034027} (\bibinfo {year}
  {2019})}\BibitemShut {NoStop}%
\bibitem [{\citenamefont {Chen}\ \emph {et~al.}(2021)\citenamefont {Chen},
  \citenamefont {Zhang}, \citenamefont {Shen}, \citenamefont {Zou},
  \citenamefont {Guo},\ and\ \citenamefont {Dong}}]{ChenY2021PRL}%
  \BibitemOpen
  \bibfield  {author} {\bibinfo {author} {\bibfnamefont {Y.}~\bibnamefont
  {Chen}}, \bibinfo {author} {\bibfnamefont {Y.-L.}\ \bibnamefont {Zhang}},
  \bibinfo {author} {\bibfnamefont {Z.}~\bibnamefont {Shen}}, \bibinfo {author}
  {\bibfnamefont {C.-L.}\ \bibnamefont {Zou}}, \bibinfo {author} {\bibfnamefont
  {G.-C.}\ \bibnamefont {Guo}},\ and\ \bibinfo {author} {\bibfnamefont {C.-H.}\
  \bibnamefont {Dong}},\ }\bibfield  {title} {\bibinfo {title} {Synthetic gauge
  fields in a single optomechanical resonator},\ }\href
  {https://doi.org/10.1103/PhysRevLett.126.123603} {\bibfield  {journal}
  {\bibinfo  {journal} {Phys. Rev. Lett.}\ }\textbf {\bibinfo {volume} {126}},\
  \bibinfo {pages} {123603} (\bibinfo {year} {2021})}\BibitemShut {NoStop}%
\bibitem [{\citenamefont {Xu}\ \emph {et~al.}(2020{\natexlab{b}})\citenamefont
  {Xu}, \citenamefont {Li}, \citenamefont {Li}, \citenamefont {Jing},\ and\
  \citenamefont {Chen}}]{XuXW2020PRAPP}%
  \BibitemOpen
  \bibfield  {author} {\bibinfo {author} {\bibfnamefont {X.-W.}\ \bibnamefont
  {Xu}}, \bibinfo {author} {\bibfnamefont {Y.}~\bibnamefont {Li}}, \bibinfo
  {author} {\bibfnamefont {B.}~\bibnamefont {Li}}, \bibinfo {author}
  {\bibfnamefont {H.}~\bibnamefont {Jing}},\ and\ \bibinfo {author}
  {\bibfnamefont {A.-X.}\ \bibnamefont {Chen}},\ }\bibfield  {title} {\bibinfo
  {title} {Nonreciprocity via nonlinearity and synthetic magnetism},\ }\href
  {https://doi.org/10.1103/PhysRevApplied.13.044070} {\bibfield  {journal}
  {\bibinfo  {journal} {Phys. Rev. Applied}\ }\textbf {\bibinfo {volume}
  {13}},\ \bibinfo {pages} {044070} (\bibinfo {year}
  {2020}{\natexlab{b}})}\BibitemShut {NoStop}%
\bibitem [{\citenamefont {Qian}\ \emph {et~al.}(2021)\citenamefont {Qian},
  \citenamefont {Lai}, \citenamefont {Chen},\ and\ \citenamefont
  {Hou}}]{QianYB2021PRA}%
  \BibitemOpen
  \bibfield  {author} {\bibinfo {author} {\bibfnamefont {Y.-B.}\ \bibnamefont
  {Qian}}, \bibinfo {author} {\bibfnamefont {D.-G.}\ \bibnamefont {Lai}},
  \bibinfo {author} {\bibfnamefont {M.-R.}\ \bibnamefont {Chen}},\ and\
  \bibinfo {author} {\bibfnamefont {B.-P.}\ \bibnamefont {Hou}},\ }\bibfield
  {title} {\bibinfo {title} {Nonreciprocal photon transmission with quantum
  noise reduction via cross-kerr nonlinearity},\ }\href
  {https://doi.org/10.1103/PhysRevA.104.033705} {\bibfield  {journal} {\bibinfo
   {journal} {Phys. Rev. A}\ }\textbf {\bibinfo {volume} {104}},\ \bibinfo
  {pages} {033705} (\bibinfo {year} {2021})}\BibitemShut {NoStop}%
\bibitem [{\citenamefont {{Seif}}\ \emph {et~al.}(2018)\citenamefont {{Seif}},
  \citenamefont {{DeGottardi}}, \citenamefont {{Esfarjani}},\ and\
  \citenamefont {{Hafezi}}}]{Seif2018NatCo}%
  \BibitemOpen
  \bibfield  {author} {\bibinfo {author} {\bibfnamefont {A.}~\bibnamefont
  {{Seif}}}, \bibinfo {author} {\bibfnamefont {W.}~\bibnamefont
  {{DeGottardi}}}, \bibinfo {author} {\bibfnamefont {K.}~\bibnamefont
  {{Esfarjani}}},\ and\ \bibinfo {author} {\bibfnamefont {M.}~\bibnamefont
  {{Hafezi}}},\ }\bibfield  {title} {\bibinfo {title} {{Thermal management and
  non-reciprocal control of phonon flow via optomechanics}},\ }\href
  {https://doi.org/10.1038/s41467-018-03624-y} {\bibfield  {journal} {\bibinfo
  {journal} {Nature Communications}\ }\textbf {\bibinfo {volume} {9}},\
  \bibinfo {eid} {1207} (\bibinfo {year} {2018})}\BibitemShut {NoStop}%
\bibitem [{\citenamefont {Barzanjeh}\ \emph {et~al.}(2018)\citenamefont
  {Barzanjeh}, \citenamefont {Aquilina},\ and\ \citenamefont
  {Xuereb}}]{Barzanjeh2018PRL}%
  \BibitemOpen
  \bibfield  {author} {\bibinfo {author} {\bibfnamefont {S.}~\bibnamefont
  {Barzanjeh}}, \bibinfo {author} {\bibfnamefont {M.}~\bibnamefont
  {Aquilina}},\ and\ \bibinfo {author} {\bibfnamefont {A.}~\bibnamefont
  {Xuereb}},\ }\bibfield  {title} {\bibinfo {title} {Manipulating the flow of
  thermal noise in quantum devices},\ }\href
  {https://doi.org/10.1103/PhysRevLett.120.060601} {\bibfield  {journal}
  {\bibinfo  {journal} {Phys. Rev. Lett.}\ }\textbf {\bibinfo {volume} {120}},\
  \bibinfo {pages} {060601} (\bibinfo {year} {2018})}\BibitemShut {NoStop}%
\bibitem [{\citenamefont {{Habraken}}\ \emph {et~al.}(2012)\citenamefont
  {{Habraken}}, \citenamefont {{Stannigel}}, \citenamefont {{Lukin}},
  \citenamefont {{Zoller}},\ and\ \citenamefont {{Rabl}}}]{Habraken2012NJPh}%
  \BibitemOpen
  \bibfield  {author} {\bibinfo {author} {\bibfnamefont {S.~J.~M.}\
  \bibnamefont {{Habraken}}}, \bibinfo {author} {\bibfnamefont
  {K.}~\bibnamefont {{Stannigel}}}, \bibinfo {author} {\bibfnamefont {M.~D.}\
  \bibnamefont {{Lukin}}}, \bibinfo {author} {\bibfnamefont {P.}~\bibnamefont
  {{Zoller}}},\ and\ \bibinfo {author} {\bibfnamefont {P.}~\bibnamefont
  {{Rabl}}},\ }\bibfield  {title} {\bibinfo {title} {{Continuous mode cooling
  and phonon routers for phononic quantum networks}},\ }\href
  {https://doi.org/10.1088/1367-2630/14/11/115004} {\bibfield  {journal}
  {\bibinfo  {journal} {New Journal of Physics}\ }\textbf {\bibinfo {volume}
  {14}},\ \bibinfo {eid} {115004} (\bibinfo {year} {2012})}\BibitemShut
  {NoStop}%
\bibitem [{\citenamefont {{Xu}}\ \emph {et~al.}(2019)\citenamefont {{Xu}},
  \citenamefont {{Jiang}}, \citenamefont {{Clerk}},\ and\ \citenamefont
  {{Harris}}}]{XuHT2019Natur}%
  \BibitemOpen
  \bibfield  {author} {\bibinfo {author} {\bibfnamefont {H.}~\bibnamefont
  {{Xu}}}, \bibinfo {author} {\bibfnamefont {L.}~\bibnamefont {{Jiang}}},
  \bibinfo {author} {\bibfnamefont {A.~A.}\ \bibnamefont {{Clerk}}},\ and\
  \bibinfo {author} {\bibfnamefont {J.~G.~E.}\ \bibnamefont {{Harris}}},\
  }\bibfield  {title} {\bibinfo {title} {{Nonreciprocal control and cooling of
  phonon modes in an optomechanical system}},\ }\href
  {https://doi.org/10.1038/s41586-019-1061-2} {\bibfield  {journal} {\bibinfo
  {journal} {\nat}\ }\textbf {\bibinfo {volume} {568}},\ \bibinfo {pages} {65}
  (\bibinfo {year} {2019})}\BibitemShut {NoStop}%
\bibitem [{\citenamefont {Lai}\ \emph {et~al.}(2022)\citenamefont {Lai},
  \citenamefont {Liao}, \citenamefont {Miranowicz},\ and\ \citenamefont
  {Nori}}]{LaiDG2022PRL}%
  \BibitemOpen
  \bibfield  {author} {\bibinfo {author} {\bibfnamefont {D.-G.}\ \bibnamefont
  {Lai}}, \bibinfo {author} {\bibfnamefont {J.-Q.}\ \bibnamefont {Liao}},
  \bibinfo {author} {\bibfnamefont {A.}~\bibnamefont {Miranowicz}},\ and\
  \bibinfo {author} {\bibfnamefont {F.}~\bibnamefont {Nori}},\ }\bibfield
  {title} {\bibinfo {title} {Noise-tolerant optomechanical entanglement via
  synthetic magnetism},\ }\href
  {https://doi.org/10.1103/PhysRevLett.129.063602} {\bibfield  {journal}
  {\bibinfo  {journal} {Phys. Rev. Lett.}\ }\textbf {\bibinfo {volume} {129}},\
  \bibinfo {pages} {063602} (\bibinfo {year} {2022})}\BibitemShut {NoStop}%
\bibitem [{\citenamefont {Sanavio}\ \emph {et~al.}(2020)\citenamefont
  {Sanavio}, \citenamefont {Peano},\ and\ \citenamefont
  {Xuereb}}]{Sanavio2020PRB}%
  \BibitemOpen
  \bibfield  {author} {\bibinfo {author} {\bibfnamefont {C.}~\bibnamefont
  {Sanavio}}, \bibinfo {author} {\bibfnamefont {V.}~\bibnamefont {Peano}},\
  and\ \bibinfo {author} {\bibfnamefont {A.}~\bibnamefont {Xuereb}},\
  }\bibfield  {title} {\bibinfo {title} {Nonreciprocal topological phononics in
  optomechanical arrays},\ }\href {https://doi.org/10.1103/PhysRevB.101.085108}
  {\bibfield  {journal} {\bibinfo  {journal} {Phys. Rev. B}\ }\textbf {\bibinfo
  {volume} {101}},\ \bibinfo {pages} {085108} (\bibinfo {year}
  {2020})}\BibitemShut {NoStop}%
\bibitem [{\citenamefont {Wilson-Rae}\ \emph {et~al.}(2007)\citenamefont
  {Wilson-Rae}, \citenamefont {Nooshi}, \citenamefont {Zwerger},\ and\
  \citenamefont {Kippenberg}}]{Wilson2007PRL}%
  \BibitemOpen
  \bibfield  {author} {\bibinfo {author} {\bibfnamefont {I.}~\bibnamefont
  {Wilson-Rae}}, \bibinfo {author} {\bibfnamefont {N.}~\bibnamefont {Nooshi}},
  \bibinfo {author} {\bibfnamefont {W.}~\bibnamefont {Zwerger}},\ and\ \bibinfo
  {author} {\bibfnamefont {T.~J.}\ \bibnamefont {Kippenberg}},\ }\bibfield
  {title} {\bibinfo {title} {Theory of ground state cooling of a mechanical
  oscillator using dynamical backaction},\ }\href
  {https://doi.org/10.1103/PhysRevLett.99.093901} {\bibfield  {journal}
  {\bibinfo  {journal} {Phys. Rev. Lett.}\ }\textbf {\bibinfo {volume} {99}},\
  \bibinfo {pages} {093901} (\bibinfo {year} {2007})}\BibitemShut {NoStop}%
\bibitem [{\citenamefont {Marquardt}\ \emph {et~al.}(2007)\citenamefont
  {Marquardt}, \citenamefont {Chen}, \citenamefont {Clerk},\ and\ \citenamefont
  {Girvin}}]{Marquardt2007PRL}%
  \BibitemOpen
  \bibfield  {author} {\bibinfo {author} {\bibfnamefont {F.}~\bibnamefont
  {Marquardt}}, \bibinfo {author} {\bibfnamefont {J.~P.}\ \bibnamefont {Chen}},
  \bibinfo {author} {\bibfnamefont {A.~A.}\ \bibnamefont {Clerk}},\ and\
  \bibinfo {author} {\bibfnamefont {S.~M.}\ \bibnamefont {Girvin}},\ }\bibfield
   {title} {\bibinfo {title} {Quantum theory of cavity-assisted sideband
  cooling of mechanical motion},\ }\href
  {https://doi.org/10.1103/PhysRevLett.99.093902} {\bibfield  {journal}
  {\bibinfo  {journal} {Phys. Rev. Lett.}\ }\textbf {\bibinfo {volume} {99}},\
  \bibinfo {pages} {093902} (\bibinfo {year} {2007})}\BibitemShut {NoStop}%
\bibitem [{\citenamefont {Duan}\ and\ \citenamefont
  {Guo}(1997)}]{DuanLM1997PRL}%
  \BibitemOpen
  \bibfield  {author} {\bibinfo {author} {\bibfnamefont {L.-M.}\ \bibnamefont
  {Duan}}\ and\ \bibinfo {author} {\bibfnamefont {G.-C.}\ \bibnamefont {Guo}},\
  }\bibfield  {title} {\bibinfo {title} {Preserving coherence in quantum
  computation by pairing quantum bits},\ }\href
  {https://doi.org/10.1103/PhysRevLett.79.1953} {\bibfield  {journal} {\bibinfo
   {journal} {Phys. Rev. Lett.}\ }\textbf {\bibinfo {volume} {79}},\ \bibinfo
  {pages} {1953} (\bibinfo {year} {1997})}\BibitemShut {NoStop}%
\bibitem [{\citenamefont {Contreras-Pulido}\ and\ \citenamefont
  {Aguado}(2008)}]{Contreras2008PRB}%
  \BibitemOpen
  \bibfield  {author} {\bibinfo {author} {\bibfnamefont {L.~D.}\ \bibnamefont
  {Contreras-Pulido}}\ and\ \bibinfo {author} {\bibfnamefont {R.}~\bibnamefont
  {Aguado}},\ }\bibfield  {title} {\bibinfo {title} {Entanglement between
  charge qubits induced by a common dissipative environment},\ }\href
  {https://doi.org/10.1103/PhysRevB.77.155420} {\bibfield  {journal} {\bibinfo
  {journal} {Phys. Rev. B}\ }\textbf {\bibinfo {volume} {77}},\ \bibinfo
  {pages} {155420} (\bibinfo {year} {2008})}\BibitemShut {NoStop}%
\bibitem [{\citenamefont {Wang}\ \emph {et~al.}(2015)\citenamefont {Wang},
  \citenamefont {Ji}, \citenamefont {Li},\ and\ \citenamefont
  {Zhou}}]{WangZH2015PRA}%
  \BibitemOpen
  \bibfield  {author} {\bibinfo {author} {\bibfnamefont {Z.~H.}\ \bibnamefont
  {Wang}}, \bibinfo {author} {\bibfnamefont {Y.~J.}\ \bibnamefont {Ji}},
  \bibinfo {author} {\bibfnamefont {Y.}~\bibnamefont {Li}},\ and\ \bibinfo
  {author} {\bibfnamefont {D.~L.}\ \bibnamefont {Zhou}},\ }\bibfield  {title}
  {\bibinfo {title} {Dissipation and decoherence induced by collective
  dephasing in a coupled-qubit system with a common bath},\ }\href
  {https://doi.org/10.1103/PhysRevA.91.013838} {\bibfield  {journal} {\bibinfo
  {journal} {Phys. Rev. A}\ }\textbf {\bibinfo {volume} {91}},\ \bibinfo
  {pages} {013838} (\bibinfo {year} {2015})}\BibitemShut {NoStop}%
\bibitem [{\citenamefont {{Barzanjeh}}\ \emph {et~al.}(2022)\citenamefont
  {{Barzanjeh}}, \citenamefont {{Xuereb}}, \citenamefont {{Gr{\"o}blacher}},
  \citenamefont {{Paternostro}}, \citenamefont {{Regal}},\ and\ \citenamefont
  {{Weig}}}]{Barzanjeh2022NatPh}%
  \BibitemOpen
  \bibfield  {author} {\bibinfo {author} {\bibfnamefont {S.}~\bibnamefont
  {{Barzanjeh}}}, \bibinfo {author} {\bibfnamefont {A.}~\bibnamefont
  {{Xuereb}}}, \bibinfo {author} {\bibfnamefont {S.}~\bibnamefont
  {{Gr{\"o}blacher}}}, \bibinfo {author} {\bibfnamefont {M.}~\bibnamefont
  {{Paternostro}}}, \bibinfo {author} {\bibfnamefont {C.~A.}\ \bibnamefont
  {{Regal}}},\ and\ \bibinfo {author} {\bibfnamefont {E.~M.}\ \bibnamefont
  {{Weig}}},\ }\bibfield  {title} {\bibinfo {title} {{Optomechanics for quantum
  technologies}},\ }\href {https://doi.org/10.1038/s41567-021-01402-0}
  {\bibfield  {journal} {\bibinfo  {journal} {Nature Physics}\ }\textbf
  {\bibinfo {volume} {18}},\ \bibinfo {pages} {15} (\bibinfo {year}
  {2022})}\BibitemShut {NoStop}%
\bibitem [{\citenamefont {{Anetsberger}}\ \emph {et~al.}(2009)\citenamefont
  {{Anetsberger}}, \citenamefont {{Arcizet}}, \citenamefont
  {{Unterreithmeier}}, \citenamefont {{Rivi{\`e}re}}, \citenamefont
  {{Schliesser}}, \citenamefont {{Weig}}, \citenamefont {{Kotthaus}},\ and\
  \citenamefont {{Kippenberg}}}]{Anetsberger2009NatPh}%
  \BibitemOpen
  \bibfield  {author} {\bibinfo {author} {\bibfnamefont {G.}~\bibnamefont
  {{Anetsberger}}}, \bibinfo {author} {\bibfnamefont {O.}~\bibnamefont
  {{Arcizet}}}, \bibinfo {author} {\bibfnamefont {Q.~P.}\ \bibnamefont
  {{Unterreithmeier}}}, \bibinfo {author} {\bibfnamefont {R.}~\bibnamefont
  {{Rivi{\`e}re}}}, \bibinfo {author} {\bibfnamefont {A.}~\bibnamefont
  {{Schliesser}}}, \bibinfo {author} {\bibfnamefont {E.~M.}\ \bibnamefont
  {{Weig}}}, \bibinfo {author} {\bibfnamefont {J.~P.}\ \bibnamefont
  {{Kotthaus}}},\ and\ \bibinfo {author} {\bibfnamefont {T.~J.}\ \bibnamefont
  {{Kippenberg}}},\ }\bibfield  {title} {\bibinfo {title} {{Near-field cavity
  optomechanics with nanomechanical oscillators}},\ }\href
  {https://doi.org/10.1038/nphys1425} {\bibfield  {journal} {\bibinfo
  {journal} {Nature Physics}\ }\textbf {\bibinfo {volume} {5}},\ \bibinfo
  {pages} {909} (\bibinfo {year} {2009})}\BibitemShut {NoStop}%
\bibitem [{\citenamefont {{Brawley}}\ \emph {et~al.}(2016)\citenamefont
  {{Brawley}}, \citenamefont {{Vanner}}, \citenamefont {{Larsen}},
  \citenamefont {{Schmid}}, \citenamefont {{Boisen}},\ and\ \citenamefont
  {{Bowen}}}]{Brawley2016NatCo}%
  \BibitemOpen
  \bibfield  {author} {\bibinfo {author} {\bibfnamefont {G.~A.}\ \bibnamefont
  {{Brawley}}}, \bibinfo {author} {\bibfnamefont {M.~R.}\ \bibnamefont
  {{Vanner}}}, \bibinfo {author} {\bibfnamefont {P.~E.}\ \bibnamefont
  {{Larsen}}}, \bibinfo {author} {\bibfnamefont {S.}~\bibnamefont {{Schmid}}},
  \bibinfo {author} {\bibfnamefont {A.}~\bibnamefont {{Boisen}}},\ and\
  \bibinfo {author} {\bibfnamefont {W.~P.}\ \bibnamefont {{Bowen}}},\
  }\bibfield  {title} {\bibinfo {title} {{Nonlinear optomechanical measurement
  of mechanical motion}},\ }\href {https://doi.org/10.1038/ncomms10988}
  {\bibfield  {journal} {\bibinfo  {journal} {Nature Communications}\ }\textbf
  {\bibinfo {volume} {7}},\ \bibinfo {eid} {10988} (\bibinfo {year}
  {2016})}\BibitemShut {NoStop}%
\bibitem [{\citenamefont {Li}\ \emph {et~al.}(2012)\citenamefont {Li},
  \citenamefont {Liu}, \citenamefont {Yi}, \citenamefont {Zou}, \citenamefont
  {Ren},\ and\ \citenamefont {Xiao}}]{LiHK2012PRA}%
  \BibitemOpen
  \bibfield  {author} {\bibinfo {author} {\bibfnamefont {H.-K.}\ \bibnamefont
  {Li}}, \bibinfo {author} {\bibfnamefont {Y.-C.}\ \bibnamefont {Liu}},
  \bibinfo {author} {\bibfnamefont {X.}~\bibnamefont {Yi}}, \bibinfo {author}
  {\bibfnamefont {C.-L.}\ \bibnamefont {Zou}}, \bibinfo {author} {\bibfnamefont
  {X.-X.}\ \bibnamefont {Ren}},\ and\ \bibinfo {author} {\bibfnamefont {Y.-F.}\
  \bibnamefont {Xiao}},\ }\bibfield  {title} {\bibinfo {title} {Proposal for a
  near-field optomechanical system with enhanced linear and quadratic
  coupling},\ }\href {https://doi.org/10.1103/PhysRevA.85.053832} {\bibfield
  {journal} {\bibinfo  {journal} {Phys. Rev. A}\ }\textbf {\bibinfo {volume}
  {85}},\ \bibinfo {pages} {053832} (\bibinfo {year} {2012})}\BibitemShut
  {NoStop}%
\bibitem [{\citenamefont {Tsang}\ and\ \citenamefont
  {Caves}(2010)}]{Tsang2010PRL}%
  \BibitemOpen
  \bibfield  {author} {\bibinfo {author} {\bibfnamefont {M.}~\bibnamefont
  {Tsang}}\ and\ \bibinfo {author} {\bibfnamefont {C.~M.}\ \bibnamefont
  {Caves}},\ }\bibfield  {title} {\bibinfo {title} {Coherent quantum-noise
  cancellation for optomechanical sensors},\ }\href
  {https://doi.org/10.1103/PhysRevLett.105.123601} {\bibfield  {journal}
  {\bibinfo  {journal} {Phys. Rev. Lett.}\ }\textbf {\bibinfo {volume} {105}},\
  \bibinfo {pages} {123601} (\bibinfo {year} {2010})}\BibitemShut {NoStop}%
\bibitem [{\citenamefont {{Chan}}\ \emph {et~al.}(2011)\citenamefont {{Chan}},
  \citenamefont {{Alegre}}, \citenamefont {{Safavi-Naeini}}, \citenamefont
  {{Hill}}, \citenamefont {{Krause}}, \citenamefont {{Gr{\"o}blacher}},
  \citenamefont {{Aspelmeyer}},\ and\ \citenamefont
  {{Painter}}}]{Chan2011Natur}%
  \BibitemOpen
  \bibfield  {author} {\bibinfo {author} {\bibfnamefont {J.}~\bibnamefont
  {{Chan}}}, \bibinfo {author} {\bibfnamefont {T.~P.~M.}\ \bibnamefont
  {{Alegre}}}, \bibinfo {author} {\bibfnamefont {A.~H.}\ \bibnamefont
  {{Safavi-Naeini}}}, \bibinfo {author} {\bibfnamefont {J.~T.}\ \bibnamefont
  {{Hill}}}, \bibinfo {author} {\bibfnamefont {A.}~\bibnamefont {{Krause}}},
  \bibinfo {author} {\bibfnamefont {S.}~\bibnamefont {{Gr{\"o}blacher}}},
  \bibinfo {author} {\bibfnamefont {M.}~\bibnamefont {{Aspelmeyer}}},\ and\
  \bibinfo {author} {\bibfnamefont {O.}~\bibnamefont {{Painter}}},\ }\bibfield
  {title} {\bibinfo {title} {{Laser cooling of a nanomechanical oscillator into
  its quantum ground state}},\ }\href {https://doi.org/10.1038/nature10461}
  {\bibfield  {journal} {\bibinfo  {journal} {\nat}\ }\textbf {\bibinfo
  {volume} {478}},\ \bibinfo {pages} {89} (\bibinfo {year} {2011})}\BibitemShut
  {NoStop}%
\bibitem [{\citenamefont {{MacCabe}}\ \emph {et~al.}(2020)\citenamefont
  {{MacCabe}}, \citenamefont {{Ren}}, \citenamefont {{Luo}}, \citenamefont
  {{Cohen}}, \citenamefont {{Zhou}}, \citenamefont {{Sipahigil}}, \citenamefont
  {{Mirhosseini}},\ and\ \citenamefont {{Painter}}}]{MacCabe2020Sci}%
  \BibitemOpen
  \bibfield  {author} {\bibinfo {author} {\bibfnamefont {G.~S.}\ \bibnamefont
  {{MacCabe}}}, \bibinfo {author} {\bibfnamefont {H.}~\bibnamefont {{Ren}}},
  \bibinfo {author} {\bibfnamefont {J.}~\bibnamefont {{Luo}}}, \bibinfo
  {author} {\bibfnamefont {J.~D.}\ \bibnamefont {{Cohen}}}, \bibinfo {author}
  {\bibfnamefont {H.}~\bibnamefont {{Zhou}}}, \bibinfo {author} {\bibfnamefont
  {A.}~\bibnamefont {{Sipahigil}}}, \bibinfo {author} {\bibfnamefont
  {M.}~\bibnamefont {{Mirhosseini}}},\ and\ \bibinfo {author} {\bibfnamefont
  {O.}~\bibnamefont {{Painter}}},\ }\bibfield  {title} {\bibinfo {title}
  {{Nano-acoustic resonator with ultralong phonon lifetime}},\ }\href
  {https://doi.org/10.1126/science.abc7312} {\bibfield  {journal} {\bibinfo
  {journal} {Science}\ }\textbf {\bibinfo {volume} {370}},\ \bibinfo {pages}
  {840} (\bibinfo {year} {2020})}\BibitemShut {NoStop}%
\bibitem [{\citenamefont {Yu}\ \emph {et~al.}(2019)\citenamefont {Yu},
  \citenamefont {Wang}, \citenamefont {Yuan},\ and\ \citenamefont
  {Xiao}}]{YuW2019PRL}%
  \BibitemOpen
  \bibfield  {author} {\bibinfo {author} {\bibfnamefont {W.}~\bibnamefont
  {Yu}}, \bibinfo {author} {\bibfnamefont {J.}~\bibnamefont {Wang}}, \bibinfo
  {author} {\bibfnamefont {H.~Y.}\ \bibnamefont {Yuan}},\ and\ \bibinfo
  {author} {\bibfnamefont {J.}~\bibnamefont {Xiao}},\ }\bibfield  {title}
  {\bibinfo {title} {Prediction of attractive level crossing via a dissipative
  mode},\ }\href {https://doi.org/10.1103/PhysRevLett.123.227201} {\bibfield
  {journal} {\bibinfo  {journal} {Phys. Rev. Lett.}\ }\textbf {\bibinfo
  {volume} {123}},\ \bibinfo {pages} {227201} (\bibinfo {year}
  {2019})}\BibitemShut {NoStop}%
\bibitem [{\citenamefont {Zhao}\ \emph {et~al.}(2020)\citenamefont {Zhao},
  \citenamefont {Liu}, \citenamefont {Wu}, \citenamefont {Duan}, \citenamefont
  {Liu},\ and\ \citenamefont {Du}}]{ZhaoJ2020PRAPP}%
  \BibitemOpen
  \bibfield  {author} {\bibinfo {author} {\bibfnamefont {J.}~\bibnamefont
  {Zhao}}, \bibinfo {author} {\bibfnamefont {Y.}~\bibnamefont {Liu}}, \bibinfo
  {author} {\bibfnamefont {L.}~\bibnamefont {Wu}}, \bibinfo {author}
  {\bibfnamefont {C.-K.}\ \bibnamefont {Duan}}, \bibinfo {author}
  {\bibfnamefont {Y.-x.}\ \bibnamefont {Liu}},\ and\ \bibinfo {author}
  {\bibfnamefont {J.}~\bibnamefont {Du}},\ }\bibfield  {title} {\bibinfo
  {title} {Observation of anti-$\mathcal{P}\mathcal{T}$-symmetry phase
  transition in the magnon-cavity-magnon coupled system},\ }\href
  {https://doi.org/10.1103/PhysRevApplied.13.014053} {\bibfield  {journal}
  {\bibinfo  {journal} {Phys. Rev. Applied}\ }\textbf {\bibinfo {volume}
  {13}},\ \bibinfo {pages} {014053} (\bibinfo {year} {2020})}\BibitemShut
  {NoStop}%
\bibitem [{\citenamefont {{Wang}}\ and\ \citenamefont
  {{Hu}}(2020)}]{WangYP2020JAP}%
  \BibitemOpen
  \bibfield  {author} {\bibinfo {author} {\bibfnamefont {Y.-P.}\ \bibnamefont
  {{Wang}}}\ and\ \bibinfo {author} {\bibfnamefont {C.-M.}\ \bibnamefont
  {{Hu}}},\ }\bibfield  {title} {\bibinfo {title} {{Dissipative couplings in
  cavity magnonics}},\ }\href {https://doi.org/10.1063/1.5144202} {\bibfield
  {journal} {\bibinfo  {journal} {Journal of Applied Physics}\ }\textbf
  {\bibinfo {volume} {127}},\ \bibinfo {eid} {130901} (\bibinfo {year}
  {2020})}\BibitemShut {NoStop}%
\bibitem [{\citenamefont {{Karg}}\ \emph {et~al.}(2020)\citenamefont {{Karg}},
  \citenamefont {{Gouraud}}, \citenamefont {{Ngai}}, \citenamefont {{Schmid}},
  \citenamefont {{Hammerer}},\ and\ \citenamefont {{Treutlein}}}]{Karg2020Sci}%
  \BibitemOpen
  \bibfield  {author} {\bibinfo {author} {\bibfnamefont {T.~M.}\ \bibnamefont
  {{Karg}}}, \bibinfo {author} {\bibfnamefont {B.}~\bibnamefont {{Gouraud}}},
  \bibinfo {author} {\bibfnamefont {C.~T.}\ \bibnamefont {{Ngai}}}, \bibinfo
  {author} {\bibfnamefont {G.-L.}\ \bibnamefont {{Schmid}}}, \bibinfo {author}
  {\bibfnamefont {K.}~\bibnamefont {{Hammerer}}},\ and\ \bibinfo {author}
  {\bibfnamefont {P.}~\bibnamefont {{Treutlein}}},\ }\bibfield  {title}
  {\bibinfo {title} {{Light-mediated strong coupling between a mechanical
  oscillator and atomic spins 1 meter apart}},\ }\href
  {https://doi.org/10.1126/science.abb0328} {\bibfield  {journal} {\bibinfo
  {journal} {Science}\ }\textbf {\bibinfo {volume} {369}},\ \bibinfo {pages}
  {174} (\bibinfo {year} {2020})}\BibitemShut {NoStop}%
\bibitem [{\citenamefont {Gardiner}\ and\ \citenamefont
  {Collett}(1985)}]{Gardiner1985PRA}%
  \BibitemOpen
  \bibfield  {author} {\bibinfo {author} {\bibfnamefont {C.~W.}\ \bibnamefont
  {Gardiner}}\ and\ \bibinfo {author} {\bibfnamefont {M.~J.}\ \bibnamefont
  {Collett}},\ }\bibfield  {title} {\bibinfo {title} {Input and output in
  damped quantum systems: Quantum stochastic differential equations and the
  master equation},\ }\href {https://doi.org/10.1103/PhysRevA.31.3761}
  {\bibfield  {journal} {\bibinfo  {journal} {Phys. Rev. A}\ }\textbf {\bibinfo
  {volume} {31}},\ \bibinfo {pages} {3761} (\bibinfo {year}
  {1985})}\BibitemShut {NoStop}%
\bibitem [{\citenamefont {Bachtold}\ \emph {et~al.}(2022)\citenamefont
  {Bachtold}, \citenamefont {Moser},\ and\ \citenamefont
  {Dykman}}]{Adrian2023RMP}%
  \BibitemOpen
  \bibfield  {author} {\bibinfo {author} {\bibfnamefont {A.}~\bibnamefont
  {Bachtold}}, \bibinfo {author} {\bibfnamefont {J.}~\bibnamefont {Moser}},\
  and\ \bibinfo {author} {\bibfnamefont {M.~I.}\ \bibnamefont {Dykman}},\
  }\bibfield  {title} {\bibinfo {title} {Mesoscopic physics of nanomechanical
  systems},\ }\href {https://doi.org/10.1103/RevModPhys.94.045005} {\bibfield
  {journal} {\bibinfo  {journal} {Rev. Mod. Phys.}\ }\textbf {\bibinfo {volume}
  {94}},\ \bibinfo {pages} {045005} (\bibinfo {year} {2022})}\BibitemShut
  {NoStop}%
\bibitem [{\citenamefont {Song}\ \emph {et~al.}(2014)\citenamefont {Song},
  \citenamefont {Oksanen}, \citenamefont {Li}, \citenamefont {Hakonen},\ and\
  \citenamefont {Sillanp\"a\"a}}]{SongX2014PRL}%
  \BibitemOpen
  \bibfield  {author} {\bibinfo {author} {\bibfnamefont {X.}~\bibnamefont
  {Song}}, \bibinfo {author} {\bibfnamefont {M.}~\bibnamefont {Oksanen}},
  \bibinfo {author} {\bibfnamefont {J.}~\bibnamefont {Li}}, \bibinfo {author}
  {\bibfnamefont {P.~J.}\ \bibnamefont {Hakonen}},\ and\ \bibinfo {author}
  {\bibfnamefont {M.~A.}\ \bibnamefont {Sillanp\"a\"a}},\ }\bibfield  {title}
  {\bibinfo {title} {Graphene optomechanics realized at microwave
  frequencies},\ }\href {https://doi.org/10.1103/PhysRevLett.113.027404}
  {\bibfield  {journal} {\bibinfo  {journal} {Phys. Rev. Lett.}\ }\textbf
  {\bibinfo {volume} {113}},\ \bibinfo {pages} {027404} (\bibinfo {year}
  {2014})}\BibitemShut {NoStop}%
\bibitem [{\citenamefont {{Weber}}\ \emph {et~al.}(2016)\citenamefont
  {{Weber}}, \citenamefont {{G{\"u}ttinger}}, \citenamefont {{Noury}},
  \citenamefont {{Vergara-Cruz}},\ and\ \citenamefont
  {{Bachtold}}}]{Weber2016NatCo}%
  \BibitemOpen
  \bibfield  {author} {\bibinfo {author} {\bibfnamefont {P.}~\bibnamefont
  {{Weber}}}, \bibinfo {author} {\bibfnamefont {J.}~\bibnamefont
  {{G{\"u}ttinger}}}, \bibinfo {author} {\bibfnamefont {A.}~\bibnamefont
  {{Noury}}}, \bibinfo {author} {\bibfnamefont {J.}~\bibnamefont
  {{Vergara-Cruz}}},\ and\ \bibinfo {author} {\bibfnamefont {A.}~\bibnamefont
  {{Bachtold}}},\ }\bibfield  {title} {\bibinfo {title} {{Force sensitivity of
  multilayer graphene optomechanical devices}},\ }\href
  {https://doi.org/10.1038/ncomms12496} {\bibfield  {journal} {\bibinfo
  {journal} {Nature Communications}\ }\textbf {\bibinfo {volume} {7}},\
  \bibinfo {eid} {12496} (\bibinfo {year} {2016})}\BibitemShut {NoStop}%
\bibitem [{\citenamefont {{Singh}}\ \emph {et~al.}(2014)\citenamefont
  {{Singh}}, \citenamefont {{Bosman}}, \citenamefont {{Schneider}},
  \citenamefont {{Blanter}}, \citenamefont {{Castellanos-Gomez}},\ and\
  \citenamefont {{Steele}}}]{Singh2014NatNa}%
  \BibitemOpen
  \bibfield  {author} {\bibinfo {author} {\bibfnamefont {V.}~\bibnamefont
  {{Singh}}}, \bibinfo {author} {\bibfnamefont {S.~J.}\ \bibnamefont
  {{Bosman}}}, \bibinfo {author} {\bibfnamefont {B.~H.}\ \bibnamefont
  {{Schneider}}}, \bibinfo {author} {\bibfnamefont {Y.~M.}\ \bibnamefont
  {{Blanter}}}, \bibinfo {author} {\bibfnamefont {A.}~\bibnamefont
  {{Castellanos-Gomez}}},\ and\ \bibinfo {author} {\bibfnamefont {G.~A.}\
  \bibnamefont {{Steele}}},\ }\bibfield  {title} {\bibinfo {title}
  {{Optomechanical coupling between a multilayer graphene mechanical resonator
  and a superconducting microwave cavity}},\ }\href
  {https://doi.org/10.1038/nnano.2014.168} {\bibfield  {journal} {\bibinfo
  {journal} {Nature Nanotechnology}\ }\textbf {\bibinfo {volume} {9}},\
  \bibinfo {pages} {820} (\bibinfo {year} {2014})}\BibitemShut {NoStop}%
\bibitem [{\citenamefont {{Dong}}\ \emph {et~al.}(2012)\citenamefont {{Dong}},
  \citenamefont {{Fiore}}, \citenamefont {{Kuzyk}},\ and\ \citenamefont
  {{Wang}}}]{DongCH2012Sci}%
  \BibitemOpen
  \bibfield  {author} {\bibinfo {author} {\bibfnamefont {C.}~\bibnamefont
  {{Dong}}}, \bibinfo {author} {\bibfnamefont {V.}~\bibnamefont {{Fiore}}},
  \bibinfo {author} {\bibfnamefont {M.~C.}\ \bibnamefont {{Kuzyk}}},\ and\
  \bibinfo {author} {\bibfnamefont {H.}~\bibnamefont {{Wang}}},\ }\bibfield
  {title} {\bibinfo {title} {{Optomechanical Dark Mode}},\ }\href
  {https://doi.org/10.1126/science.1228370} {\bibfield  {journal} {\bibinfo
  {journal} {Science}\ }\textbf {\bibinfo {volume} {338}},\ \bibinfo {pages}
  {1609} (\bibinfo {year} {2012})}\BibitemShut {NoStop}%
\bibitem [{\citenamefont {Wang}\ and\ \citenamefont
  {Clerk}(2012)}]{WangYD2012PRL}%
  \BibitemOpen
  \bibfield  {author} {\bibinfo {author} {\bibfnamefont {Y.-D.}\ \bibnamefont
  {Wang}}\ and\ \bibinfo {author} {\bibfnamefont {A.~A.}\ \bibnamefont
  {Clerk}},\ }\bibfield  {title} {\bibinfo {title} {Using interference for high
  fidelity quantum state transfer in optomechanics},\ }\href
  {https://doi.org/10.1103/PhysRevLett.108.153603} {\bibfield  {journal}
  {\bibinfo  {journal} {Phys. Rev. Lett.}\ }\textbf {\bibinfo {volume} {108}},\
  \bibinfo {pages} {153603} (\bibinfo {year} {2012})}\BibitemShut {NoStop}%
\bibitem [{\citenamefont {Tian}(2012)}]{TianL2012PRL}%
  \BibitemOpen
  \bibfield  {author} {\bibinfo {author} {\bibfnamefont {L.}~\bibnamefont
  {Tian}},\ }\bibfield  {title} {\bibinfo {title} {Adiabatic state conversion
  and pulse transmission in optomechanical systems},\ }\href
  {https://doi.org/10.1103/PhysRevLett.108.153604} {\bibfield  {journal}
  {\bibinfo  {journal} {Phys. Rev. Lett.}\ }\textbf {\bibinfo {volume} {108}},\
  \bibinfo {pages} {153604} (\bibinfo {year} {2012})}\BibitemShut {NoStop}%
\bibitem [{\citenamefont {Lai}\ \emph {et~al.}(2020)\citenamefont {Lai},
  \citenamefont {Huang}, \citenamefont {Yin}, \citenamefont {Hou},
  \citenamefont {Li}, \citenamefont {Vitali}, \citenamefont {Nori},\ and\
  \citenamefont {Liao}}]{LaiDG2020PRA}%
  \BibitemOpen
  \bibfield  {author} {\bibinfo {author} {\bibfnamefont {D.-G.}\ \bibnamefont
  {Lai}}, \bibinfo {author} {\bibfnamefont {J.-F.}\ \bibnamefont {Huang}},
  \bibinfo {author} {\bibfnamefont {X.-L.}\ \bibnamefont {Yin}}, \bibinfo
  {author} {\bibfnamefont {B.-P.}\ \bibnamefont {Hou}}, \bibinfo {author}
  {\bibfnamefont {W.}~\bibnamefont {Li}}, \bibinfo {author} {\bibfnamefont
  {D.}~\bibnamefont {Vitali}}, \bibinfo {author} {\bibfnamefont
  {F.}~\bibnamefont {Nori}},\ and\ \bibinfo {author} {\bibfnamefont {J.-Q.}\
  \bibnamefont {Liao}},\ }\bibfield  {title} {\bibinfo {title} {Nonreciprocal
  ground-state cooling of multiple mechanical resonators},\ }\href
  {https://doi.org/10.1103/PhysRevA.102.011502} {\bibfield  {journal} {\bibinfo
   {journal} {Phys. Rev. A}\ }\textbf {\bibinfo {volume} {102}},\ \bibinfo
  {pages} {011502} (\bibinfo {year} {2020})}\BibitemShut {NoStop}%
\bibitem [{\citenamefont {Zhao}\ \emph {et~al.}(2022)\citenamefont {Zhao},
  \citenamefont {Zhang}, \citenamefont {Artoni}, \citenamefont {La~Rocca},\
  and\ \citenamefont {Wu}}]{ZhaoHM2022PRA}%
  \BibitemOpen
  \bibfield  {author} {\bibinfo {author} {\bibfnamefont {H.-M.}\ \bibnamefont
  {Zhao}}, \bibinfo {author} {\bibfnamefont {X.-J.}\ \bibnamefont {Zhang}},
  \bibinfo {author} {\bibfnamefont {M.}~\bibnamefont {Artoni}}, \bibinfo
  {author} {\bibfnamefont {G.~C.}\ \bibnamefont {La~Rocca}},\ and\ \bibinfo
  {author} {\bibfnamefont {J.-H.}\ \bibnamefont {Wu}},\ }\bibfield  {title}
  {\bibinfo {title} {Photon-pair generation on resonance via a dark state},\
  }\href {https://doi.org/10.1103/PhysRevA.106.023711} {\bibfield  {journal}
  {\bibinfo  {journal} {Phys. Rev. A}\ }\textbf {\bibinfo {volume} {106}},\
  \bibinfo {pages} {023711} (\bibinfo {year} {2022})}\BibitemShut {NoStop}%
\bibitem [{\citenamefont {Doolin}\ \emph {et~al.}(2014)\citenamefont {Doolin},
  \citenamefont {Hauer}, \citenamefont {Kim}, \citenamefont {MacDonald},
  \citenamefont {Ramp},\ and\ \citenamefont {Davis}}]{Doolin2014PRA}%
  \BibitemOpen
  \bibfield  {author} {\bibinfo {author} {\bibfnamefont {C.}~\bibnamefont
  {Doolin}}, \bibinfo {author} {\bibfnamefont {B.~D.}\ \bibnamefont {Hauer}},
  \bibinfo {author} {\bibfnamefont {P.~H.}\ \bibnamefont {Kim}}, \bibinfo
  {author} {\bibfnamefont {A.~J.~R.}\ \bibnamefont {MacDonald}}, \bibinfo
  {author} {\bibfnamefont {H.}~\bibnamefont {Ramp}},\ and\ \bibinfo {author}
  {\bibfnamefont {J.~P.}\ \bibnamefont {Davis}},\ }\bibfield  {title} {\bibinfo
  {title} {Nonlinear optomechanics in the stationary regime},\ }\href
  {https://doi.org/10.1103/PhysRevA.89.053838} {\bibfield  {journal} {\bibinfo
  {journal} {Phys. Rev. A}\ }\textbf {\bibinfo {volume} {89}},\ \bibinfo
  {pages} {053838} (\bibinfo {year} {2014})}\BibitemShut {NoStop}%
\bibitem [{\citenamefont {{Asano}}\ \emph {et~al.}(2020)\citenamefont
  {{Asano}}, \citenamefont {{Zhang}}, \citenamefont {{Tawara}}, \citenamefont
  {{Yamaguchi}},\ and\ \citenamefont {{Okamoto}}}]{Asano2020CmPhy}%
  \BibitemOpen
  \bibfield  {author} {\bibinfo {author} {\bibfnamefont {M.}~\bibnamefont
  {{Asano}}}, \bibinfo {author} {\bibfnamefont {G.}~\bibnamefont {{Zhang}}},
  \bibinfo {author} {\bibfnamefont {T.}~\bibnamefont {{Tawara}}}, \bibinfo
  {author} {\bibfnamefont {H.}~\bibnamefont {{Yamaguchi}}},\ and\ \bibinfo
  {author} {\bibfnamefont {H.}~\bibnamefont {{Okamoto}}},\ }\bibfield  {title}
  {\bibinfo {title} {{Near-field cavity optomechanical coupling in a compound
  semiconductor nanowire}},\ }\href
  {https://doi.org/10.1038/s42005-020-00498-9} {\bibfield  {journal} {\bibinfo
  {journal} {Communications Physics}\ }\textbf {\bibinfo {volume} {3}},\
  \bibinfo {eid} {230} (\bibinfo {year} {2020})}\BibitemShut {NoStop}%
\bibitem [{\citenamefont {Grudinin}\ \emph {et~al.}(2010)\citenamefont
  {Grudinin}, \citenamefont {Lee}, \citenamefont {Painter},\ and\ \citenamefont
  {Vahala}}]{Grudinin2010PRL}%
  \BibitemOpen
  \bibfield  {author} {\bibinfo {author} {\bibfnamefont {I.~S.}\ \bibnamefont
  {Grudinin}}, \bibinfo {author} {\bibfnamefont {H.}~\bibnamefont {Lee}},
  \bibinfo {author} {\bibfnamefont {O.}~\bibnamefont {Painter}},\ and\ \bibinfo
  {author} {\bibfnamefont {K.~J.}\ \bibnamefont {Vahala}},\ }\bibfield  {title}
  {\bibinfo {title} {Phonon laser action in a tunable two-level system},\
  }\href {https://doi.org/10.1103/PhysRevLett.104.083901} {\bibfield  {journal}
  {\bibinfo  {journal} {Phys. Rev. Lett.}\ }\textbf {\bibinfo {volume} {104}},\
  \bibinfo {pages} {083901} (\bibinfo {year} {2010})}\BibitemShut {NoStop}%
\bibitem [{\citenamefont {Wang}\ \emph {et~al.}(2014)\citenamefont {Wang},
  \citenamefont {Wang}, \citenamefont {Zhang}, \citenamefont {\"Ozdemir},
  \citenamefont {Yang},\ and\ \citenamefont {Liu}}]{WangH2014PRA}%
  \BibitemOpen
  \bibfield  {author} {\bibinfo {author} {\bibfnamefont {H.}~\bibnamefont
  {Wang}}, \bibinfo {author} {\bibfnamefont {Z.}~\bibnamefont {Wang}}, \bibinfo
  {author} {\bibfnamefont {J.}~\bibnamefont {Zhang}}, \bibinfo {author}
  {\bibfnamefont {S.~K.}\ \bibnamefont {\"Ozdemir}}, \bibinfo {author}
  {\bibfnamefont {L.}~\bibnamefont {Yang}},\ and\ \bibinfo {author}
  {\bibfnamefont {Y.-x.}\ \bibnamefont {Liu}},\ }\bibfield  {title} {\bibinfo
  {title} {Phonon amplification in two coupled cavities containing one
  mechanical resonator},\ }\href {https://doi.org/10.1103/PhysRevA.90.053814}
  {\bibfield  {journal} {\bibinfo  {journal} {Phys. Rev. A}\ }\textbf {\bibinfo
  {volume} {90}},\ \bibinfo {pages} {053814} (\bibinfo {year}
  {2014})}\BibitemShut {NoStop}%
\bibitem [{\citenamefont {Mahmoodian}\ \emph {et~al.}(2016)\citenamefont
  {Mahmoodian}, \citenamefont {Lodahl},\ and\ \citenamefont
  {S\o{}rensen}}]{Mahmoodian2016PRL}%
  \BibitemOpen
  \bibfield  {author} {\bibinfo {author} {\bibfnamefont {S.}~\bibnamefont
  {Mahmoodian}}, \bibinfo {author} {\bibfnamefont {P.}~\bibnamefont {Lodahl}},\
  and\ \bibinfo {author} {\bibfnamefont {A.~S.}\ \bibnamefont {S\o{}rensen}},\
  }\bibfield  {title} {\bibinfo {title} {Quantum networks with
  chiral-light--matter interaction in waveguides},\ }\href
  {https://doi.org/10.1103/PhysRevLett.117.240501} {\bibfield  {journal}
  {\bibinfo  {journal} {Phys. Rev. Lett.}\ }\textbf {\bibinfo {volume} {117}},\
  \bibinfo {pages} {240501} (\bibinfo {year} {2016})}\BibitemShut {NoStop}%
\bibitem [{\citenamefont {{Lodahl}}\ \emph {et~al.}(2017)\citenamefont
  {{Lodahl}}, \citenamefont {{Mahmoodian}}, \citenamefont {{Stobbe}},
  \citenamefont {{Rauschenbeutel}}, \citenamefont {{Schneeweiss}},
  \citenamefont {{Volz}}, \citenamefont {{Pichler}},\ and\ \citenamefont
  {{Zoller}}}]{Lodahl2017Natur}%
  \BibitemOpen
  \bibfield  {author} {\bibinfo {author} {\bibfnamefont {P.}~\bibnamefont
  {{Lodahl}}}, \bibinfo {author} {\bibfnamefont {S.}~\bibnamefont
  {{Mahmoodian}}}, \bibinfo {author} {\bibfnamefont {S.}~\bibnamefont
  {{Stobbe}}}, \bibinfo {author} {\bibfnamefont {A.}~\bibnamefont
  {{Rauschenbeutel}}}, \bibinfo {author} {\bibfnamefont {P.}~\bibnamefont
  {{Schneeweiss}}}, \bibinfo {author} {\bibfnamefont {J.}~\bibnamefont
  {{Volz}}}, \bibinfo {author} {\bibfnamefont {H.}~\bibnamefont {{Pichler}}},\
  and\ \bibinfo {author} {\bibfnamefont {P.}~\bibnamefont {{Zoller}}},\
  }\bibfield  {title} {\bibinfo {title} {{Chiral quantum optics}},\ }\href
  {https://doi.org/10.1038/nature21037} {\bibfield  {journal} {\bibinfo
  {journal} {\nat}\ }\textbf {\bibinfo {volume} {541}},\ \bibinfo {pages} {473}
  (\bibinfo {year} {2017})}\BibitemShut {NoStop}%
\bibitem [{\citenamefont {{Hill}}\ \emph {et~al.}(2012)\citenamefont {{Hill}},
  \citenamefont {{Safavi-Naeini}}, \citenamefont {{Chan}},\ and\ \citenamefont
  {{Painter}}}]{Hill2012NatCo}%
  \BibitemOpen
  \bibfield  {author} {\bibinfo {author} {\bibfnamefont {J.~T.}\ \bibnamefont
  {{Hill}}}, \bibinfo {author} {\bibfnamefont {A.~H.}\ \bibnamefont
  {{Safavi-Naeini}}}, \bibinfo {author} {\bibfnamefont {J.}~\bibnamefont
  {{Chan}}},\ and\ \bibinfo {author} {\bibfnamefont {O.}~\bibnamefont
  {{Painter}}},\ }\bibfield  {title} {\bibinfo {title} {{Coherent optical
  wavelength conversion via cavity optomechanics}},\ }\href
  {https://doi.org/10.1038/ncomms2201} {\bibfield  {journal} {\bibinfo
  {journal} {Nature Communications}\ }\textbf {\bibinfo {volume} {3}},\
  \bibinfo {eid} {1196} (\bibinfo {year} {2012})}\BibitemShut {NoStop}%
\bibitem [{\citenamefont {{Bochmann}}\ \emph {et~al.}(2013)\citenamefont
  {{Bochmann}}, \citenamefont {{Vainsencher}}, \citenamefont {{Awschalom}},\
  and\ \citenamefont {{Cleland}}}]{Bochmann2013NatPh}%
  \BibitemOpen
  \bibfield  {author} {\bibinfo {author} {\bibfnamefont {J.}~\bibnamefont
  {{Bochmann}}}, \bibinfo {author} {\bibfnamefont {A.}~\bibnamefont
  {{Vainsencher}}}, \bibinfo {author} {\bibfnamefont {D.~D.}\ \bibnamefont
  {{Awschalom}}},\ and\ \bibinfo {author} {\bibfnamefont {A.~N.}\ \bibnamefont
  {{Cleland}}},\ }\bibfield  {title} {\bibinfo {title} {{Nanomechanical
  coupling between microwave and optical photons}},\ }\href
  {https://doi.org/10.1038/nphys2748} {\bibfield  {journal} {\bibinfo
  {journal} {Nature Physics}\ }\textbf {\bibinfo {volume} {9}},\ \bibinfo
  {pages} {712} (\bibinfo {year} {2013})}\BibitemShut {NoStop}%
\bibitem [{\citenamefont {{Andrews}}\ \emph {et~al.}(2014)\citenamefont
  {{Andrews}}, \citenamefont {{Peterson}}, \citenamefont {{Purdy}},
  \citenamefont {{Cicak}}, \citenamefont {{Simmonds}}, \citenamefont
  {{Regal}},\ and\ \citenamefont {{Lehnert}}}]{Andrews2014NatPh}%
  \BibitemOpen
  \bibfield  {author} {\bibinfo {author} {\bibfnamefont {R.~W.}\ \bibnamefont
  {{Andrews}}}, \bibinfo {author} {\bibfnamefont {R.~W.}\ \bibnamefont
  {{Peterson}}}, \bibinfo {author} {\bibfnamefont {T.~P.}\ \bibnamefont
  {{Purdy}}}, \bibinfo {author} {\bibfnamefont {K.}~\bibnamefont {{Cicak}}},
  \bibinfo {author} {\bibfnamefont {R.~W.}\ \bibnamefont {{Simmonds}}},
  \bibinfo {author} {\bibfnamefont {C.~A.}\ \bibnamefont {{Regal}}},\ and\
  \bibinfo {author} {\bibfnamefont {K.~W.}\ \bibnamefont {{Lehnert}}},\
  }\bibfield  {title} {\bibinfo {title} {{Bidirectional and efficient
  conversion between microwave and optical light}},\ }\href
  {https://doi.org/10.1038/nphys2911} {\bibfield  {journal} {\bibinfo
  {journal} {Nature Physics}\ }\textbf {\bibinfo {volume} {10}},\ \bibinfo
  {pages} {321} (\bibinfo {year} {2014})}\BibitemShut {NoStop}%
\bibitem [{\citenamefont {{Bagci}}\ \emph {et~al.}(2014)\citenamefont
  {{Bagci}}, \citenamefont {{Simonsen}}, \citenamefont {{Schmid}},
  \citenamefont {{Villanueva}}, \citenamefont {{Zeuthen}}, \citenamefont
  {{Appel}}, \citenamefont {{Taylor}}, \citenamefont {{S{\o}rensen}},
  \citenamefont {{Usami}}, \citenamefont {{Schliesser}},\ and\ \citenamefont
  {{Polzik}}}]{Bagci2014Natur}%
  \BibitemOpen
  \bibfield  {author} {\bibinfo {author} {\bibfnamefont {T.}~\bibnamefont
  {{Bagci}}}, \bibinfo {author} {\bibfnamefont {A.}~\bibnamefont {{Simonsen}}},
  \bibinfo {author} {\bibfnamefont {S.}~\bibnamefont {{Schmid}}}, \bibinfo
  {author} {\bibfnamefont {L.~G.}\ \bibnamefont {{Villanueva}}}, \bibinfo
  {author} {\bibfnamefont {E.}~\bibnamefont {{Zeuthen}}}, \bibinfo {author}
  {\bibfnamefont {J.}~\bibnamefont {{Appel}}}, \bibinfo {author} {\bibfnamefont
  {J.~M.}\ \bibnamefont {{Taylor}}}, \bibinfo {author} {\bibfnamefont
  {A.}~\bibnamefont {{S{\o}rensen}}}, \bibinfo {author} {\bibfnamefont
  {K.}~\bibnamefont {{Usami}}}, \bibinfo {author} {\bibfnamefont
  {A.}~\bibnamefont {{Schliesser}}},\ and\ \bibinfo {author} {\bibfnamefont
  {E.~S.}\ \bibnamefont {{Polzik}}},\ }\bibfield  {title} {\bibinfo {title}
  {{Optical detection of radio waves through a nanomechanical transducer}},\
  }\href {https://doi.org/10.1038/nature13029} {\bibfield  {journal} {\bibinfo
  {journal} {\nat}\ }\textbf {\bibinfo {volume} {507}},\ \bibinfo {pages} {81}
  (\bibinfo {year} {2014})}\BibitemShut {NoStop}%
\bibitem [{\citenamefont {{Dong}}\ \emph
  {et~al.}(2015{\natexlab{b}})\citenamefont {{Dong}}, \citenamefont {{Fiore}},
  \citenamefont {{Kuzyk}}, \citenamefont {{Tian}},\ and\ \citenamefont
  {{Wang}}}]{Dong2015AnP}%
  \BibitemOpen
  \bibfield  {author} {\bibinfo {author} {\bibfnamefont {C.}~\bibnamefont
  {{Dong}}}, \bibinfo {author} {\bibfnamefont {V.}~\bibnamefont {{Fiore}}},
  \bibinfo {author} {\bibfnamefont {M.~C.}\ \bibnamefont {{Kuzyk}}}, \bibinfo
  {author} {\bibfnamefont {L.}~\bibnamefont {{Tian}}},\ and\ \bibinfo {author}
  {\bibfnamefont {H.}~\bibnamefont {{Wang}}},\ }\bibfield  {title} {\bibinfo
  {title} {{Optical wavelength conversion via optomechanical coupling in a
  silica resonator}},\ }\href {https://doi.org/10.1002/andp.201400110}
  {\bibfield  {journal} {\bibinfo  {journal} {Annalen der Physik}\ }\textbf
  {\bibinfo {volume} {527}},\ \bibinfo {pages} {100} (\bibinfo {year}
  {2015}{\natexlab{b}})}\BibitemShut {NoStop}%
\bibitem [{\citenamefont {{Tian}}(2015)}]{Tian2015AnP}%
  \BibitemOpen
  \bibfield  {author} {\bibinfo {author} {\bibfnamefont {L.}~\bibnamefont
  {{Tian}}},\ }\bibfield  {title} {\bibinfo {title} {{Optoelectromechanical
  transducer: Reversible conversion between microwave and optical photons}},\
  }\href {https://doi.org/10.1002/andp.201400116} {\bibfield  {journal}
  {\bibinfo  {journal} {Annalen der Physik}\ }\textbf {\bibinfo {volume}
  {527}},\ \bibinfo {pages} {1} (\bibinfo {year} {2015})}\BibitemShut {NoStop}%
\bibitem [{\citenamefont {{Forsch}}\ \emph {et~al.}(2020)\citenamefont
  {{Forsch}}, \citenamefont {{Stockill}}, \citenamefont {{Wallucks}},
  \citenamefont {{Marinkovi{\'c}}}, \citenamefont {{G{\"a}rtner}},
  \citenamefont {{Norte}}, \citenamefont {{van Otten}}, \citenamefont
  {{Fiore}}, \citenamefont {{Srinivasan}},\ and\ \citenamefont
  {{Gr{\"o}blacher}}}]{Forsch2020NatPh}%
  \BibitemOpen
  \bibfield  {author} {\bibinfo {author} {\bibfnamefont {M.}~\bibnamefont
  {{Forsch}}}, \bibinfo {author} {\bibfnamefont {R.}~\bibnamefont
  {{Stockill}}}, \bibinfo {author} {\bibfnamefont {A.}~\bibnamefont
  {{Wallucks}}}, \bibinfo {author} {\bibfnamefont {I.}~\bibnamefont
  {{Marinkovi{\'c}}}}, \bibinfo {author} {\bibfnamefont {C.}~\bibnamefont
  {{G{\"a}rtner}}}, \bibinfo {author} {\bibfnamefont {R.~A.}\ \bibnamefont
  {{Norte}}}, \bibinfo {author} {\bibfnamefont {F.}~\bibnamefont {{van
  Otten}}}, \bibinfo {author} {\bibfnamefont {A.}~\bibnamefont {{Fiore}}},
  \bibinfo {author} {\bibfnamefont {K.}~\bibnamefont {{Srinivasan}}},\ and\
  \bibinfo {author} {\bibfnamefont {S.}~\bibnamefont {{Gr{\"o}blacher}}},\
  }\bibfield  {title} {\bibinfo {title} {{Microwave-to-optics conversion using
  a mechanical oscillator in its quantum ground state}},\ }\href
  {https://doi.org/10.1038/s41567-019-0673-7} {\bibfield  {journal} {\bibinfo
  {journal} {Nature Physics}\ }\textbf {\bibinfo {volume} {16}},\ \bibinfo
  {pages} {69} (\bibinfo {year} {2020})}\BibitemShut {NoStop}%
\bibitem [{\citenamefont {{Hu}}\ \emph {et~al.}(2013)\citenamefont {{Hu}},
  \citenamefont {{Xiao}}, \citenamefont {{Liu}},\ and\ \citenamefont
  {{Gong}}}]{HuYW2013FrPhy}%
  \BibitemOpen
  \bibfield  {author} {\bibinfo {author} {\bibfnamefont {Y.-W.}\ \bibnamefont
  {{Hu}}}, \bibinfo {author} {\bibfnamefont {Y.-F.}\ \bibnamefont {{Xiao}}},
  \bibinfo {author} {\bibfnamefont {Y.-C.}\ \bibnamefont {{Liu}}},\ and\
  \bibinfo {author} {\bibfnamefont {Q.}~\bibnamefont {{Gong}}},\ }\bibfield
  {title} {\bibinfo {title} {{Optomechanical sensing with on-chip
  microcavities}},\ }\href {https://doi.org/10.1007/s11467-013-0384-y}
  {\bibfield  {journal} {\bibinfo  {journal} {Frontiers of Physics}\ }\textbf
  {\bibinfo {volume} {8}},\ \bibinfo {pages} {475} (\bibinfo {year}
  {2013})}\BibitemShut {NoStop}%
\bibitem [{\citenamefont {{Metcalfe}}(2014)}]{Metcalfe2014ApPRv}%
  \BibitemOpen
  \bibfield  {author} {\bibinfo {author} {\bibfnamefont {M.}~\bibnamefont
  {{Metcalfe}}},\ }\bibfield  {title} {\bibinfo {title} {{Applications of
  cavity optomechanics}},\ }\href {https://doi.org/10.1063/1.4896029}
  {\bibfield  {journal} {\bibinfo  {journal} {Applied Physics Reviews}\
  }\textbf {\bibinfo {volume} {1}},\ \bibinfo {eid} {031105} (\bibinfo {year}
  {2014})}\BibitemShut {NoStop}%
\bibitem [{\citenamefont {{Li}}\ \emph {et~al.}(2021)\citenamefont {{Li}},
  \citenamefont {{Ou}}, \citenamefont {{Lei}},\ and\ \citenamefont
  {{Liu}}}]{LiBB2021Nanop}%
  \BibitemOpen
  \bibfield  {author} {\bibinfo {author} {\bibfnamefont {B.-B.}\ \bibnamefont
  {{Li}}}, \bibinfo {author} {\bibfnamefont {L.}~\bibnamefont {{Ou}}}, \bibinfo
  {author} {\bibfnamefont {Y.}~\bibnamefont {{Lei}}},\ and\ \bibinfo {author}
  {\bibfnamefont {Y.-C.}\ \bibnamefont {{Liu}}},\ }\bibfield  {title} {\bibinfo
  {title} {{Cavity optomechanical sensing}},\ }\href
  {https://doi.org/10.1515/nanoph-2021-0256} {\bibfield  {journal} {\bibinfo
  {journal} {Nanophotonics}\ }\textbf {\bibinfo {volume} {10}},\ \bibinfo {eid}
  {256} (\bibinfo {year} {2021})}\BibitemShut {NoStop}%
\end{thebibliography}%

\end{document}